\documentclass[manuscript,screen]{acmart}
\AtBeginDocument{%
  }
\newcommand{\gray}{\cellcolor{gray!20}}

\setcopyright{acmlicensed}
\copyrightyear{2018}
\acmYear{2018}
\acmDOI{XXXXXXX.XXXXXXX}
\acmISBN{978-1-4503-XXXX-X/18/06}
\usepackage{pifont}

\usepackage{array}
\usepackage{wrapfig}
\newcommand{\fully}{\ding{52}}
\newcommand{\partly}{\ding{115}}
\newcommand{\notso}{\ding{56}}

\newcolumntype{P}[1]{>{\centering\arraybackslash}p{#1}}

\usepackage{colortbl}

\usepackage{enumitem}
\setlist[itemize]{leftmargin=*}

\begin{document}

\title{Advancing Human-Machine Teaming: Concepts, Challenges, and Applications}

\author{Dian Chen}
\email{dianc@vt.edu}
\orcid{0009-0000-7641-454X}
\author{Han Jun Yoon}
\email{godzmdi93@vt.edu}
\orcid{0009-0006-0499-1245}
\author{Zelin Wan}
\email{zelin@vt.edu}
\orcid{0000-0001-5293-0363}
\affiliation{%
  \institution{Virginia Tech}
  \city{Arlington}
  \state{VA}
  \country{USA}
}

\author{Nithin Alluru}
\email{nithin@vt.edu}
\orcid{0009-0007-3494-5693}
\author{Sang Won Lee}
\email{sangwonlee@vt.edu}
\orcid{0000-0002-1026-315X}
\affiliation{%
  \institution{Virginia Tech}
  \city{Blacksburg}
 \state{VA}
  \country{USA}
}

\author{Richard He}
\email{richardhe789@gmail.com}
\orcid{0009-0008-1440-7764}
\affiliation{%
  \institution{C. G. Woodson High School}
  \city{Fairfax}
 \state{VA}
  \country{USA}}

\author{Terrence J. Moore, Frederica F. Nelson}
\email{terrence.j.moore.civ@army.mil, frederica.f.nelson.civ@army.mil}
\orcid{0000-0003-3279-2965, 0000-0001-8641-384X}
\affiliation{%
 \institution{US DEVCOM Army Research Laboratory}
 \city{Adelphi}
 \state{MD}
 \country{USA}}

\author{Seunghyun Yoon}
\email{syoon@kentech.ac.kr}
\orcid{0000-0001-6264-976X}
\author{Hyuk Lim}
\email{hlim@kentech.ac.kr}
\orcid{0000-0002-9926-3913}
\affiliation{%
 \institution{Korea Institute of Energy Technology (KENTECH)}
 \city{Naju-si,}
 \state{Jeollanam-do}
 \country{Republic of Korea}}

\author{Dan Dongseong Kim}
\email{dan.kim@uq.edu.au}
\orcid{0000-0003-2605-187X}
\affiliation{%
 \institution{The University of Queensland}
 \city{Brisbane, Queensland}
 \country{Australia}}

\author{Jin-Hee Cho}
\email{jicho@vt.edu}
\orcid{0000-0002-5908-4662}
\affiliation{%
 \institution{Virginia Tech}
 \city{Arlington}
 \state{VA}
 \country{USA}}
\renewcommand{\shortauthors}{Chen et al.}

\begin{abstract}
Human-Machine Teaming (HMT) is revolutionizing collaboration across domains such as defense, healthcare, and autonomous systems by integrating AI-driven decision-making, trust calibration, and adaptive teaming. This survey presents a comprehensive taxonomy of HMT, analyzing theoretical models, including reinforcement learning, instance-based learning, and interdependence theory, alongside interdisciplinary methodologies. Unlike prior reviews, we examine team cognition, ethical AI, multi-modal interactions, and real-world evaluation frameworks. Key challenges include explain-ability, role allocation, and scalable benchmarking. We propose future research in cross-domain adaptation, trust-aware AI, and standardized testbeds. By bridging computational and social sciences, this work lays a foundation for resilient, ethical, and scalable HMT systems.
\end{abstract}




\maketitle

\section{Introduction}
\label{sec:intro}

Human-Machine Teaming (HMT) is transforming technology, enabling seamless collaboration between humans and intelligent systems in defense, healthcare, autonomous systems, and industrial automation~\cite{christopher2018navigating, henry2022human}. For example, in autonomous military operations, HMT enables AI-driven decision support for surveillance, target recognition, and real-time strategy adjustments while maintaining human oversight. In healthcare, AI-assisted diagnostic systems and robotic surgical assistants enhance precision and efficiency, improving patient outcomes through adaptive human-AI collaboration~\cite{gillespie2022building}. As AI advances, machines must act as effective teammates rather than mere tools~\cite{greenberg2023foundational}. Despite its potential, HMT research faces challenges, including trust calibration, decision-making autonomy, adaptability, and ethical considerations~\cite{dreslin2023shoulda}. Existing studies focus on specific HMT aspects like multi-robot control, human-agent trust, and cognitive load, but a comprehensive interdisciplinary understanding remains fragmented~\cite{vats2024survey, patel2024give}.

{\bf Primary goal of this survey.}  This work aims to provide a structured and holistic analysis of HMT research by integrating insights from both computational and social sciences. Unlike prior reviews that focus on individual components, such as metrics, safety, or specific teaming models, this work systematically categorizes HMT concepts, challenges, and applications~\cite{damacharla2018common, cleland2022extending}. This survey establishes a taxonomy of HMT systems, analyzes empirical studies on team performance, and examines key theoretical frameworks that drive human-machine collaboration. Furthermore, it presents an evaluation framework encompassing testbeds, performance metrics, and datasets, addressing the need for standardized benchmarking in HMT research~\cite{damacharla2018common}.

\vspace{1mm}
\textbf{Scope of this survey.} This work primarily emphasizes trust, decision-making, cognitive alignment, and evaluation methodologies in human-machine collaboration. The survey focuses on HMT applications in defense, healthcare, robotics, and human-autonomy teaming systems while not covering purely technological advancements in AI algorithms or robotics hardware, as these aspects are extensively reviewed in engineering literature. Instead, this work highlights the interdisciplinary and human-centered dimensions of HMT to guide future research directions and foster interdisciplinary collaboration. The insights provided in this survey will help researchers, engineers, and policymakers design more effective, resilient, and trustworthy HMT systems for real-world applications.

\subsection{Comparison with Existing Surveys}
\label{subsec:similar-survey-papers}

\begin{table*}[t]
\centering
\footnotesize 
\caption{Comparison of our Survey Paper with the Existing Surveys of Human-Machine Teaming Networks}
\label{tab:Comparison}
\vspace{-5mm}
\begin{center}
\begin{tabular}{|p{3.5cm}|P{0.5cm}|P{0.5cm}|P{0.5cm}|P{0.5cm}|P{0.5cm}|P{0.5cm}|P{0.5cm}|P{0.5cm}|P{0.5cm}|P{0.5cm}|P{0.5cm}|P{0.5cm}|P{0.5cm}|}
\hline
\multicolumn{1}{|c|}{\bf Key Criteria} & \cite{murphy2013survey} (2013) & \cite{chen2014human} (2014) & \cite{lasota2017survey} (2017) & \cite{damacharla2018common} (2018) &\cite{gomez2019considerations} (2019) & \cite{warren2020friend} (2020) & \cite{kashima2022trustworthy} (2022)  & \cite{Yang22-thms-review-mhrobotics} (2022)    &   \cite{gebru2022review} (2022)    & \cite{o2022human} (2022) & \cite{berretta2023defining} (2023) & \cite{vats2024survey} (2024) & {\bf Ours} \\
\hline
Key concepts & \partly & \partly & \partly & \partly & \notso & \partly & \notso & \notso & \partly & \partly & \partly & \partly & \fully\\ 
\hline
Key performance factors & \partly & \partly & \partly & \notso & \notso & \partly & \notso & \partly & \partly & \fully & \partly & \partly &  \fully\\ 
\hline
Relationships of the key components & \notso & \partly & \notso & \fully & \partly & \partly & \fully & \fully & \partly & \fully & \partly & \partly & \fully\\ 
\hline
HMT architecture & \notso & \partly  & \notso & \partly & \partly & \partly & \partly & \partly & \notso & \notso & \partly & \partly & \fully \\ 
\hline
Trustworthiness of an HMT system & \notso & \partly & \notso & \notso & \partly & \partly & \fully & \partly & \fully & \partly & \notso & \partly & \fully\\ 
\hline
Extensive review of diverse HMT research and its analysis & \notso  & \partly  & \partly  & \partly  & \partly  & \partly  & \partly  & \partly  & \partly  & \partly  & \partly & \partly & \fully\\ 
\hline
Key theories for HMT research & \partly  & \partly  & \fully  &  \partly  &  \partly  &\fully   &\fully   & \partly  & \partly  & \partly  & \partly & \partly & \fully\\ 
\hline
Concepts, challenges, and applications of an HMT system & \partly & \partly & \partly & \partly & \notso & \partly & \partly & \partly & \partly & \partly & \partly & \partly & \fully \\ 
\hline
Metrics for HMT research & \fully & \notso & \partly & \fully & \partly & \notso & \partly & \partly & \partly & \notso & \notso & \notso & \fully \\ 
\hline
Testbeds for HMT research & \notso & \notso & \partly & \notso & \notso & \notso & \notso & \notso & \partly & \partly & \notso & \notso & \fully \\ 
\hline
Limitations/Insights & \notso & \partly & \partly & \partly & \partly & \partly & \partly & \partly & \partly & \partly & \partly & \partly & \fully \\ 
\hline
Future Research Directions & \notso & \partly & \partly & \partly & \partly & \partly & \fully & \partly & \partly & \partly & \partly & \partly & \fully \\ 
\hline
\end{tabular}

\vspace{1mm}
\fully: Fully addressed; \partly: Partially addressed; \notso: Not addressed at all.
\end{center}
\vspace{-5mm}
\end{table*}
\citet{murphy2013survey} reviewed 42 human-robot interaction (HRI) metrics across human, robot, and system categories, identifying gaps in measuring specialized features that limit generalizability and scalability. Conducted in 2013, their survey excludes recent HRI metrics and focuses on agents' cognitive and autonomous capabilities. \citet{chen2014human} surveyed human performance issues in multi-robot control, emphasizing user-interface design for enhancing human-agent teaming (HAT). However, their findings, derived from controlled lab settings, may not fully generalize complex real-world scenarios. \citet{lasota2017survey} examined safety and robustness strategies in HRI, highlighting the impact of unpredictable environments, human dynamics, and sensing limitations on system states. While technical safety was the focus, the study largely omitted psychological and social aspects like user experience and trust, which are salient or influential elements when humans are involved.

\citet{damacharla2018common} categorized HMT system metrics but did not address evaluation methods, testbeds, or broader applications. \citet{gomez2019considerations} reviewed approaches for optimizing cyber task performance, covering automatic platform assessments, shared mental models, provenance tracking, and user-performance modeling, offering a research roadmap but predating recent developments.  \citet{warren2020friend} analyzed Lethal Autonomous Weapons Systems (LAWS) and HMT in military contexts, focusing on trust, reliability, and AI regulation but lacking empirical validation. \citet{kashima2022trustworthy} examined human-AI trust via Reliability, Availability, and Serviceability (RAS) and ethical considerations but primarily addressed one-way trust in AI, overlooking bidirectional trust and AI’s ethical obligations.

\citet{Yang22-thms-review-mhrobotics} provided a comprehensive review of perception, decision-making, and execution approaches in HMC for robotics, emphasizing safety, efficiency, and collaboration in remote operations, surgery, and assembly. However, evaluation methodologies, such as testbeds and simulations, were not covered. \citet{gebru2022review} examined trust measurement in humans (e.g., heart-rate variability and brain activity) and machines (e.g., robustness, explainability, verifiability) but did not explore broader system resilience. \citet{o2022human} reviewed 76 human-autonomy teaming studies, analyzing research settings, variables, and team composition but found most studies were lab-based with limited real-world applicability. Their survey lacked resilience, safety metrics, and cybersecurity concerns. \citet{berretta2023defining} identified five HAT research clusters, i.e., human factors, task variables, AI explainability, AI-driven robotics, and AI’s impact on human perception, but omitted technology-centered perspectives on HMT advancements. \citet{vats2024survey} explored Large Pre-trained Models in HAT, focusing on AI model improvements, ethical considerations, and applications, but did not cover broader HMT studies.

Table~\ref{tab:Comparison} compares existing HMT surveys based on key criteria. \citet{murphy2013survey} covered metrics but omitted architecture, resilience, and safety. \citet{chen2014human} partially addressed trust and applications but lacked testbeds, while \citet{lasota2017survey} focused on safety without discussing cybersecurity. \citet{damacharla2018common} and \citet{gomez2019considerations} offered limited insights into resilience and cybersecurity. \citet{warren2020friend} and \citet{kashima2022trustworthy} prioritized trust over technical evaluations. \citet{Yang22-thms-review-mhrobotics} covered architecture but omitted trust and resilience, whereas \citet{gebru2022review} emphasized trust measurement over system resilience and safety. \citet{o2022human} analyzed human-autonomy teaming but overlooked resilience and cybersecurity. \citet{berretta2023defining} examined human-AI teaming from a human-centric perspective, while \citet{vats2024survey} focused solely on Large Pre-trained Models in HAT.

Our review fully addresses these dimensions, covering architecture, trust, resilience, safety, metrics, testbeds, multi-domain applicability, and cybersecurity, providing a unified framework. We offer in-depth architectural insights, robust evaluation methods, and emerging threat analyses for HMT systems, delivering actionable recommendations for real-world implementations and future research.

\subsection{Key Contributions} \label{subsec:key-contributions}
Our survey paper made the following {\bf key contributions}:

\textbf{Comprehensive Taxonomy.} This survey establishes a structured taxonomy of Human-Machine Teaming (HMT), categorizing key components, architectures, and trustworthiness factors. Unlike prior reviews focused on specific HMT subdomains or applications, our taxonomy provides a unified framework integrating diverse HMT systems, including human-unmanned vehicle teaming, human-robot collaboration, and AI-driven social media interactions. This broader, systematic classification enables researchers to uncover cross-disciplinary insights that were previously overlooked.

\textbf{Empirical Study Insights.} We analyze empirical studies across multiple HMT dimensions, including team training, autonomy, trust, and shared mental models (SMMs). Compared to existing reviews that primarily focus on conceptual discussions or theoretical models, our work synthesizes findings from both experimental and real-world deployments. By identifying critical success factors such as communication effectiveness, situational awareness, and ethical considerations, this review highlights research gaps and calls for standardized methodologies to improve reproducibility across studies.

\textbf{Evaluation Framework.} This survey presents an in-depth discussion of HMT evaluation methodologies, covering testbeds, performance metrics, and dataset availability. Unlike prior reviews emphasizing qualitative or quantitative assessment techniques, we integrate both perspectives, proposing a holistic evaluation framework. Our review uniquely emphasizes the role of trust, resilience, and ethical constraints in performance assessments, ensuring a more robust and context-aware evaluation of HMT implementations.

\textbf{Multidisciplinary Perspective.} Unlike many prior surveys focusing predominantly on computational or human-centered aspects, our review bridges both perspectives by incorporating social science theories and AI-driven frameworks. We explore how Human-in-the-Loop Reinforcement Learning (HRL) and Instance-Based Learning Theory (IBLT) can enhance adaptive teaming mechanisms. By synthesizing insights from cognitive science, machine learning, and human factors research, this survey offers a more integrative approach to fostering synergistic human-AI collaboration.

\begin{wrapfigure}{r}{0.7\textwidth}
\vspace{-5mm}
    \centering
    \includegraphics[width=\linewidth]{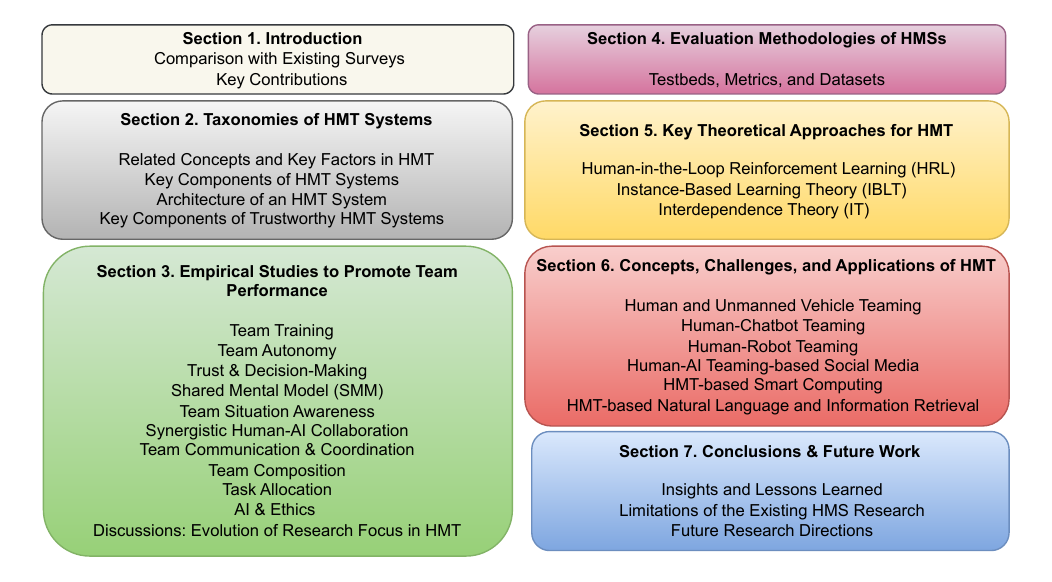}
    \vspace{-8mm}
    \caption{Structure of this review paper.}
    \label{fig:table-of-contents}
    \vspace{-4mm}
\end{wrapfigure}

\textbf{Comparative Strength over Prior Surveys.} Unlike existing HMT reviews, such as \cite{murphy2013survey} on metrics, \cite{chen2014human} on multi-robot control, and \cite{lasota2017survey} on safety, our work provides a broader, more inclusive analysis covering architecture, performance metrics, trust dynamics, and interdisciplinary evaluation. While surveys like \cite{warren2020friend, kashima2022trustworthy} explore trust, they do not fully examine its relationship with team coordination and task allocation in HMT. Our review fills these gaps by systematically comparing diverse HMT applications and integrating key concepts into a unified research framework.

\textbf{Future Research Roadmap.} We outline key future research directions, including enhancing explainability in HMT systems, improving AI-human trust calibration, and mitigating ethical and bias-related challenges. Unlike previous reviews that primarily summarize past findings, our work proposes concrete steps for advancing HMT research. We emphasize the need for cross-sector collaboration between academia, industry, and government to develop standardized benchmarks and evaluation protocols. This roadmap serves as a guideline for ensuring the practical deployment of HMT across various domains, including defense, healthcare, and industrial automation.

The remainder of this paper follows the structure outlined in Fig.~\ref{fig:table-of-contents}.

\section{Taxonomies of HMT Systems} \label{sec:taxonomies}

\subsection{Related Concepts and Key Factors in HMT}
\label{subsec:related-concepts-key-factors-hmt}

{\bf Concept of HMT.}  HMT has been discussed across various related domains, with their definitions discussed below.

\begin{itemize}
    \item \textit{Human-Machine Interaction (HMI).} The interaction and communication between humans and machines (e.g., industrial or transportation systems) via an interface~\cite{johannsen2009human, hoc2000human}, where the human can act as a decision-maker.

    \item \textit{Human-Machine Collaboration (HMC).} A collaborative form of HMI where humans and machines work toward shared goals through interaction, common in health management, surveillance, and manufacturing~\cite{pizon2023human}.

    \item \textit{Human-Machine Teaming (HMT).} The partnership of humans and machines collaborating to achieve a shared objective through interaction and communication~\cite{lyn2019opportunities}.

    \item \textit{Human-Machine Teaming Network (HMTN).} A network of HMTs collaborating under a structured framework (e.g., Cloud-Sharing Network)~\cite{longa2022human}, with each HMT serving as a specialized team member rather than a simple swarm.

    \item \textit{Human-Machine Symbiosis (HMS).} A mutually beneficial relationship between humans, offering subjective expertise, and machines, providing objective computational abilities~\cite{gill2012human}. This collaboration unites complementary strengths to support complex decision-making in human-centered systems.

    \item \textit{Human-Autonomy Teaming (HAT).} The interdependent collaboration of humans and autonomous agents, with distinct roles, working toward shared goals~\cite{o2022human}. HAT agents independently interact with other team members.

    \item \textit{Human-AI Teaming Systems (HATS).} An integrated system of humans and AI models collaborating to achieve shared objectives by combining their respective strengths~\cite{andrews2023role}.

    \item \textit{Human-Robot Systems (HRS).} The physical collaboration between humans and intelligent devices, such as sensors and actuators, combining human dexterity and robotic precision for improved productivity~\cite{ogenyi2021robotcollab}.

    \item \textit{Autonomous Systems.} Systems capable of adapting to unforeseen events during operation~\cite{watson2005autonomous}, featuring self-perception, self-planning, and self-learning. A joint unmanned air combat system exemplifies these capabilities~\cite{automation2021chen}.

    \item \textit{Automation \& Autonomy.} Autonomy, a higher level of automation, is defined by self-determination and independent operation without human intervention~\cite{automation2021chen}, unlike automation, which depends on external control.
\end{itemize}

{\bf Hierarchical Relationships in HMT Systems.}  
\\ \vspace{-4mm}
\begin{wrapfigure}{r}{0.35\textwidth} 
\vspace{-10mm}
    \centering
    \includegraphics[width=\linewidth]{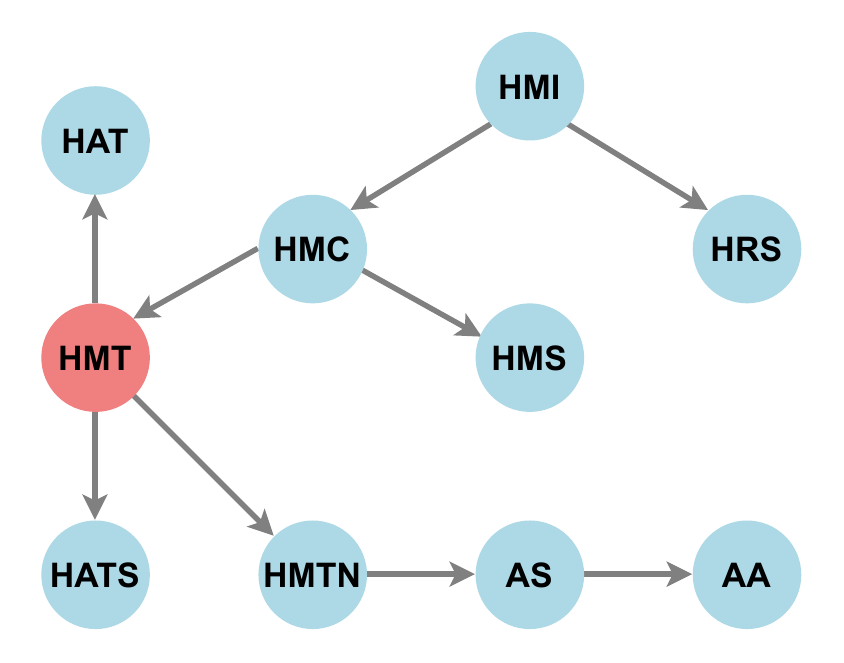} 
    \vspace{-4mm}
    \caption{Hierarchical diagram illustrating relationships and dependencies among HMT-related concepts.}
    \label{fig:hierarchical_diagram_hmt}
\end{wrapfigure}
Fig.~\ref{fig:hierarchical_diagram_hmt} illustrates the interconnections among key HMT concepts. At the center, \textbf{HMT} represents collaborative human-machine interactions for shared objectives. Surrounding it, \textbf{HMI}, \textbf{HMC}, \textbf{HAT}, \textbf{HMS}, \textbf{HATS}, and \textbf{HRS} contribute to different collaboration aspects. Broader concepts like \textbf{HMTN}, \textbf{AS}, and \textbf{AA} provide higher-level structures. The diagram illustrates the hierarchical relationships, providing a clearer understanding of their roles in HMT. \\
{\bf Comparative Analysis of HMT Concepts.} Table~\ref{tab:hmt_comparison} compares key HMT concepts based on collaboration level, autonomy, human involvement, and system focus. Each concept is briefly described, with corresponding attributes in other columns. The table spans basic HMI to advanced human-autonomy and HAT systems (HATS), illustrating varying levels of collaboration, autonomy, and human involvement. The system focus column identifies primary applications, such as decision-making, autonomous systems, or robotics.

\begin{table}[t]
\footnotesize 
\centering
\caption{Comparison of Human-Machine Teaming Concepts}
\label{tab:hmt_comparison}
\vspace{-3mm}
\begin{tabular}{|P{3cm}|P{4cm}|P{2cm}|P{2cm}|P{2cm}|}
\hline
\textbf{Concept} & \textbf{Description} & \textbf{Collaboration Level} & \textbf{Autonomy} & \textbf{Human Involvement} \\ \hline
\textbf{Human-Machine Interaction (HMI)} & Interaction between human and machine via interface. & Low & Low & Decision-making \\ \hline
\textbf{Human-Machine Collaboration (HMC)} & Humans and machines working together for shared goals. & Medium & Medium & Active \\ \hline
\textbf{Human-Machine Teaming (HMT)} & Partnership between humans and machines working toward a shared objective. & High & Low to Medium & High \\ \hline
\textbf{HMT Network (HMTN)} & Network of HMTs collaborating under a framework. & High & Low to Medium & High \\ \hline
\textbf{Human-Machine Symbiosis (HMS)} & Mutually beneficial collaboration leveraging human expertise and machine precision. & High & Low & High \\ \hline
\textbf{Human-Autonomy Teaming (HAT)} & Collaboration between humans and autonomous agents. & High & High & High \\ \hline
\textbf{Human-AI Teaming Systems (HATS)} & Integrated system of humans and AI models working toward shared objectives. & High & High & High \\ \hline
\textbf{Human-Robot Systems (HRS)} & Physical collaboration between humans and robots. & Medium & Medium & Medium \\ \hline
\textbf{Autonomous Systems (AS)} & Systems that operate independently and adapt to events. & Low to Medium & High & Low \\ \hline
\textbf{Automation \& Autonomy (AA)} & Automation is external control, while autonomy is a self-determined operation. & Low & Medium to High & Low \\ \hline
\end{tabular}
\vspace{-7mm}
\end{table}
Key insights from Table~\ref{tab:hmt_comparison} include the increasing complexity of systems as collaboration, autonomy, and human involvement grow. While HMI centers on human control, HMC, HMT, and HAT introduce progressively higher collaboration and autonomy, with HAT and HATS integrating human and autonomous agents. In addition, Table~\ref{tab:hmt_comparison} highlights how these concepts apply across domains, from human-robot collaboration in HRS to decision support in HMS, showcasing the broad scope of human-machine systems.

{\bf Key Factors for Enhancing HMT Systems.} We identify the following for enhancing HMT systems' performance:

\textbf{\em Leadership}: In HMTS, human team leaders must communicate effectively to manage diverse teams~\cite{flathmann2021fostering}. Strong communication aids information search, analysis, and resource management, ensuring tasks are completed on time.

\textbf{\em Team Diversity}: Trustworthy AI systems require diverse teams unified by shared ethics~\cite{smith2019designing}. Diversity in gender, race, education, skills, and problem-solving reduces bias and unintended consequences, leading to fairer solutions.
 \\ \vspace{-4mm}  \begin{wrapfigure}{r}{0.4\textwidth} 
\vspace{-7mm}
\centering
\includegraphics[width=0.4\textwidth]{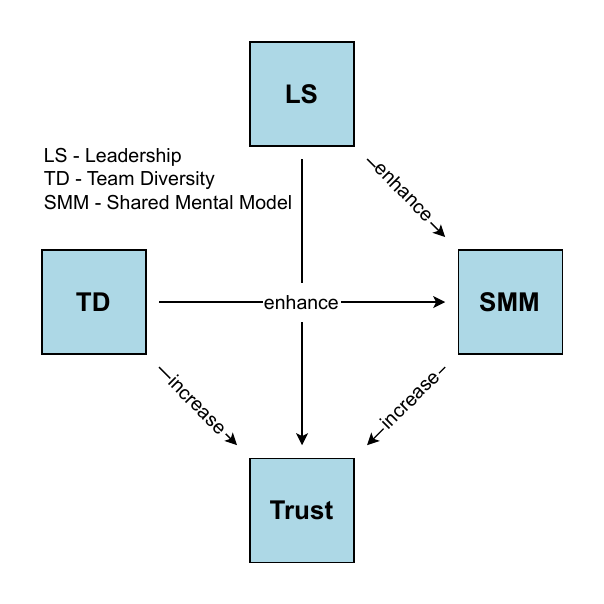} 
\vspace{-7mm}
\caption{Conceptual relationships of the key factors influencing the performance of HMT systems.}    \label{fig:key_factors_hmt}
\vspace{-4mm}
\end{wrapfigure}
\indent \textbf{\em Shared Mental Model (SMM)}: An SMM represents a team's shared understanding of system behavior, enhancing coordination~\cite{andrews2023role}. It consists of (1) the {\em task model}, which defines goals and processes, and (2) the {\em team model}, which outlines roles and expectations. \textit{\em SMM benefits} include minimizing miscommunication, fostering trust, and improving coordination.

\textbf{\em Trust}: Trust in HMTS is studied in: (1) \textit{\em Trust in automation}, an agent’s reliability in uncertain conditions, assessed through trust metrics~\cite{jiunyinin2000}. (2) \textit{\em Mutual trust}, which enhances communication, consistency, shared knowledge, and adaptability for efficient teamwork.

Fig.~\ref{fig:key_factors_hmt} represents a conceptual relationship map of key factors influencing the performance of HMT systems.  It illustrates the interactions between four main concepts: Leadership (LS), Team Diversity (TD), Shared Mental Model (SMM), and Trust. Leadership (LS) influences both SMM and Trust, promoting team alignment and transparency. Team Diversity (TD) enhances SMM by integrating diverse perspectives while increasing {\em trust} by fostering mutual respect among team members. A well-established SMM improves Trust by reducing misunderstandings and increasing predictability in interactions. The arrows labeled {\em enhance} show that Leadership and Diversity contribute to a stronger Shared Mental Model, while the arrows labeled {\em increase} indicate that Leadership and Diversity also increase Trust. This map highlights the interdependencies of these factors and their combined effect on HMT system performance.

\subsection{Key Components of HMT Systems} \label{sec:texonomies-key-components}

To develop an HMT system, recent studies have identified the following core components: 
\begin{itemize}

\item {\em  Agents}: Members of a team consisting of humans and machines where (1) {\em human agents} are categorized as operators, engineers, maintenance personnel, or managers, responsible for controlling machines or making decisions~\cite{johannsen2009human}, and (2) {\em machine agents} are also called a bot or AI agent, refers to any dynamic technical system (e.g., autonomous systems, decision-making, or decision support systems)~\cite{johannsen2009human}.

\item {\em  Role}: A container for an agent with the required qualifications~\cite{madni2018architectural}, assigned based on task-specific requirements.

\item {\em  Inputs}: Categorized by their focus, such as humans, machines, or contexts~\cite{stowers2017framework}.
(1) {\em  Human input}: Includes cognitive competencies (e.g., skills, memory, expertise) and interpersonal traits (e.g., personality, values), impacting interaction quality.  
(2) {\em  Machine input}: Includes automation level, reliability, and transparency, influencing trust, workload, and system performance.  
(3) {\em  Contextual input}: Covers environmental factors and task variables, including external conditions, task type, complexity, and load.

\item {\em  Human-Machine Interface (HMI)}: Enables interaction between humans and cyber-physical elements through displays, controls, and information exchange. Effective HMIs support situational awareness, transparent machine operations, appropriate intervention, and manageable cognitive workload~\cite{madni2018architectural, wachs2010analytical}.

\item {\em  Task}: A job performed by an agent assigned to a role. Humans and machines share task loads and switch roles as needed. Tasks are reallocated based on outcomes and evaluations in dynamic environments~\cite{madni2018architectural, hu2017optimal}.

\item {\em  Environment}: Contains factors and conditions affecting task performance. In dynamic environments, task allocation depends on context, availability of agents, and human cognitive and emotional states~\cite{madni2018architectural}.

\item {\em  Disrupting Events}: External or internal disturbances requiring human-machine teams to adapt. Such events can impact environments, altering consequences and task sharing. Models must account for these disruptions to manage task allocation effectively~\cite{madni2018architectural, connors1998teaming}.

\item {\em  Adapting Actions}: Modifications to structure, sequence, or HMT parameters in response to disruptions. Effective adaptation prevents poor outcomes and improves task evaluations, ensuring robust performance~\cite{madni2018architectural, dubois2020adaptive}.

\end{itemize}

\subsection{Architecture of an HMT System} \label{subsec:architecture-hmts}

{\bf Queuing Network-Model Human Processor (QN-MHP).} This is designed to model multitask human performance by integrating queuing networks with symbolic cognitive models. It effectively models dual tasks, such as in-vehicle map reading and steering, demonstrating its potential for cognitive modeling. Despite its novel approach, QN-MHP lacks accuracy in predicting task durations under certain conditions, such as varying glance durations on sharp curves, and does not address speed control or complex road geometry adjustments \cite{liu2006multitask}.

{\bf Cyber-Physical Human System (CPHS).} This explored options in simulated contexts. The architecture integrates human electrophysiological signals to infer intent and sensors to capture environmental changes processed by a controller. This feedback loop mirrors an adaptive CPHS with tight coupling and strict temporal constraints \cite{madni2018architectural}.

{\bf HMI for Active and Assisted Living Solutions.} This approach emphasizes HMI as a critical factor in adopting active and assisted living solutions. Using a context-based approach, it aims to enhance HMS efficacy, adaptability, and performance. Experiments with the INRIA dataset demonstrated that adaptable actions to dynamic contexts improve system performance when algorithms are sequenced. However, the approach is impractical for complex systems requiring expert analysis and lacks details on distributed programming for large-scale systems \cite{quintas2018architecture}.

{\bf Assistive Robotic System for People with Physical Disabilities.} This system enables simple tasks through a semi-autonomous decision-and-control architecture. It features soft robotics, combinable skill libraries, assistive sensing, and multimodal feedback. Components include lightweight design, impedance control, collision detection, and virtual workspace boundaries. However, it is limited to simple tasks and cannot handle complex scenarios \cite{vogel2015assistive}.

{\bf \underline{PRE}dictive \underline{SEN}sorimotor \underline{C}ontrol and \underline{E}mulation Architecture (PRESENCE).} The PRESENCE architecture was introduced for natural speech interactions in HMI. Inspired by neurobiology, it integrates hierarchical feedback control based on \emph{perceptual control theory} (PCT), where speakers and listeners align perceptions and intentions through closed-loop feedback. While innovative at the time, the study's settings are outdated compared to modern HMI research \cite{moore2007presence}.

\begin{figure}[t]
\centering
\includegraphics[width=0.9\textwidth]{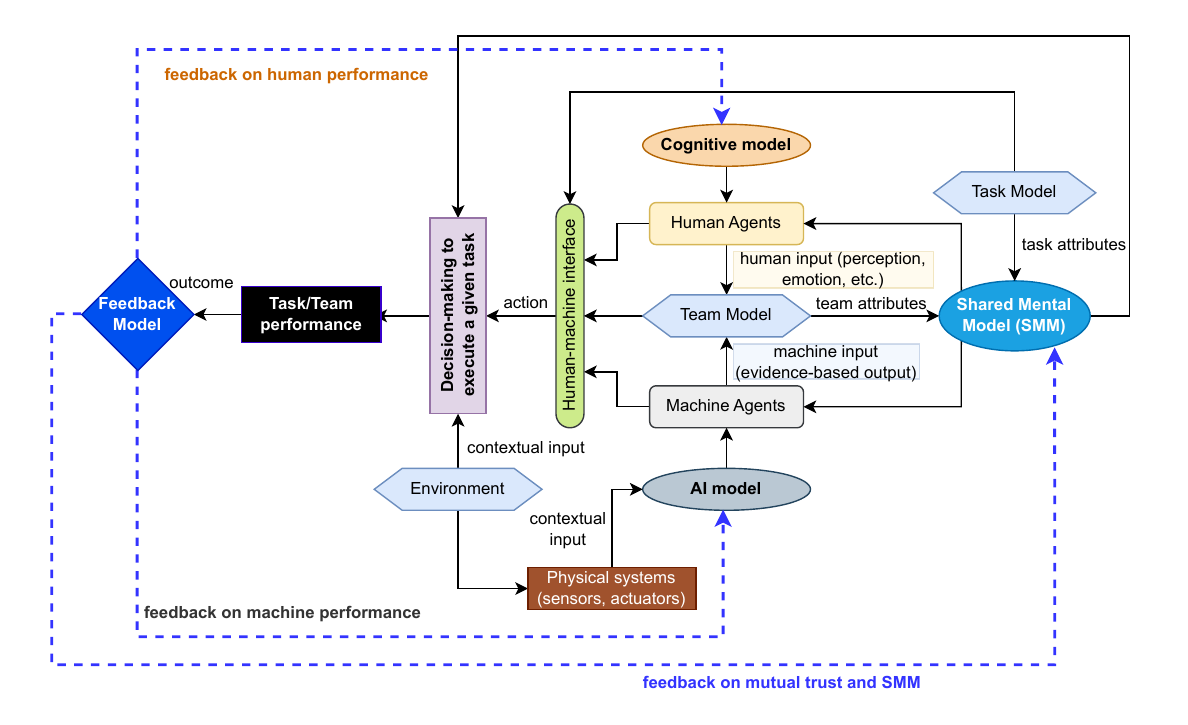}
\caption{Architecture of a human-machine teaming system based on its core components.}
\label{fig:hmt-architecture}
\vspace{-5mm}
\end{figure}
{\bf Proposed HMT Architecture.} Building on the core components of HMT systems, we designed an HMT architecture, illustrated in Fig.~\ref{fig:hmt-architecture}, where each component is essential for optimizing task and team performance. This architecture integrates human and machine agents, an HMI for bidirectional communication, and SMMs to align team roles and task objectives. Inputs from human, machine, and contextual sources and continuous feedback loops enable dynamic adaptability to environmental and task changes. The architecture emphasizes trustworthy interactions, leveraging the synergy of human cognitive capabilities and machine-driven precision to achieve effective task execution. Fig.~\ref{fig:hmt-architecture} demonstrates a comprehensive framework for enhancing HMT systems' operational efficiency and resilience by addressing adaptability and trust as central elements.

\subsection{Key Components of Trustworthy HMT Systems}  \label{sec:key-components-trustworthy-HMTSs}

Developing trustworthy HMT systems requires addressing key challenges~\cite{lyn2019opportunities}. Understanding user needs in dynamic contexts is essential, particularly for complex tasks like cybersecurity. Seamless collaboration depends on integrating human decision-making models with machine analytics. AI security and reliability must be ensured by mitigating adversarial attacks and biases. Transparency and explainability foster trust between human and machine agents. Successful collaboration in high-pressure environments depends on strong interactions and a shared team framework to improve communication and coordination. Building on these challenges, we discuss the key factors necessary for developing trustworthy HMT systems, focusing on their role in enhancing reliability and collaboration.

\begin{wrapfigure}{r}{0.6\textwidth}
\vspace{-5mm}
    \centering
    \includegraphics[width=\linewidth]{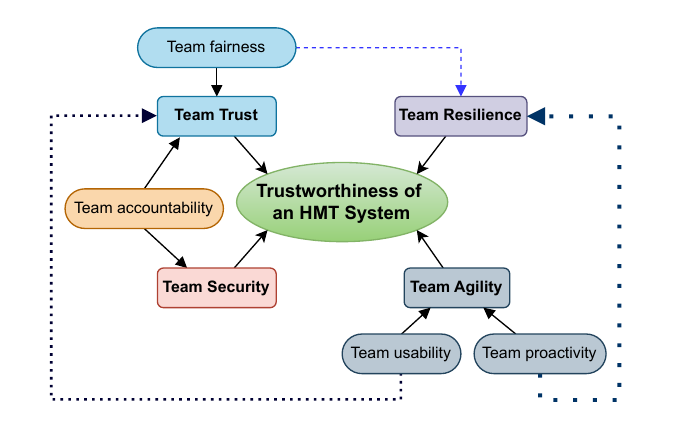}
\vspace{-10mm}    
    \caption{Attributes of a trustworthy HMT system based on the STRAM framework.}
    \label{fig:hmt-trustworthiness}
\vspace{-5mm}    
\end{wrapfigure}

{\bf Key Properties of Trustworthy HMT Systems.}  Via extensive literature review, we identified the following properties essential for building a trustworthy HMT system:
\begin{itemize}
\item {\em Team Resilience}: Resilience ensures robustness in industrial production and preparedness for disruptions. Adaptability is key, requiring smooth adjustments in responsibilities and work divisions. Effective HMT systems must support flexible task allocations and role changes while remaining resilient to environmental and system changes~\cite{kaasinen2022smooth}.

\item {\em Team Fairness}: Fairness in HMT systems emphasizes non-discrimination and unbiased AI models that are inclusive and representative~\cite{pflanzer2023ethics}. Fairness metrics for human-robot teams include equality of workload (equal subtasks regardless of type or capability), equality of capability (matching subtasks by time, accuracy, and difficulty), and equality of task type (equal distribution of subtasks within the same category)~\cite{fairness_chang}.

\item {\em Team Proactivity}: Machine behavior can be reactive, active, or proactive~\cite{edgar2023improving}. Reactive agents respond to requests, active agents provide information as needed, and proactive agents initiate tasks, remind teammates, and adapt to situations. Proactive robots improve efficiency and performance in demanding environments~\cite{edgar2023improving}.

\item {\em Team Usability}: Team usability refers to how effectively humans collaborate with AI to achieve goals~\cite{smith2019designing}. High usability fosters trust and willingness to engage with AI teammates, enhancing human-machine collaboration.

\item {\em Team Trust}: Team trust reflects humans’ confidence in autonomous systems and willingness to interact~\cite{ibrahim2022trust}. Trust is influenced by system reliability, transparency, predictability, and proactive assistance~\cite{lyons2019trust}. Mutual trust requires machines to estimate human trust levels and adapt behavior to ensure safety and interaction quality~\cite{alhaji2020toward}.

\item {\em Team Accountability}: Accountability entails shared responsibility between humans and AI, ensuring transparency and enabling evaluation of decisions and actions~\cite{chen2014human, lopez2023complex}.

\item {\em Team Security}: Team security is the ability of human and machine teammates to counter attacks while maintaining performance. Key aspects include machine involvement, role transparency, and action traceability~\cite{assaad2023ethics, brown2019designing}. Simple decision rules and high machine involvement enhance security, but poor coordination weakens it.

\item {\em Team Agility}: Agility measures a team’s ability to adapt to changing task requirements and unforeseen obstacles~\cite{demir2021modeling}. Agile teams emphasize collaboration over individual assignments, enabling interactive decision-making and adaptability without requiring a designated leader.
\end{itemize}
According to the metric framework {\tt STRAM}~\cite{cho2019stram}, which defines the key metric attributes of a computer-based trustworthy system ({\em \underline{S}ecurity}, {\em \underline{T}rust}, {\em \underline{R}esilience}, and {\em \underline{A}gility}), we define the trustworthiness of an HMT system based on team security, trust, resilience, and agility, as shown in Fig.~\ref{fig:hmt-trustworthiness}. These attributes form the backbone of a reliable and effective human-machine teaming framework. Meanwhile, auxiliary attributes like team usability, fairness, accountability, and proactivity complement and enhance these core elements through specific interdependencies.

Team usability enhances agility by ensuring intuitive human-machine interactions that adapt quickly to changing tasks while supporting resilience by reducing cognitive load and improving coordination under stress. Fairness bolsters trust and resilience through equitable task allocation and unbiased decision-making, preventing overload and maintaining balance. \\ \vspace{-4mm} 
Accountability reinforces trust and security by enabling transparent, traceable decisions, mitigating errors, and enhancing reliability. Proactivity drives agility and resilience by dynamically anticipating changes and addressing challenges, fostering trust through foresight and reliability.

These auxiliary attributes collectively strengthen the core dimensions of team security, trust, resilience, and agility. For example, usability and proactivity enhance agility, while fairness and accountability reinforce trust and resilience. Recognizing these interdependencies is key to designing HMT systems that combine technical robustness with ethical integrity, enabling effective human-machine collaboration in complex environments.

\section{Empirical Studies to Promote Team Performance}\label{sec:topic-based-research}
This section categorizes subsections by key HMT topics that enhance team performance. Detailed discussions of reviewed papers are in the supplement (Appendix A) due to space constraints.

\subsection{Team Training}\label{subsec:team-training}

\textbf{Key Ideas.} \citet{johnson2021impact} examined the impact of communication and calibration on HAT performance using a mixed testbed of real and simulated platforms. They tested three training approaches, including control, coordination, and calibration, showing that coordination improved trust but did not outperform control-trained teams. Limited statistical power constrained conclusions. \citet{harpstead2023speculative} applied speculative game design to study HMT dynamics, focusing on roles, structures, and coordination. While effective for exploring team interactions, the approach lacked details on AI training and relied on simplified game environments.  

\citet{myers2019} introduced autonomous synthetic teammates (ASTs) to reduce scheduling and training costs, balancing human-like behavior with task performance. However, cognitive fidelity limitations affected their effectiveness in complex tasks. \citet{o2023human} proposed the Input-Mediator-Output (IMO) model to structure HMT research, though it overlooked bidirectional feedback loops essential in dynamic interactions. \citet{endsley2024taxonomy} developed an HMT taxonomy advocating for collaborative autonomous systems over autonomy-centric designs. \citet{cleland2024human} extended the MAPE-K feedback loop to MAPE-KHMT, integrating adaptive human-machine interactions. \citet{lematta2024practical} highlighted HMT assurance challenges, using General Motors Cruise robotaxis to illustrate trust and safety failures due to poor design, impacting emergency operations and human interaction reliability.  

\textbf{Merits.}  The existing approaches to team training using HMT have advanced HMT across multiple domains, focusing on collaboration, communication, and decision-making. A key theme is structured communication and calibration, where training enhances situational awareness and trust in AI teammates~\cite{johnson2021impact}. Similarly, role allocation and task adaptability have been explored in speculative game design and asymmetric cooperative tasks, providing insights into team coordination and knowledge distribution~\cite{harpstead2023speculative}. Integrating synthetic agents, such as Autonomous Synthetic Teammates (ASTs), optimizes training efficiency while balancing performance and human-like behavior~\cite{myers2019}. Methodologies for categorizing HMT interactions have been refined, such as the Input-Mediator-Output (IMO) model~\cite{o2023human} and MAPE-KHMT feedback integration~\cite{cleland2024human}.  Advancements in HMT assurance and real-world validation are evident in AI deployment failures, such as the General Motors Cruise robotaxis case~\cite{lematta2024practical}. These studies highlight the importance of transparent interaction mechanisms, system reliability, and robust human-AI collaboration in safety-critical applications.  

\textbf{Limitations.} Despite advancements, key challenges remain. Many studies rely on simulated environments like Minecraft-based experiments \cite{chiou2021towards} or grid-world tasks \cite{harpstead2023speculative}, limiting real-world applicability. Empirical testing in operational environments is needed to assess scalability and effectiveness.  While AI transparency and trust calibration are emphasized, studies often focus on conceptual models without experimental validation. Frameworks like \cite{o2023human} and MAPE-KHMT \cite{cleland2024human} improve HMT modeling but lack systematic deployment in real-world decision-making.

Human-centered training remains underexplored. Studies on coordination and training \cite{johnson2021impact} do not fully address non-expert adaptation, raising concerns about accessibility and usability in dynamic environments. HMT failure analysis is often case-specific without scalable solutions. \cite{lematta2024practical} highlights trust and safety failures in autonomous vehicles but lacks a structured framework for improving AI-driven systems. Addressing these challenges requires empirical validation, better AI training, and human-centered frameworks to enhance HMT systems' robustness and viability.

\subsection{Team Autonomy} \label{subsec:team-performance}

\textbf{Key Ideas.}  Recent studies examine the role of AI teammates in HMT, emphasizing performance improvement, cognitive adaptation, and optimized task execution. AI teammates have been shown to improve team coordination, mediate interactions, and positively influence human perceptions of human-AI teams~\cite{schelble2023investigating}. Spatial awareness also plays a crucial role in team cognition, as varying levels of awareness impact mental model similarity, decision-making efficiency, and overall team performance~\cite{schelble2022see}. Beyond team cognition, adaptability is an essential trait for AI teammates, where autonomous agents must adjust their behaviors dynamically based on evolving task demands and human interaction patterns~\cite{Hauptman23-autonomy}. AI-driven automation has also been leveraged in large-scale data analysis, such as in cybersecurity, where machine learning and natural language processing (NLP) techniques enable efficient taxonomy construction with reduced human effort~\cite{mahaini2019building}. These studies collectively highlight the increasing integration of AI teammates in diverse domains, from emergency response coordination to cognitive adaptability and automated data classification.  

\textbf{Merits.}  The reviewed works demonstrate AI's potential to enhance HMT performance by improving task execution, cognitive adaptability, and coordination efficiency. Studies on AI teammates reveal their ability to mediate interactions and foster human-AI collaboration, indicating that well-integrated AI systems can support human decision-making and situational awareness~\cite{schelble2023investigating}. The examination of spatial awareness provides valuable insights into the cognitive processes underlying human-agent interactions, emphasizing the importance of shared mental models in decision-making~\cite{schelble2022see}. Further, research on autonomy levels in AI teammates highlights the need for adaptive decision-making models that align AI behavior with evolving team dynamics~\cite{Hauptman23-autonomy}. Automated data analysis in cybersecurity showcases the benefits of integrating AI for large-scale information processing, demonstrating how human expertise and NLP techniques can collaboratively generate structured taxonomies with improved accuracy and efficiency~\cite{mahaini2019building}. These contributions collectively advance the understanding of how AI teammates can improve human performance, decision-making, and large-scale knowledge organization in complex environments.  

\textbf{Limitations.}  Despite their contributions, the reviewed works face notable limitations. Many studies rely on controlled simulations with simplified tasks, such as abstract game-based environments or pre-defined operational conditions, which limit their applicability to real-world HMT scenarios~\cite{schelble2023investigating, schelble2022see}. The reliance on experimental settings without explicit communication between AI teammates and humans also constrains the generalizability of these findings, as real-world tasks often involve more complex, bidirectional interactions. In addition, research on AI adaptability and autonomy levels is often based on hypothetical questionnaire-based studies, which may not fully capture the complexities of real-time decision-making~\cite{Hauptman23-autonomy}. While AI-driven taxonomy construction in cybersecurity presents promising applications, its effectiveness in other domains remains uncertain, and its reliance on structured datasets raises concerns about adaptability to unstructured or rapidly changing information landscapes~\cite{mahaini2019building}. These limitations highlight the need for empirical validation in real-world environments, better integration of communication mechanisms in human-agent teaming, and broader assessments of AI-driven automation across multiple domains.

\subsection{Trust \& Decision-Making} \label{subsec:trust}

\textbf{Key Ideas.}  Recent studies explore the role of trust, decision-making, and interaction dynamics in HMT, particularly in high-stakes environments like medical care, emergency response, and military operations. Machine learning-based systems such as the {\em Targeted Real-time Warning System} (TREWS) demonstrate effectiveness in clinical decision-making but suffer from low user trust due to a lack of explainability \cite{henry2022human}. To address such trust issues, architectures like the EASE framework \cite{bersani2023towards} integrate statistical and machine learning models to enhance transparency, while frameworks like ITA \cite{centeio2024interdependence} evaluate trust along competence, willingness, and external factors to optimize task allocation. Trust formation in HMT is also influenced by perception, cognitive alignment, and system reliability \cite{zhang2020anideal, Demir2021Exploration}. Studies on "swift trust" \cite{haring2021applying} suggest that appearance and functionality play crucial roles in early trust formation in human-robot teams, though long-term trust requires robust interactions. Furthermore, trust levels impact team performance, as in RPAS simulations where low trust in autonomous agents created negative feedback loops and hindered collaboration \cite{mcneese2021trust}. 

\textbf{Merits.}  The reviewed works emphasize the complex interplay between trust, decision-making, and team cognition in HMT. Machine learning models such as TREWS demonstrate how AI-driven systems can enhance decision-making in critical environments \cite{henry2022human}. Studies on interaction dynamics reveal how trust evolves over time, linking positive AI experiences to increased human trust and collaboration \cite{zhang2020anideal, Demir2021Exploration}. Trust assessment frameworks, including ITA and the military-oriented HAT framework \cite{centeio2024interdependence, maathuis2024trustworthy}, provide structured approaches to designing reliable HMT systems by integrating ethical, societal, and operational considerations. Research on trust in team composition further highlights how mixed human-AI teams influence cognitive alignment and coordination \cite{MUSICK2021106852, schelble2022let}, offering insights into the impact of agent presence on human cognition. These studies collectively advance the field by identifying key factors that shape trust in HMT and proposing methodologies to improve human-agent collaboration.  

\textbf{Limitations.}  Despite their contributions, these studies face notable limitations. Many rely on controlled simulations with simplified tasks, such as emergency response simulations \cite{schelble2022let} and RPAS scenarios \cite{mcneese2021trust}, which limit real-world applicability. Small sample sizes, as seen in studies on medical AI adoption \cite{henry2022human} and sentiment in multi-agent teams \cite{MUSICK2021106852}, reduce generalizability. Trust formation research often lacks empirical validation beyond theoretical models, as demonstrated in studies on swift trust \cite{haring2021applying}. Moreover, some approaches focus on static trust measurement rather than evolving trust dynamics, overlooking how trust adapts to system learning and behavioral changes over time. The challenge of generalizing findings across domains is evident in cybersecurity-focused models like EASE \cite{bersani2023towards} and AI perception studies with unbalanced participant demographics \cite{zhang2020anideal}. Future research must expand empirical validation, integrate dynamic trust modeling, and assess trust frameworks across diverse operational settings to ensure HMT systems are both adaptable and practically viable.

\subsection{Shared Mental Model (SMM)} \label{subsec:smm}

\textbf{Key Ideas.}  SMMs play a crucial role in aligning team understanding, facilitating effective communication, and enhancing HMT performance. SMMs can be categorized into task models, which focus on goals, required skills, and workload distribution, and role models, which define team members' capabilities and responsibilities~\cite{andrews2023role}. Empirical studies have shown that shared task and team mental models improve team dynamics and interactions, as demonstrated in flight-combat simulations~\cite{mathieu2000influence}. Research on HMT applications in USAR teams highlights the importance of shared cognition and communication efficiency, with natural language improving performance in Minecraft simulations~\cite{demir2020understanding}. Industry 4.0 adaptation strategies further reinforce the role of design thinking and strategic alignment in team performance~\cite{de2023managerial}. Proactive behaviors in robots, such as information sharing and joint task engagement, enhance team effectiveness under high cognitive loads~\cite{edgar2023improving}. Computational frameworks have also been proposed to enable artificial agents to develop and maintain SMMs, thereby improving coordination and team coherence~\cite{scheutz2017framework}.  

\textbf{Merits.}  Studies consistently demonstrate that well-structured SMMs enhance team coordination, communication, and performance across various HMT applications. Flight-combat simulations validate the influence of shared task and role models in improving team efficiency~\cite{mathieu2000influence}. The integration of proactive robot behaviors with SMMs in virtual reality simulations has shown promising results in mitigating cognitive overload and aligning human-robot goals~\cite{edgar2023improving}. In human-agent teams, planning has been shown to improve SMMs, enhancing coordination and reducing errors under high workloads~\cite{stout1999planning}. Furthermore, frameworks enabling artificial agents to maintain SMMs offer computational approaches for optimizing mixed human-agent teams~\cite{scheutz2017framework}. Studies on human-agent communication highlight the role of iterative cognition development and structured dialogue in improving decision-making and task execution~\cite{schelble2022let}.  

\textbf{Limitations.}  Despite strong theoretical and experimental support, research on SMMs in HMT faces notable limitations. Many studies rely on controlled simulations, such as Minecraft-based USAR tests~\cite{demir2020understanding} and virtual reality environments~\cite{edgar2023improving}, limiting their generalizability to real-world applications. Behavioral strategies for Industry 4.0 adaptation are based primarily on qualitative data, which lacks empirical validation~\cite{de2023managerial}. The proposed computational frameworks for artificial agents do not fully capture the complexities of human cognition, which may hinder team coherence in real-world scenarios~\cite{scheutz2017framework}. In addition, human-agent team studies often fail to assess agents' mental models directly, restricting deeper insights into HAT cognition and decision-making processes~\cite{schelble2022let}. To address these gaps, future research must integrate real-world testing, refine computational models for SMM adaptation, and explore scalable strategies for improving shared cognition in dynamic environments.

\subsection{Team Situation Awareness} \label{subsec:tsa}

\textbf{Key Ideas.} Team Situation Awareness (TSA) is crucial for HMT performance, shaping communication, knowledge sharing, and coordination in dynamic environments. It aggregates individual situational awareness, where shared knowledge enhances stability, while dynamic settings require timely coordination \cite{salas2017situation}. Measuring TSA remains challenging, as methods like SAGAT provide robust assessments but risk disrupting task dynamics \cite{endsley1995measurement}. UAV-based HMT studies show proactive information sharing improves TSA, though synthetic environments and language constraints limit applicability \cite{demir2019evolution, demir2017tsa}. RPAS-based research highlights TSA’s role in performance, though machines struggle with non-textual data \cite{mcneese2021team}. Transparency models like the SAT framework enhance human-AI collaboration but introduce trade-offs in workload and decision latency \cite{chen2017tsa}. TSA models for collaborative driving, such as the SSA framework, improve trust alignment but assume hierarchical relationships, limiting real-world applicability \cite{rinta2020tsa}.

\textbf{Merits.} TSA research offers valuable insights into how shared cognition and proactive communication enhance coordination in HMT. Transparency in AI decision-making improves human understanding, as demonstrated by the SAT model’s impact on trust and collaboration \cite{chen2017tsa}. Proactive information sharing strengthens TSA effectiveness in UAV- and RPAS-based HMTs, reinforcing structured communication’s role in team performance \cite{demir2019evolution, mcneese2021tsa}. Empirical findings highlight that TSA evolution in synthetic RPAS teams improves coordination over time \cite{mcneese2021tsa}. Collaborative driving research reduces automation over-reliance by aligning human trust with system capabilities \cite{rinta2020tsa}. TSA research continues to refine decision-support frameworks, laying the foundation for scalable, real-time solutions.

\textbf{Limitations.} TSA research in HMT faces key limitations. Many studies rely on controlled simulations, such as UAV-based synthetic environments \cite{demir2019evolution} and RPAS-based testing \cite{mcneese2021tsa}, limiting real-world applicability. Text-based machine agents struggle with multimodal interactions, reducing effectiveness in dynamic teams \cite{mcneese2021team}. TSA measurement methods like SAGAT must balance accuracy with operational flow, raising ecological validity concerns \cite{salas2017situation}. Transparency models such as SAT enhance understanding but increase cognitive workload and decision latency, potentially impacting real-time performance \cite{chen2017tsa}. The SSA model assumes hierarchical team structures, oversimplifying automation trust dynamics where goal alignment and conflict resolution play critical roles \cite{rinta2020tsa}. Future research should integrate real-time TSA assessment, improve machine adaptability, and ensure transparency mechanisms support rather than burden human decision-making.

\subsection{Synergistic Human-AI Collaboration} \label{subsec:synergy-collab}

\textbf{Key Ideas.}  HMT enhances cyber malware detection by integrating human cognition with machine learning for pattern recognition. Expert knowledge is converted into visual ontologies and graph structures, enabling collaborative analysis of malware distribution networks (MDNs)~\cite{cai2020perceiving}. Human feedback refines machine learning algorithms, improving malware detection, mapping, and prediction while identifying critical vulnerabilities like hubs and bridges. This approach leverages human expertise and AI analytics to strengthen cyber defense strategies and adaptive threat mitigation.  

\textbf{Merits.}  The study demonstrates the benefits of combining human intelligence with AI to enhance malware detection and analysis. Visual ontologies and graph structures offer an intuitive framework for malware analysis, improving interpretability for both humans and machines~\cite{cai2020perceiving}. Human-in-the-loop feedback enables adaptive learning, allowing systems to refine detection capabilities over time. This integration supports proactive cybersecurity measures, highlighting HMT's potential in complex threat environments.  

\textbf{Limitations.}  Despite its strengths, this approach faces scalability challenges due to reliance on high-quality human input, limiting automation and requiring expert involvement~\cite{cai2020perceiving}. Dependence on structured ontologies and graph-based models may hinder adaptability to evolving malware techniques. Data privacy concerns and computational costs further constrain implementation. Future research should reduce human dependency while maintaining accuracy, improve AI adaptability, and optimize computational efficiency for large-scale cybersecurity applications.

\subsection{Team Communication \& Coordination} \label{subsec:team-communication}

\textbf{Key Ideas.} Effective communication is critical for collaboration in HMT, enhancing situational awareness, coordination, and adaptability~\cite{endsley1999level}. Transparent interactions foster trust and improves system performance~\cite{lee2004trust}, but existing strategies often fail to adapt to evolving technologies and user needs~\cite{parasuraman2008situation}. Research on search and rescue missions indicates that structured, on-demand explanations are as effective as frequent communication in improving situational awareness~\cite{chiou2021towards}. NLP-enhanced autonomous agents improve operational interactions but remain vulnerable to failures such as automation breakdowns and cyber-attacks~\cite{demir2019evolution}. Studies on verbal interactions in HATs suggest that increased exchanges improve performance; however, these findings are primarily based on controlled simulations and need validation in real‐world settings~\cite{demir2017team}.

\textbf{Merits.} Research consistently highlights communication as a key factor in trust, situational awareness, and coordination. Optimized communication strategies enhance team efficiency, with moderate information exchange yielding the best performance~\cite{demir2019evolution}. Machine learning-driven interfaces improve interaction quality and reduce cognitive burdens on operators~\cite{moore2007presence}. In high-stakes environments, selective communication strategies balance information delivery, preventing overload while maintaining awareness~\cite{chiou2021towards}. Advances in speech-based HMIs enable cooperative behavior through mutual intention recognition, addressing limitations in conventional systems~\cite{moore2007presence}.  

\textbf{Limitations.} Many studies rely on simulations, limiting applicability to dynamic real-world environments~\cite{chiou2021towards, demir2017team}. Research often assumes rigid hierarchical automation structures, failing to address adaptive interactions~\cite{endsley1999level}. Although NLP-enhanced systems improve interactions, they present usability challenges for untrained users—necessitating improved interface design and additional training~\cite{demir2019evolution}. Speech-based HMIs demand precise parameter tuning, complicating large-scale deployment~\cite{moore2007presence}. Verbal behavior studies lack validation across diverse scenarios, necessitating empirical research on adaptive communication frameworks and scalable solutions for varying expertise levels.

\subsection{Team Composition} \label{subsec:team-composition}

\textbf{Key Ideas.} Human-autonomy teaming (HAT) research highlights challenges in communication, coordination, and team cognition. AI struggles with natural language processing and does not effectively leverage nonverbal cues, which humans rely on heavily~\cite{mcneese2021my}. NeoCITIES experiments show AI-inclusive teams achieve higher TSA, while all-human teams exhibit superior cognition~\cite{mcneese2021my}. Coordination effectiveness varies by environment, with synthetic teams showing performance fluctuations under shifting conditions~\cite{demir2018team}. UAS task studies confirm all-human teams often outperform AI-integrated teams due to stronger coordination~\cite{mcneese2018teaming}. Team behaviors shift when humans are informed of AI roles, influencing social interactions but not necessarily improving performance~\cite{walliser2019team}. AI's impact depends on context, shaping team preferences and decision-making~\cite{flathmann2023examining}.  

\textbf{Merits.} Research shows AI enhances TSA and structured coordination in HATs, particularly with reinforcement learning-based decision support~\cite{mcneese2021my}. RPAS and CERTT-II studies confirm AI improves coordination in structured teams under stable conditions~\cite{demir2018team}. Human perceptions of AI shape interaction strategies, reinforcing the importance of trust calibration in HMT~\cite{walliser2019team}. AI’s adaptability in shared resource management influences decision-making~\cite{flathmann2023examining}, improving understanding of team composition, communication, and coordination in human-machine collaboration.

\textbf{Limitations.} AI struggles with naturalistic communication, lacking the ability to process implicit nonverbal cues~\cite{mcneese2021my}. While AI improves TSA, it does not necessarily enhance cognition or performance in dynamic environments~\cite{mcneese2018teaming}. Many studies rely on controlled test beds, limiting real-world applicability~\cite{demir2018team}. The impact of team-building interventions on HAT performance remains unclear, focusing more on behavior than quantifiable performance gains~\cite{walliser2019team}. The absence of standardized metrics for assessing AI’s influence on shared decision-making complicates generalization across domains~\cite{flathmann2023examining}. Future research should integrate real-world validation, refine AI communication strategies to include nonverbal dynamics, and develop robust frameworks for long-term AI impact assessment in HMT.

\subsection{Task Allocation} \label{subsec:task-allocation}

\textbf{Key Ideas.} Research on function allocation in HMT explores workload distribution, situational awareness, and responses to automation failures to enhance collaboration~\cite{roth2019function}. While traditional frameworks guide allocation, they struggle with highly autonomous systems. Studies debunk full automation myths, emphasizing human roles in AI-assisted tasks such as content moderation and image labeling~\cite{gray2019ghost}. This {\em ghost work} reveals how automation reshapes jobs rather than eliminating them, challenging conventional Human-Centered AI (HCAI) models. Task planning frameworks integrating human and robotic roles focus on balanced workload models and decision-making, optimizing hybrid workplace designs~\cite{tsarouchi2017human}. These approaches automate task allocation while incorporating user-defined criteria.

\textbf{Merits.} Task allocation HMT research aims to optimize human-machine role distribution via workload balance and adaptability for efficiency~\cite{roth2019function}. Recognizing human involvement in AI-assisted work reframes automation’s impact, showing AI transforms rather than replaces jobs~\cite{gray2019ghost}. Task planning frameworks integrate human decision-making into automated processes, ensuring adaptability in dynamic settings~\cite{tsarouchi2017human}. Multicriteria decision-making in task allocation enables scalable, user-driven hybrid workplace designs, boosting productivity while maintaining human oversight.

\textbf{Limitations.} Traditional function allocation models struggle to keep pace with autonomous systems, limiting long-term applicability~\cite{roth2019function}. The persistence of human roles in automation underscores the need for transparency in AI decision-making and ethical considerations of {\em ghost work}~\cite{gray2019ghost}. Existing task planning frameworks often lack scalability, requiring customization for different operational settings~\cite{tsarouchi2017human}. Balancing human and machine contributions remains challenging, as rigid role allocation frameworks may reduce system flexibility. Future research should focus on adaptable, human-centric design principles to sustain function allocation in HMT environments.

\subsection{AI \& Ethics} \label{subsec:ai-ethics}

\textbf{Key Ideas.} Ethical AI in HMT relies on responsibility, trustworthiness, transparency, and fairness to align with human values~\cite{assaad2023ethics, smith2019designing}. While ethical compliance fosters trust, violations, especially regarding proportionality, undermine it~\cite{text2022Exploring}. Trust in AI’s ethical decisions strengthens team cohesion and risk-taking, yet unethical behaviors disrupt coordination~\cite{lopez2023complex}. Research highlights the difficulty of trust repair in HATs, where even corrective actions like apologies fail to restore confidence~\cite{schelble2022ethical}. Ethical frameworks such as Utilitarianism and Deontology guide AI decision-making but lack proven applicability in operational settings~\cite{steen2023meaningful}. While these frameworks define ethical AI interactions and transparency, limited empirical validation raises concerns about scalability and adaptability across HMT contexts~\cite{smith2019designing, assaad2023ethics}.

\textbf{Merits.} Integrating ethical principles enhances AI accountability and ensures alignment with human values, reinforcing trust in HMT interactions~\cite{assaad2023ethics}. Ethical compliance sustains confidence in AI teammates, fostering collaboration and improving team dynamics~\cite{text2022Exploring, lopez2023complex}. Transparent, interpretable AI decision-making benefits from natural language processing, strengthening human-AI communication~\cite{smith2019designing}. Studies on MHC frameworks provide insights into balancing AI autonomy with human oversight in decision-making processes~\cite{steen2023meaningful}. 

\textbf{Limitations.} Despite advances in ethical AI principles, translating them into scalable, validated strategies remains challenging across diverse HMT scenarios~\cite{smith2019designing, assaad2023ethics}. Many studies rely on conceptual models and simulations, limiting real-world applicability~\cite{steen2023meaningful, text2022Exploring}. Ensuring explainability and mitigating bias remain challenging, as current techniques focus on post‐hoc explanations rather than proactive, fairness‐aware AI design. Bias in AI teammates can erode trust, particularly in high-stakes applications affecting safety and autonomy. Ethical trade-offs arise where AI autonomy for efficiency conflicts with human oversight, accountability, and fairness, especially in the military and healthcare. Trust repair remains difficult, as AI apologies fail to restore confidence in unethical AI teammates~\cite{schelble2022ethical}. The lack of standardized evaluation metrics hampers benchmarking, raising concerns about adaptability in dynamic environments~\cite{lopez2023complex}. Future research must emphasize real-world validation, develop fairness-aware bias mitigation, and establish ethical frameworks that balance autonomy, accountability, and oversight in AI-driven HMT systems.

\subsection{Discussions: Evolution of Research Focus in HMT}

\begin{wrapfigure}{r}{0.7\textwidth}
\vspace{-5mm}
    \centering
    \includegraphics[width=\linewidth]{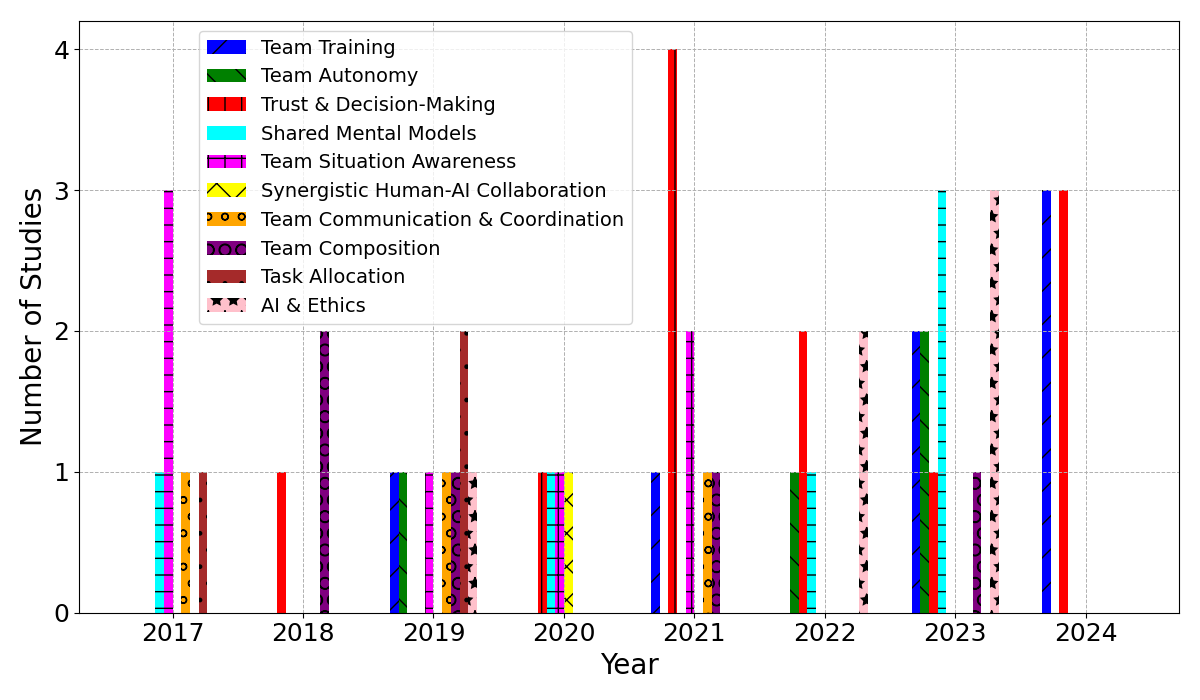}
    \vspace{-8mm}
    \caption{Frequency of ten categories of HMT research from 2017 to 2024.}
    \label{fig:research-focus-10}
    \vspace{-5mm}
\end{wrapfigure}

Fig.~\ref{fig:research-focus-10} shows research trends across ten HMT categories from 2017 to 2024, based on 93 studies. The distribution highlights a steady rise in activity across all areas, reflecting the growing interest in HMT systems.

\textit{Team Training} consistently leads as the most researched area, emphasizing its central role in building effective HMT systems. \textit{Trust \& Decision-Making} is another prominent focus, highlighting its critical importance in fostering reliable collaboration between humans and machines. Other areas, such as \textit{Shared Mental Models}, \textit{Team Autonomy}, and \textit{Team Communication \& Coordination}, have shown steady growth, particularly in their contributions to improving interaction quality and alignment in team dynamics. \\ \vspace{-4mm}
Emerging areas like \textit{AI \& Ethics} and \textit{Synergistic Human-AI Collaboration} have gained traction in recent years, addressing the increasing need for ethical considerations and user-centric design in HMT systems.  The interconnectedness of these subsections is evident. Advancements in \textit{Team Training} often support \textit{Trust \& Decision-Making} and \textit{Shared Mental Models}, which influence \textit{Team Autonomy} and \textit{Task Allocation}. Likewise, ethical considerations in \textit{AI \& Ethics} impact \textit{Synergistic Human-AI Collaboration} and \textit{Trust}, highlighting the integrative nature of HMT research. These trends reflect evolving priorities and the synergy between HMT's core areas, driving comprehensive system development.  For reference, the contributions and limitations of each work are summarized in Table I of Appendix A (Empirical Studies to Promote Team Performance: Summary).

\section{Evaluation Methodologies of Human-Machine Teaming Systems (HMTSs)} 
\label{sec:evaluation-methodologies-HMSs}

\subsection{Evaluation Testbeds} \label{subsec:evaluation-testbeds}

Table~\ref{tab:hmt_evaluation_methods} summarizes four evaluation methods commonly used for Human-Machine Teaming (HMT) systems: Statistics-based, Simulation Model-based, Emulation Testbed-based, and Real Testbed-based evaluations. Each method is described alongside examples and relevant references.

The \textit{Statistics-based Evaluation} uses methods like Mixed Analysis of Variance (ANOVA) and Multivariate Analysis of Variance (MANOVA) to analyze data, focusing on performance measurement, trend analysis, error rates, and hypothesis testing~\cite{chen2014human,demir2020understanding,fotouhi2019survey,
javaid2023communication,mcneese2021team, mcneese2021my, mcneese2018teaming, walliser2019team,
murphy2013survey, Demir2021Exploration,damacharla2018common, gomez2019considerations, kashima2022trustworthy, Yang22-thms-review-mhrobotics, o2022human, berretta2023defining}. This approach excels in control but is less realistic due to its reliance on mathematical models rather than real-world scenarios.

The \textit{Simulation Model-based Evaluation} employs computational models to mimic real-world processes, allowing assessment of behavior and performance under various conditions~\cite{demir2018team, 
cummings2010role,
cummings2011impact, mcneese2021team, fotouhi2019survey,
javaid2023communication,
gay2019operator,
johnson2020understanding,  lawless2021towards, lawless2023interdisciplinary, liang2017human,
mcdermott2005effective,bersani2023towards, flathmann2023examining, nourbakhsh2005human, talamadupula2010planning, wu2022survey,
murphy2013survey, Yang22-thms-review-mhrobotics}. Examples include testbeds like CERTT-II, NeoCITIES, self-driving car simulation platforms, and Minecraft. This method balances scalability and control but may lack the realism of physical environments.  JADE-based frameworks like HACO provided development environments for simulating human-AI team interactions and testing teaming concepts~\cite{dubey2020haco}. The \textit{Emulation Testbed-based Evaluation} creates controlled, realistic conditions for testing system performance and human-machine interactions. Examples include RPAS-STE, Malware Distribution Network (i.e., MDN) Graphs~\cite{ cai2020perceiving}, RoboLeader~\cite{chen2014human}, Security Information and Event Management (SIEM) platforms~\cite{gomez2019considerations}, Amazon Mechanical Turk (MTurk)~\cite{kashima2022trustworthy}, CERTT-UAS-STE (i.e., Cognitive Engineering Research on Team Tasks - Unmanned Aerial System Synthetic Task Environment)~\cite{o2022human}, NASA-TLX, and COBALT~\cite{johnson2021impact,  hart2006nasa, walliser2019team, knox2009interactively,  leon2013human, nagy2021interdependence, blaha2020cognitive}. This approach combines realism with control, making it effective for detailed studies of HMT systems.  The \textit{Real Testbed-based Evaluation} focuses on real-world scenarios to examine human-AI interactions, such as Voice AI Agent and Aircraft with Embedded AI~\cite{moore2007presence, fotouhi2019survey, lopez2023complex, chen2010ai, sheridan2016human}. While highly realistic and applicable, this method can be limited by scalability and logistical constraints.  Table~\ref{tab:hmt_evaluation_methods} outlines the methods, aiding in selecting suitable evaluation strategies for HMT systems based on research objectives and constraints. Detailed explanations of the cited works are available in the supplementary document (see Appendix B: Evaluation Methods for Human-Machine Teaming Systems).
\\ \vspace{-4mm} 
\begin{wrapfigure}{r}{0.45\textwidth} 
\vspace{-5mm}
    \centering
    \includegraphics[width=0.43\textwidth]{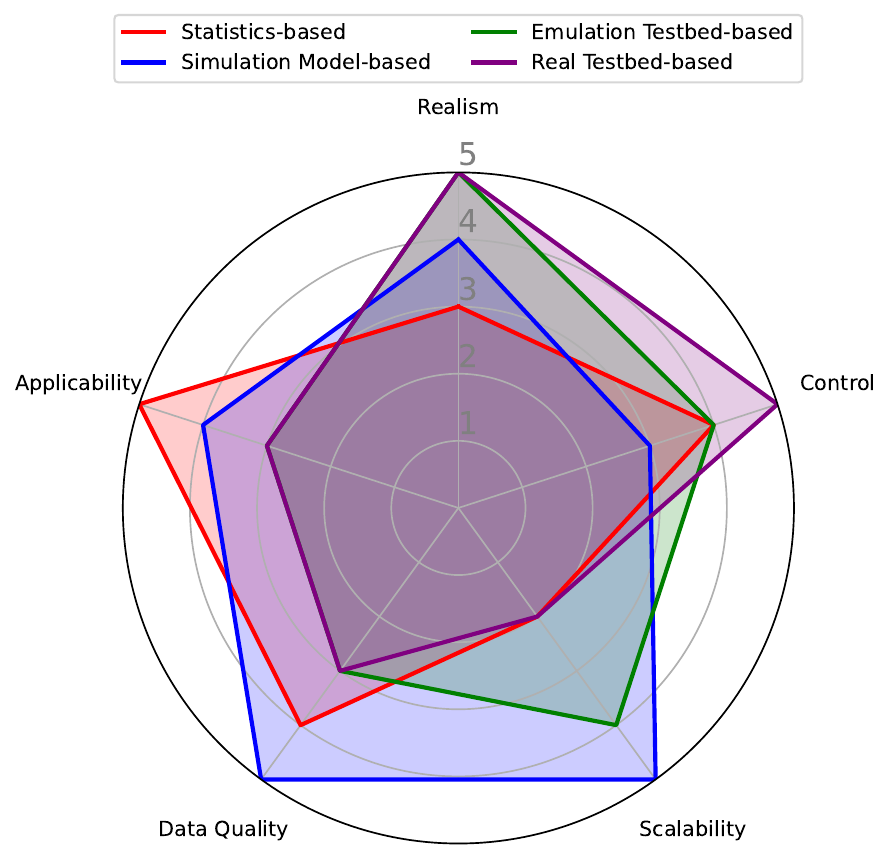} 
    \caption{Key attributes of HMT evaluation methods.}
    \label{fig:hmt_evaluation}
    \vspace{-3mm}
\end{wrapfigure}
{\bf Comparative Analysis of the Four Evaluation Methods}.  Fig.~\ref{fig:hmt_evaluation} illustrates a comparative analysis of four evaluation methods used in Human-Machine Teaming (HMT) systems -- Statistics-based, Simulation Model-based, Emulation Testbed-based, and Real Testbed-based -- across five key criteria: realism, control, scalability, data quality, and applicability. Each evaluation method is represented as a distinct colored area, showcasing its strengths and trade-offs. Statistics-based evaluation scores are highly in control (5) due to their precise management of variables but have lower realism (2), reflecting its limited ability to replicate real-world scenarios. Simulation Model-based evaluation excels in scalability (5), making it suitable for diverse experimental conditions while maintaining moderate realism (3). Emulation Testbed-based evaluation effectively balances multiple criteria, achieving high data quality scores (5) and realism (4), allowing for controlled yet realistic assessments. Real Testbed-based evaluation provides maximum realism (5) and applicability (5) by operating in real-world settings, though it faces challenges in scalability (2) due to logistical constraints. Quantitative examples highlighted on the chart further illustrate these trends, emphasizing strengths such as high data quality in emulation-based methods and the unmatched realism of real testbeds. This visualization underscores the trade-offs inherent in each method, aiding researchers in selecting the most appropriate approach based on their evaluation objectives.

\begin{table}[t]
\footnotesize
\centering
\caption{Summary of HMT Evaluation Methods}
\label{tab:hmt_evaluation_methods}
\vspace{-3mm}
\begin{tabular}{|p{3cm}|p{5.5cm}|p{5.5cm}|}
\hline
\multicolumn{1}{|c|}{\bf Evaluation Method} & \multicolumn{1}{c|}{\bf Description} & \multicolumn{1}{c|}{\bf Examples with References} \\ \hline
\textit{Statistics-based Evaluation} & 
Uses statistical methods like Mixed ANOVA and MANOVA for analyzing data, measuring performance, and testing hypotheses. &
Mixed ANOVA~\cite{demir2020understanding, mcneese2021team, mcneese2021my, mcneese2018teaming, walliser2019team}, 
MANOVA~\cite{demir2020understanding, mcneese2021team, Demir2021Exploration} \\ \hline

\textit{Simulation Model-based Evaluation} & 
Simulates real-world processes using computational models, assessing behavior and performance under varied conditions. &
CERTT-II Testbed~\cite{demir2018team, mcneese2018teaming, demir2017team, mcneese2021team}, 
NeoCITIES~\cite{mcneese2021my, schelble2020designing, schelble2022see, schelble2022let}, 
ML-based Simulation~\cite{bersani2023towards}, 
Rocket League Platform~\cite{flathmann2023examining}, 
QN-MHP Testbed~\cite{liu2006multitask}, 
Minecraft~\cite{chiou2021towards, demir2020understanding}, 
USAR Arenas~\cite{nourbakhsh2005human, talamadupula2010planning} \\ \hline

\textit{Emulation Testbed-Based Evaluation} & 
Simulates realistic conditions using testbeds for controlled assessments of system performance and human-machine interactions. &
RPAS-STE~\cite{johnson2021impact, demir2019evolution, myers2019, mcneese2021trust, Demir2021Exploration}, 
NASA-TLX~\cite{hart2006nasa, gay2019operator}, 
Strike Group Defender~\cite{walliser2019team}, 
F-16 Fighter Aircraft~\cite{mathieu2000influence}, 
COBALT~\cite{blaha2020cognitive} \\ \hline

\textit{Real Testbed-based Evaluation} & 
Tests HMT applications in real-world environments to study human-AI interactions under practical conditions. &
Voice AI Agent~\cite{moore2007presence}, 
Aircraft with Embedded AI~\cite{lopez2023complex}, 
AI and Opinion Mining~\cite{chen2010ai} \\ \hline

\end{tabular}
\vspace{-5mm}
\end{table}

\subsection{Metrics} \label{subsec:evaluation-metrics}

Table~\ref{tab:hmt_metrics} presents an overview of the metrics used to evaluate HMT performance, categorized into three main types: human-centric, machine-centric, and team-centric metrics. Each category focuses on different aspects of HMT systems, providing complementary insights into their performance and dynamics.

\textit{Human-Centric Metrics} emphasize evaluating human factors, including trust, collaboration, cognitive processes, and emotional responses. These metrics, such as perceived trust in AI, workload (e.g., NASA-TLX), and electrodermal activity (EDA), are crucial for understanding how humans interact with and perceive AI systems. They help identify factors influencing human engagement, workload, situational awareness, human reliability, and trust, which are critical for successful HMT operations~\cite{chen2014human,lopez2023complex, walliser2019team, 
cummings2010role,
cummings2011impact, ciechanowski2019shades,
gay2019operator,
johnson2020understanding,lawless2023interdisciplinary,  leon2013human,  chiou2021towards,
mcdermott2005effective,nagy2021interdependence,  sheridan2016human, murphy2013survey,
edwards2023advise,damacharla2018common,gomez2019considerations, kashima2022trustworthy, gebru2022review, o2022human}.  Error rates in AI decision-making and fault tolerance during system operation are often used as machine-centric metrics~\cite{nagy2021interdependence}. The ProtoEM method employs confusion matrices and the EM algorithm to estimate human uncertainty when integrating human and machine decisions in legal case matching tasks~\cite{huang2024co}. Moreover, \citet{knox2009interactively} evaluate the effectiveness of human trainers in shaping an agent’s behavior through reinforcement signals. These metrics primarily assess how human feedback influences the agent's learning process, performance, and human-machine collaboration.

\textit{Machine-Centric Metrics} assess the performance and reliability of machines or AI systems in HMT. Metrics such as response time, mission success rate, prediction accuracy, agent reliability, task allocation efficiency, transparency and explainability (i.e., how well AI explains its decisions), and task interpretation accuracy provide objective and reproducible measures of a machine's capabilities. These metrics ensure that AI systems meet functional requirements and operate effectively in dynamic environments~\cite{moore2007presence,bersani2023towards,cai2020perceiving,cummings2010role,cummings2011impact,javaid2023communication,liang2017human,patel2018anomaly,beetz2017guidelines, wu2022survey,murphy2013survey,chen2014human}. 

\textit{Team-Centric Metrics} evaluate the collective performance and dynamics of human-machine teams. Metrics like situation awareness, team communication flow, shared mental models, human-agent collaboration efficiency, collaboration effectiveness, and creativity measure the synergy between human and AI agents, offering insights into collaboration, adaptability, and overall team efficiency. These metrics are essential for understanding how well the team functions as an integrated unit~\cite{damacharla2018common, gay2019operator,Hauptman23-autonomy, schelble2022let, lawless2021towards,  stout1999planning, hwang2021ideabot,murphy2013survey, kashima2022trustworthy}.

Table~\ref{tab:hmt_metrics} highlights the diverse range of metrics available for evaluating HMT systems, emphasizing their complementary roles in providing a comprehensive assessment. Combining these metrics can enhance the evaluation process, ensuring a balanced analysis of human, machine, and team performance. Detailed explanations of each work utilizing the metrics mentioned above are provided in the supplementary document (see Appendix C: Metrics for HMT Systems).

\begin{wrapfigure}{r}{0.4\textwidth}
    \centering
    \includegraphics[width=0.38\textwidth]{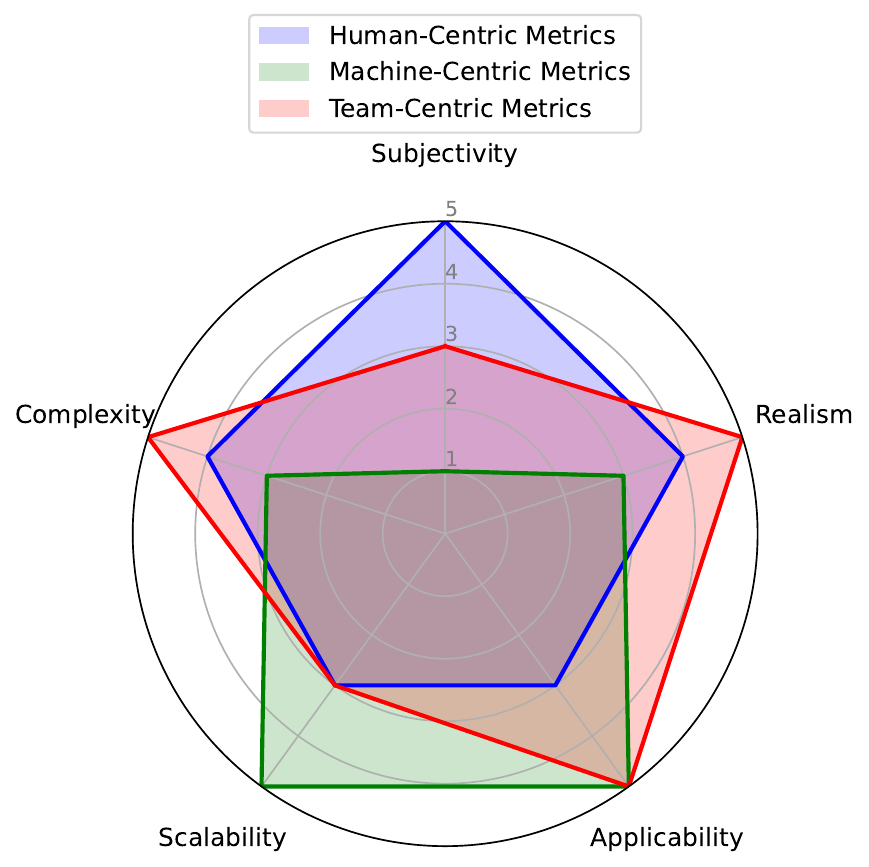} 
\vspace{-3mm}
    \caption{Key attributes of the three primary metrics for HMT systems.}
    \label{fig:hmt_metrics_comparison}
\vspace{-5mm}
\end{wrapfigure}

{\bf Comparative Analysis of Metrics for Evaluating HMT Systems.} Fig.~\ref{fig:hmt_metrics_comparison} comparatively analyzes the three categories of metrics used for evaluating HMT systems. \\ \vspace{-4mm} 
Fig.~\ref{fig:hmt_metrics_comparison} visualizes their performance across five key evaluation criteria: realism, subjectivity, complexity, scalability, and applicability. In {\em realism}, {\em Team-Centric Metrics} excel due to their focus on collaborative dynamics, while {\em Human-Centric Metrics} score high by capturing human behavioral nuances. Regarding {\em subjectivity}, {\em Human-Centric Metrics} rely heavily on subjective assessments, whereas {\em Machine-Centric Metrics} are largely objective, offering consistent and reproducible results. {\em Complexity} is another critical factor for both Human-Centric and Team-Centric metrics because they rely on sophisticated data collection and analysis methodologies. {\em Scalability} varies significantly across the categories; Machine-Centric metrics are the most scalable because of their quantitative nature. Human-Centric and Team-Centric metrics often face challenges when applied to diverse scenarios. {\em Applicability} is a strength for both Machine-Centric and Team-Centric metrics, as they are versatile across various HMT contexts. In contrast, Human-Centric metrics tend to be more specific to the context of human-AI interactions. This visualization underscores the complementary strengths and trade-offs of these metric categories, highlighting the importance of integrating them for a comprehensive evaluation of HMT systems.

\subsection{Datasets} \label{subsec:datasets}

 We highlight several notable trends and limitations in the datasets utilized for HMT research, as reviewed in this work. First, {\bf structured text datasets} dominate~\cite{cai2020perceiving, fotouhi2019survey, 
javaid2023communication, gay2019operator, johnson2020understanding, lawless2021towards,   leon2013human,
mcdermott2005effective,nagy2021interdependence}, often capturing survey or interview-based insights into human-machine interactions. \\ \vspace{-4mm}
\begin{wrapfigure}{r}{0.5\textwidth} 
\vspace{-5mm}
    \centering
    \includegraphics[width=0.48\textwidth]{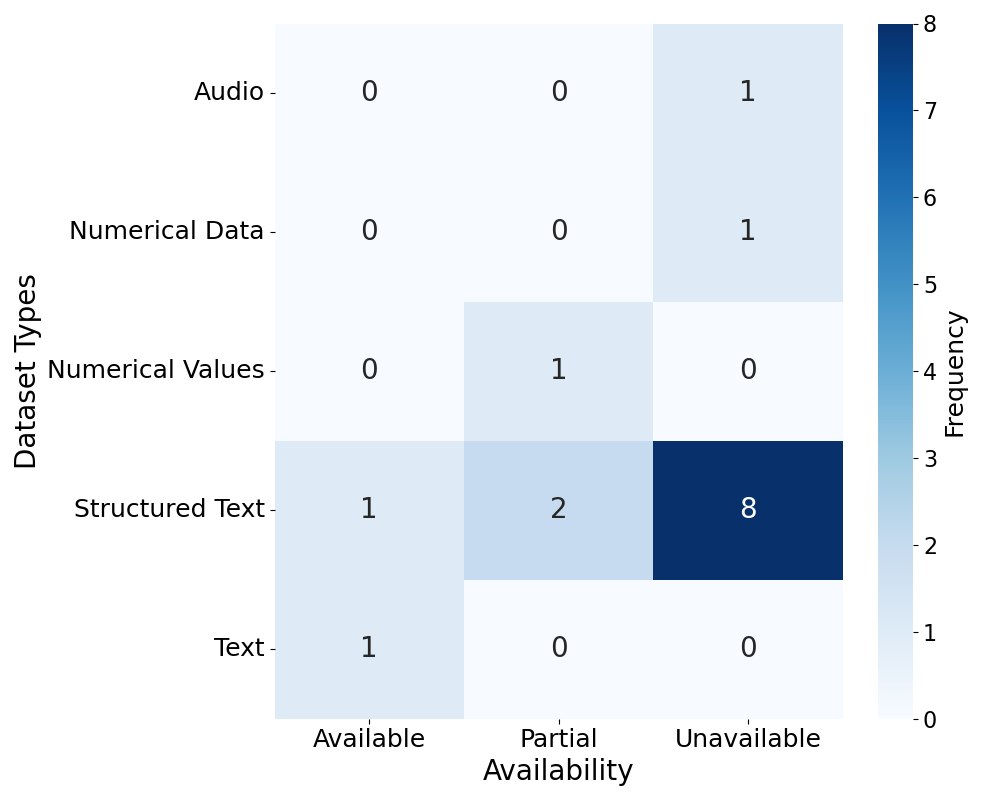} 
\vspace{-3mm}
    \caption{Heatmap showing the distribution of dataset types and their availability in studies related to HMT systems.}
    \label{fig:dataset_heatmap}
    \vspace{-5mm}
\end{wrapfigure}However, many of these datasets remain unavailable or restricted, hindering reproducibility and progress in the field.  Second, {\bf text datasets} from sources like ENISA and NCSC~\cite{mahaini2019building} and audio datasets~\cite{moore2007presence} are underrepresented. Text datasets are often secondary and collected from public repositories, while audio datasets are rarely used, limiting multi-modal interaction analysis.  Third, {\bf numerical datasets}, such as Lyapunov exponents~\cite{demir2018team}, COBALT task data~\cite{blaha2020cognitive}, simulation data from RoboLeader~\cite{chen2014human}, crowdsourcing performance data (i.e., MTurk)~\cite{kashima2022trustworthy}, physiological and psychological data (i.e., electroencephalography)~\cite{ho2018psychological}, and human-automation interaction data~\cite{cummings2010role, cummings2011impact}, are application-specific and cater to tasks requiring precise quantitative analysis. Speech and language processing datasets are vital for HMT systems, enabling the evaluation of human-machine communication and interaction quality~\cite{Berkol23-hmt-nlp}.  However, their restricted accessibility limits the replication and validation of results.   Fourth, most datasets, including structured text and numerical data, are unavailable or partially accessible~\cite{zhang2020anideal}. {\bf This lack of data availability} challenges reproducibility and cross-study comparisons, restricting broader adoption of methodologies.  Lastly, the reliance on structured text and numerical data reveals {\bf a lack of diverse datasets}, such as video or sensor data, which could provide richer insights into complex human-machine interactions.

Making datasets publicly available fosters reproducibility and addresses these issues. Expanding to multi-modal data, such as video and audio, broadens research capabilities. Standardized formats and repositories tailored to HMT research could greatly enhance accessibility and advance the field.

To facilitate easy reference to the datasets used in each HMT study, we summarize their types and corresponding descriptions in Table~2 of Appendix B (Metrics for HMT Systems). 

{\bf Dataset Availability and Distribution in HMT Research.} Fig.~\ref{fig:dataset_heatmap} presents dataset distribution and availability across HMT studies, categorizing Text, Audio, Structured Text, and Numerical Data as Available, Partially Available, or Unavailable. Structured text is the most used but often unavailable, limiting reproducibility and adoption. Numerical and text datasets contribute but remain less accessible. The "Low" and "High" labels on the color bar indicate dataset frequency, with darker shades representing higher occurrences. This visualization underscores dataset usage trends, emphasizing the need for improved accessibility and shared repositories.

\begin{table}[t]
\centering
\footnotesize
\caption{Metrics for Evaluating Human-Machine Teaming (HMT) Performance}
\label{tab:hmt_metrics}
\vspace{-3mm}
\begin{tabular}{|p{3cm}|p{8cm}|p{3cm}|}
\hline
\multicolumn{1}{|c|}{\bf Metric Type} & \multicolumn{1}{c|}{\bf Description} & \multicolumn{1}{c|}{\bf References} \\ \hline

\textit{Human-Centric Metrics} & 
Metrics focusing on human factors in HMT performance evaluation, including trust, collaboration, and cognitive measures. & 
\cite{lopez2023complex, walliser2019team, damacharla2018common, cummings08, ciechanowski2019shades, gay2019operator, knox2009interactively, liu2006multitask, schelble2022let, fallon2018improving, sheridan2016human, chiou2021towards} \\ \hline

\textit{Machine-Centric Metrics} & 
Metrics assessing machine or AI system performance, such as response time, mission success rate, and accuracy in classification tasks. & 
\cite{moore2007presence, bersani2023towards, fallon2018improving, cai2020perceiving, chiou2021towards, damacharla2018common, yildirim2022smt, knox2009interactively, blaha2020cognitive, liang2017human, patel2018anomaly, lee2015testing, wu2022survey, beetz2017guidelines} \\ \hline

\textit{Team-Centric Metrics} & 
Metrics evaluating the performance and dynamics of human-machine teams, including situation awareness, communication flow, and shared mental models. & 
\cite{damacharla2018common, Hauptman23-autonomy, demir2017team, schelble2022let, stout1999planning, gay2019operator, hwang2021ideabot, mcneese2021team, national2021human} \\ \hline

\end{tabular}
\vspace{-6mm}
\end{table}

\section{Key Theoretical Approaches for Human-Machine Teaming} \label{sec:key-theoretical-approaches-hmts}

Through an extensive review of HMT systems, we have identified three widely utilized primary theories: Human-in-the-Loop Reinforcement Learning (HRL), Instance-Based Learning Theory (IBLT), and Interdependence Theory (IT).

\subsection{Human-in-the-Loop Reinforcement Learning (HRL)} \label{subsec:hrl}

{\bf Key Ideas.} AI alone is often less effective than human expertise in solving complex problems~\cite{wu2022survey}. To address this, Human-in-the-Loop Reinforcement Learning (HRL) integrates human knowledge into AI decision systems to enhance prediction accuracy at minimal cost. HRL research addresses (1) improving data processing by selecting relevant samples, (2) enhancing model performance via human-provided high-dimensional knowledge, and (3) designing efficient systems to incorporate human input without compromising performance or robustness.

{\bf HRL Approaches in HMT Systems.} \citet{liang2017human} proposed an HRL framework for vehicle decision-making, integrating human-vehicle interactions with auxiliary driving equipment. They introduced a hybrid Markov decision process model leveraging human intuition. \citet{leon2013human} developed a reinforcement learning framework combining human demonstrations and feedback, enabling non-experts to train robots. When applied to a ``pick and place" task, it improved learning efficiency, robustness to noisy data, and task performance. \citet{knox2009interactively} introduced the Tamer framework, using human reinforcement to train agents without environmental rewards, reducing sample complexity and enabling rapid policy learning. While Tamer agents excel initially, autonomous agents outperform them with extensive training. A hybrid approach balancing human reinforcement and environmental rewards enhances efficiency and long-term performance in delayed-feedback scenarios.

{\bf Limitations.}  These HRL-based studies show promise in integrating human knowledge into AI systems but face challenges like scalability across diverse tasks, biases from human input, and limited validation in dynamic real-world environments. Future research should prioritize scalable frameworks, reducing biases, enhancing adaptability, and validating performance in realistic settings to ensure trustworthiness in HMT systems.

\subsection{Instance-based Learning Theory (IBLT)} 

{\bf Key Ideas.} \textit{Instance-Based Learning Theory} (IBLT)~\cite{gonzalez2003instance} explains learning in dynamic decision-making (DDM) through accumulating, recognizing, and refining instances. Knowledge is stored as \textit{Stored Decision Units} (SDUs), capturing context, actions, and expected outcomes. Decision-making relies on \textit{recognition-based retrieval}, where memory retrieves SDUs based on situational similarity. Learning adapts through \textit{adaptive strategies}, alternating between heuristics for unfamiliar situations and instance-based approaches for familiar ones. \textit{Necessity-based choice} balances exploring alternatives and committing to the best-known option based on constraints like time or preference. \textit{Feedback updates} refine SDU utilities by linking actions to observed outcomes, enhancing decision accuracy over time.

{\bf IBLT Approaches in HMT Research.} \citet{gonzalez2022adaptive} tackled static cyber defense limitations, proposing an adaptive IBLT framework for dynamic deception using signaling, masking, and decaying. Integrating game theory and machine learning, this model refined cognitive strategies but lacked empirical validation for safety and trust. Instance-Based Learning (IBL) generates predictions from individual instances without abstraction~\cite{aha1991instance}. \citet{gupta2023fostering} extended the TSM-CI framework, identifying IBLT as a suitable AI architecture. \citet{dong2022toward} introduced the Dispatching Multitasking Trust Paradigm (DMTP) to enhance dispatcher trust in Fully Automated Operation (FAO) metro systems, though its IBLT-based trust model lacked validation. \citet{yang2023inner} explored human-machine communication via {\em Machine Theory of Mind}, proposing three MToM approaches: modeling human inference, simulating AI-like behavior, and integrating domain-specific actions. Their IBLT scenario demonstrated effective collaboration. \citet{blaha2020cognitive} developed an IBLT-based computational model of human decision-making, showing how trust calibration evolves via task experience.

{\bf Limitations.}  While IBLT offers valuable mechanisms for dynamic decision-making and trust calibration, its application in developing trustworthy HMT systems faces notable limitations. Existing studies often lack empirical validation, particularly in real-world dynamic environments, which undermines their scalability and practical applicability. Further, the reliance on predefined instances limits adaptability in unpredictable scenarios, where novel situations demand innovative solutions. Future research should incorporate empirical evaluations, refining feedback mechanisms, and enhancing adaptability by integrating IBLT with advanced cognitive architectures to address the complexities of HMT.

\subsection{Interdependence Theory (IT)} \label{subsec:interdependence-theory}

{\bf Key Ideas.}  {\em Independence theory} (IT) in HMT emphasizes balancing human and machine autonomy to optimize collaboration by delineating decision-making and task boundaries, leveraging strengths, and minimizing conflicts~\cite{sheridan2016human}. Combined with interdependence theory, it identifies tasks best suited for humans or machines, fostering efficiency and synergy for enhanced team performance beyond individual capabilities.

{\bf IT Approach in HMT Research.} \citet{johnson2020understanding} emphasized interdependence analysis to enhance human-autonomy teaming, addressing the limitations of technology-centric tools and promoting system adaptability. \citet{lawless2023interdisciplinary} criticized traditional team science for treating individuals as independent, proposing state dependency to resolve validation and replication crises by examining interconnected team dynamics. \citet{nagy2021interdependence} highlighted the challenge of diagnosing ML algorithms due to limited code transparency, advocating for human interventions via interdependence analysis. \citet{lawless2021towards} argued interdependence improves team performance, countering the tendency to oversimplify data. However, \citet{lawless2020quantum} noted the lack of a formal mathematical foundation for designing and operating Autonomous HMTs (A-HMTs), calling for mathematical models to enhance development and reliability. 
\\ \vspace{-4mm}
\begin{wrapfigure}{r}{0.5\textwidth}
\vspace{-5mm}
    \centering
    \includegraphics[width=\linewidth]{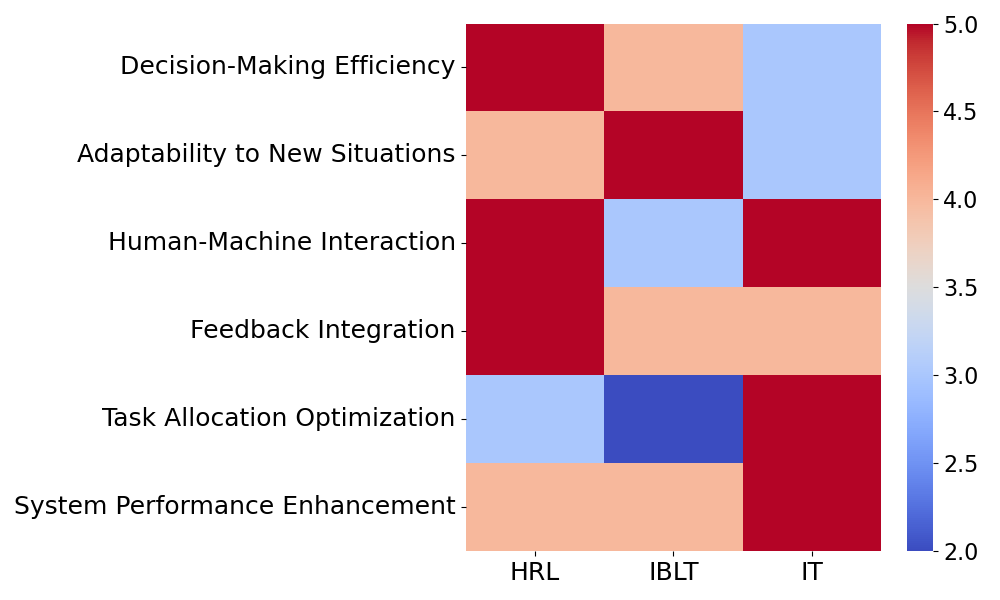}
    \vspace{-7mm}
    \caption{Key benefits of HRL, IBLT, and IT for HMT performance.}
    \label{fig:hrl-iblt-it-comparison-heatmap}
    \vspace{-7mm}
\end{wrapfigure}
{\bf Limitations.} Existing HMT studies using independence theory rely on static task boundaries and insufficiently integrate dynamic interdependence in human-machine interactions~\cite{sheridan2016human, lawless2023interdisciplinary}. They often overlook adaptive mechanisms for evolving team dynamics, ML safety concerns, and a mathematical foundation for interdependence modeling~\cite{nagy2021interdependence, lawless2020quantum}. Future research should develop formalized interdependence models to enhance adaptability, safety, and performance in A-HMTs, incorporating real-time feedback to adjust task allocations and roles based on team and context.

{\bf Comparative Analysis of HRL, IBLT, and IT Frameworks.}  Fig.~\ref{fig:hrl-iblt-it-comparison-heatmap} compares the strengths of HRL, IBLT, and IT across six key factors. HRL excels in decision-making efficiency, feedback integration, and human-machine interaction, making it ideal for real-time systems like autonomous vehicles or disaster response. IBLT offers superior adaptability by leveraging prior cases, making it suitable for healthcare diagnostics and training. IT is strongest in task allocation and overall performance, highlighting its relevance in collaborative applications like project management or organizational planning.  HRL emphasizes real-time adaptability with human oversight, IBLT leverages historical knowledge, and IT focuses on interdependencies and coordination. HRL suits environments requiring continuous feedback, while IBLT is preferable for case-based reasoning. IT thrives in collaborative settings where role optimization is key. Each framework has distinct advantages, suggesting domain-specific applications with potential synergies for comprehensive solutions.

\section{Concepts, Challenges, and Applications of Human-Machine Teaming} \label{sec:applications-hmtns}

\subsection{Human and Unmanned Vehicle Teaming} \label{subsec: humans-uvt}

\indent {\bf Concept of UAV Teaming.}  
Unmanned Aerial Vehicle (UAV) teaming involves collaborative operations between UAVs or between UAVs and human operators to achieve shared objectives~\cite{fotouhi2019survey}. This concept leverages UAVs' strengths, such as agility, autonomy, and scalability, to perform complex tasks efficiently. Effective UAV teaming requires seamless integration of communication, coordination, and control to adapt to dynamic and uncertain environments~\cite{javaid2023communication}.  \citet{mcdermott2005effective} and \citet{cummings2011impact} emphasized the importance of optimizing human-automation interaction for unmanned vehicle operations, highlighting how operator proximity and UV type influence mission efficiency and how decentralized planning can enhance effectiveness in dynamic scenarios.

{\bf Challenges in UAV Teaming Systems.}  
Key challenges in UAV teaming include ensuring reliable communication, balancing workload between humans and automation, and maintaining coordination in dynamic environments. \citet{cummings2010role} highlighted that high automation replanning rates increase operator workload, while \citet{calhoun2018human} proposed adaptive interfaces to improve decision-making, though both faced real-world validation issues. Additional challenges include scalability, adapting to unpredictable conditions, and resilience against cyber threats~\cite{gay2019operator}.

{\bf UAV Teaming Applications and Their Limitations.}  
UAV teaming has been explored as follows:
\vspace{-2mm}
\begin{itemize}
    \item \textit{Military Operations:} UAV teaming supports surveillance, reconnaissance, and tactical missions. \citet{gay2019operator} studied cyberattack detection in unmanned ground vehicles (UGVs), introducing ``Suspicion Theory'' to structure operator suspicion responses. While promising, this approach needs further adaptation to complex military environments.
    \item \textit{Disaster Response:} UAV teams enhance situational awareness and coordination in search-and-rescue missions. However, scalability and challenges in real-time coordination limit their broader applicability~\cite{mcdermott2005effective}.
    \item \textit{Logistics and Supply Chain:} UAVs are employed for package delivery and supply chain optimization. Yet, regulatory constraints, battery limitations, and collision avoidance remain significant hurdles~\cite{cummings2011impact}.
    \item \textit{Automation Workload Balance:} Research has proposed strategies to balance automation rates and operator workload in UAV networks~\cite{cummings2010role, calhoun2018human}. However, these strategies require empirical validation in diverse, dynamic contexts.
\end{itemize}

Despite advancements, reliance on simulations that miss real-world complexities remains a key limitation. Limited focus on communication or automation rates underscores the need for integrated, scalable UAV teaming systems.

\subsection{Human-Chatbot Teaming} \label{subsec:va}

{\bf Concept of Human-Chatbot Teaming (HCT).}  
Human-Chatbot Teaming (HCT) refers to collaborative interactions where humans and chatbots work together to achieve shared objectives. This concept leverages chatbots' strengths, such as accessibility, consistency, and rapid information processing, to enhance team productivity and decision-making~\cite{chen2023human, ho2018psychological}. Effective HCT relies on integrating conversational interfaces with contextual understanding and adaptability to foster seamless collaboration in diverse environments~\cite{mygland2021affordances, al2024navigating}.

{\bf Challenges in HCT Systems.}  
HCT systems face significant challenges, including achieving natural and context-aware conversations, managing user trust, and ensuring equitable workload distribution~\cite{ho2018psychological, al2024navigating}. Developing chatbots that can adapt to individual user preferences and maintain engagement in long-term interactions is particularly difficult~\cite{mygland2021affordances, chen2023human}. Furthermore, technical challenges like integrating with diverse platforms, addressing ethical concerns such as bias, and ensuring data privacy and security add to the complexity of designing robust HCT systems~\cite{al2024navigating}. These challenges are amplified in dynamic and high-stakes scenarios where accuracy and reliability are critical~\cite{ho2018psychological, chen2023human}.

{\bf HCT Applications and Their Limitations.}  
HCT has been explored across various domains, with notable studies highlighting the following applications and limitations:  
\begin{itemize}
    \item \textit{Emotional and Collaborative Responses:} \citet{ciechanowski2019shades} investigated human-chatbot interactions, revealing that simpler chatbots elicited fewer emotional responses, with users favoring less human-like features. This suggests that chatbot mimicry of human traits may not be essential for effective collaboration.
    \item \textit{Task Performance Enhancement:} \citet{sowa2020digital} showed that chat-based virtual assistants enhance productivity in business simulations, highlighting chatbots' potential to improve team outcomes through assistance.
    \item \textit{Security Task Collaboration:} \citet{nagy2021interdependence} developed a framework to improve vulnerability detection in software, finding that teams with virtual assistants outperformed standalone tools in error identification.
    \item \textit{Creativity and Team Perception:} \citet{hwang2021ideabot} explored creativity in brainstorming, showing that conversational styles and perceived identities affect task performance, team dynamics, and creative self-efficacy.
\end{itemize}

Despite these advancements, common limitations persist. Many studies rely on controlled environments and simplified tasks, reducing real-world applicability. The lack of face-to-face communication in some experiments further limits generalizability. Besides that, framework-specific constraints and limited tool integration hinder broader adoption and scalability, emphasizing the need for more adaptable and comprehensive HCT systems.

\subsection{Human-Robot Teaming}
\label{subsec:human-robot}

{\bf Concept of HRT.}  \emph{Human-Robot Teaming} (HRT) refers to the collaboration between humans and robots to achieve shared goals by leveraging their complementary strengths~\cite{natarajan2023human, ciocirlan2019human}. Effective HRT relies on communication, coordination, and trust to ensure seamless interaction and optimize performance in dynamic environments~\cite{natarajan2023human, ciocirlan2019human}.

Designing effective HRT systems involves key elements: understanding agent abilities, taskwork, metrics, and peer-to-peer interactions. Agent abilities enable effective task execution, taskwork focuses on specific activities, metrics evaluate performance, and peer-to-peer interactions ensure communication and coordination. Balancing these factors is essential for collaboration and optimal outcomes~\cite{mingyue2018human}.

{\bf Challenges in HRT Systems.}
HRT faces challenges in developing robust communication, fostering trust, and adapting to dynamic tasks. Robots must predict human actions, and algorithms must handle evolving conditions. Trust is essential for human confidence in robot capabilities~\cite{natarajan2023human}.

{\bf HRT Applications and Their Limitations.} HRT applications cover diverse domains, including:
\vspace{-2mm}
\begin{itemize}
    \item \textit{Humanoid Robots}: \citet{hoffman2004collaboration} introduced \emph{Robonaut}, a humanoid robot designed for space missions. It utilized the Joint Intention Theory for collaborative tasks but focused on theoretical aspects with limited empirical testing. Reliance on complex social cues may restrict applicability in non-social settings.
    \item \textit{Simulation Testbed for HRT Systems}: \citet{nourbakhsh2005human} and NIST developed a testbed for search-and-rescue operations using the multi-agent system \emph{Retsina}. While effective for training, challenges included spatiotemporal coordination and complex control interfaces, hindering multi-robot management.
    \item \textit{Planning Systems for HRT}: \citet{talamadupula2010planning} proposed planning systems for dynamic environments, integrating counterfactual reasoning and conditional goals. However, scalability, reliance on accurate environmental models, and robust communication protocols remain hurdles.
    \item \textit{Team Coordination}: \citet{gombolay2015coordination} explored human-robot collaboration with the Willow Garage PR2 platform. While prioritizing human preferences improved perception, it reduced overall task efficiency, highlighting trade-offs between satisfaction and performance.
\end{itemize}

\subsection{Human-AI Teaming-based Social Media} \label{subsec:ai-social-media}

{\bf Concept of HAT-based Social Media.}  
HAT (Human-AI Teaming)-based Social Media refers to the collaborative integration of AI technologies and human decision-making processes to enhance the efficiency, accuracy, and ethical standards of social media systems. This concept leverages AI's ability to process vast amounts of data, provide contextual insights, and support human judgment while maintaining critical oversight and ethical considerations~\cite{brundage2019toward, hagendorff2023ai}. Effective HAT-based social media systems aim to address challenges such as misinformation, content moderation, and user engagement by combining AI's computational strengths with human creativity and ethical sensibility~\cite{tufchi2023comprehensive, balayogi2025human}.

{\bf Challenges in HAT-based Social Media.}  
HAT-based social media systems face numerous challenges, including balancing automation and human oversight, ensuring transparency in AI decision-making, and addressing biases in AI models. The complexity of social media data, such as informal language, cultural nuances, and real-time dynamics, complicates the development of robust AI systems. Besides that, ethical concerns like user privacy, algorithmic fairness, and the potential misuse of AI for malicious purposes amplify the difficulty of creating trustworthy HAT systems. Regulatory and governance challenges also emerge, requiring international cooperation and adaptable policies to ensure compliance with ethical standards and data protection laws.

{\bf HAT-based Social Media Applications and Their Limitations.}  
Applications of HAT-based social media span a wide range of domains, with notable studies highlighting both their potential and limitations:
\begin{itemize}
    \item \textit{Fake News Detection:} A Hybrid Multi-thread Metaheuristic method has been introduced for detecting fake news, demonstrating competitive performance on COVID-19 and Syrian war datasets~\cite{yildirim2022smt}. However, hardware constraints and limited dataset applicability present significant challenges.
    
    \item \textit{Opinion Analysis:} AI frameworks integrating sentiment analysis and stock modeling have been developed to assess user opinions on world events~\cite{chen2010ai}. Despite their innovative approach, external variables and unreliable sentiment extraction limit their accuracy.
    
    \item \textit{Health Issue Detection:} Ethical AI-integrated social media has been explored to detect health-related issues, ensuring fairness and privacy~\cite{liu2023application}. However, findings remain constrained by small sample sizes and qualitative data collection.
    
    \item \textit{Cultural Ecosystem Services (CES):} AI-driven analysis of Flickr images has been used to assess CESs and identify key hotspots~\cite{egarter2021harnessing}. Broader applicability is limited by geographic constraints and ethical concerns surrounding data use.
    
    \item \textit{Crisis Response:} AI-based disaster-related social media analysis has been applied to enhance crisis response, proposing methods for improving detection and intervention~\cite{imran2020using}. Challenges persist in detecting minor disasters and mitigating risks posed by manipulated images.
    
    \item \textit{Social Media Marketing for SMEs:} AI-driven social media marketing has been studied for its impact on customer retention and business growth~\cite{basri2020examining}. However, outdated analytics and vague survey methods limit the insights gained.

    \item \textit{Regulation of AI Content:} Regulatory solutions for AI-generated social media content have been proposed, including traceability logs for private data to enhance accountability~\cite{lewis2020rights}. However, governance receptivity and legal complexities continue to hinder implementation.
\end{itemize}

These applications demonstrate the potential of HAT-based social media systems to address critical issues, but limitations such as scalability, ethical challenges, and data biases underscore the need for continued research and development to achieve reliable and equitable outcomes.

\subsection{HMT-based Smart Computing} \label{subsec:hmt-smart-computing}

{\bf Concept of HAuT-based Smart Computing.}  
{\em Human-Autonomy Teaming (HAuT)-based Smart Computing} refers to the integration of autonomous systems and human decision-making to address complex computational tasks~\cite{lyons2021human}. This concept leverages the strengths of autonomy, such as efficiency, scalability, and data-driven insights, while maintaining critical human oversight for ethical, adaptive, and context-aware decision-making~\cite{o2022human}. By combining human creativity with machine precision, HAuT-based smart computing aims to enhance productivity, reliability, and situational awareness across diverse domains like manufacturing, IoT, and swarm robotics~\cite{abbass2021smart}.

{\bf Challenges in HAuT-based Smart Computing.}  
HAuT-based smart computing systems face significant challenges in scalability, adaptability, and maintaining mutual trust between humans and autonomous systems. Ensuring real-time data processing and decision-making in dynamic and unpredictable environments is particularly demanding~\cite{abbass2021smart, haindl2022towards}. The computational demands of relational machine learning and the intricacies of monitoring and validating ethical policies further complicate the implementation of robust systems~\cite{haindl2022towards, haindl2022quality}. Moreover, the lack of adaptability to rapidly evolving technologies and maintaining effective communication in complex scenarios remain pressing issues~\cite{haindl2022quality, oh2024link}.

{\bf HAuT-based Smart Computing Applications and Their Limitations.}  
Applications of HAuT-based smart computing demonstrate significant potential but also reveal inherent limitations:
\begin{itemize}
    \item \textit{Human-Swarm Teaming:} \citet{abbass2021smart} introduced intelligent shepherding agents to manage human-swarm teams, emphasizing contextual awareness for autonomous decision-making. While promising, the architecture requires validation using real data and precise metrics, especially in dynamic and unpredictable environments.
    \item \textit{Smart Manufacturing:} \citet{haindl2022towards} proposed a reference architecture for AI-enabled smart manufacturing, focusing on enhancing mutual trust and situational awareness. However, scalability for real-time data processing and the complexity of implementing and validating ethical policies remain challenges.
    \item \textit{Software Platforms for HAuT:} \citet{haindl2022quality} identified software platform characteristics critical for effective human-autonomy teaming, such as explicability and trustworthiness. However, scalability and adaptability issues hinder deployment in rapidly evolving manufacturing technologies.
    \item \textit{IoT Environments:} \citet{oh2024link} developed a rule-based adaptive system for improved human-machine collaboration in IoT setups, validated in simulated environments. However, maintaining effective communication and scalability in real-world scenarios remains a significant limitation.
\end{itemize}

These applications demonstrate the potential of HAuT-based smart computing across fields. However, ensuring scalability, adaptability, and ethical integrity is essential for robust and trustworthy implementations.

\subsection{HMT-based Natural Language and Information Retrieval}

{\bf Concept of HMT-based NLP and IR.} HMT-based NLP and Information Retrieval (IR) integrate human expertise with machine intelligence for complex language understanding, extraction, and retrieval~\cite{Berkol23-hmt-nlp}. By combining human decision-making and contextual insights with machine-driven data processing, this approach enhances accuracy, efficiency, and adaptability in applications such as taxonomy development, rescue missions, and legal analysis~\cite{wan2022user, safdar2024human, mahaini2019building, beetz2017guidelines, huang2024co}.

{\bf Challenges in HMT-based NLP and IR.}  
HMT-based NLP and IR face challenges like scalability, adaptability to dynamic data, and balancing human input with machine automation. Manual data collection and refinement limit scalability~\cite{mahaini2019building}, while unpredictable language expressions pose obstacles in real-world scenarios~\cite{beetz2017guidelines}. Ensuring robustness in diverse applications and integrating these systems into workflows adds complexity~\cite{dubey2020haco, edwards2023advise}. Capturing tacit knowledge and addressing decision-making uncertainties remain critical hurdles~\cite{huang2024co}.

{\bf HMT-based NLP and IR Applications and Their Limitations.}  
Applications of HMT-based NLP and IR demonstrate significant potential but also reveal notable limitations:
\begin{itemize}
    \item \textit{Cybersecurity Taxonomy Development:} \citet{mahaini2019building} used HMT to streamline cybersecurity taxonomy creation by integrating NLP and IR tools for term extraction, guided by human decision-making. However, reliance on manual processes limits scalability and adaptability to new domains.
    \item \textit{Natural Language Command Interpretation in Rescue Missions:} \citet{beetz2017guidelines} developed NLP systems to help robots interpret human instructions in simulated rescue scenarios. While effective in controlled settings, the system's performance is limited by unpredictable real-world elements and diverse language expressions.
    \item \textit{Framework for Human-AI Collaboration:} \citet{dubey2020haco} extended the Java Agent Development Framework (JADE) to improve human-AI interactions through a structured taxonomy and graphical interfaces, reducing development time. However, scalability and robustness in diverse contexts remain key challenges.
    \item \textit{Evidence Gap Map (EGM) Creation:} \citet{edwards2023advise} integrated AI, specifically a BERT-based model, to automate the initial screening in evidence synthesis processes, significantly reducing human workload. Yet, the approach faces challenges with larger datasets, dependency on high-quality training data, and workflow integration.
    \item \textit{Legal Case Matching:} \citet{huang2024co} proposed the Co-Matching framework, integrating human tacit knowledge and machine analysis to enhance legal document matching accuracy. Articulating tacit knowledge and estimating human decision uncertainty are key limitations, especially with limited behavioral data.
\end{itemize}

These applications highlight the promise of HMT-based NLP and IR in addressing complex challenges across various domains. However, scalability, adaptability, and the integration of human expertise into machine-readable processes remain critical areas for future research and development.

\section{Conclusions \& Future Work}
\label{sec:conclusions-future-work}

\subsection{Key Findings}
\label{subsec:insights-lessons}

This survey provides a comprehensive examination of HMT by integrating insights from computational, cognitive, and social sciences. Our findings highlight several critical advancements and challenges shaping the field.

\textbf{Advancements in HMT.} HMT research has significantly evolved across domains such as defense, healthcare, industrial automation, and autonomous systems. Advances in \textit{team training}, \textit{shared mental models (SMMs)}, and \textit{trust calibration} have contributed to improved team performance, decision-making, and adaptability. Studies demonstrate that AI-driven collaboration frameworks, such as \textit{human-in-the-loop reinforcement learning (HRL)} and \textit{instance-based learning theory (IBLT)}, enhance adaptive teaming mechanisms and efficiency. Moreover, the development of explainable AI (XAI) techniques has strengthened human trust in automated decision-making.

\textbf{Challenges in Trust, Transparency, and Ethical Considerations.} Trust remains a key issue in HMT, particularly in explainability, reliability, and ethical AI decision-making. Research shows that \textit{trust calibration} between human and AI teammates is dynamic and difficult to restore once compromised. Transparency models like \textit{situation awareness-based agent transparency (SAT)} improve trust but increase cognitive workload. In addition, ethical frameworks guiding AI behavior remain largely theoretical, with limited empirical validation in real-world HMT applications.

\textbf{Task Allocation and Decision-Making.} Effective task allocation in HMT depends on balancing automation with human oversight. While \textit{reinforcement learning} and \textit{multi-criteria decision-making} enable scalable human-AI collaboration, challenges remain in role adaptability and real-time adjustments. Research shows that \textit{team composition}, especially in mixed human-agent teams, impacts task performance, coordination, and cognitive alignment.

\textbf{Evaluation Frameworks and Benchmarking.} The need for standardized evaluation methodologies in HMT remains a critical challenge. While existing testbeds and simulations provide controlled insights into team dynamics, real-world applicability is limited. Studies leveraging \textit{NASA-TLX}, \textit{CERTT-II}, and \textit{NeoCITIES} testbeds highlight gaps in scalability and generalizability. Moreover, dataset availability remains an obstacle, as structured text and numerical datasets dominate, restricting access to multi-modal and real-world interaction data.

\subsection{Future Research Directions}
\label{subsec:future-research}
Based on the identified challenges, future research in HMT should focus on the following key areas:

\textbf{Enhancing Explainability and Trust Calibration.} Future work should explore advanced \textit{XAI} mechanisms that integrate transparency with adaptive learning models to improve user trust. Developing \textit{real-time trust assessment} methods will enable HMT systems to dynamically adjust interactions based on evolving human-agent relationships.

\textbf{Human-Centered AI Design.} Research should emphasize \textit{user-centered design principles} to improve HMT effectiveness by ensuring AI teammates align with human cognitive models, decision-making patterns, and workload considerations. Future studies should explore \textit{adaptive interfaces} that enhance usability and foster intuitive collaboration.

\textbf{Scalability and Real-World Deployment.} Transitioning HMT systems from controlled settings to real-world applications remains challenging. Future research should explore \textit{scalable architectures} that balance performance and adaptability while ensuring robust human-agent interaction within domain-specific constraints.

\textbf{Standardized Evaluation Metrics and Testbeds.} Establishing \textit{benchmarking standards} for HMT systems is essential to ensure comparability across studies. Developing \textit{large-scale empirical testbeds} that integrate real-world HMT scenarios will provide more robust performance evaluations. Future research should also address the reproducibility crisis in HMT by improving dataset accessibility and adopting \textit{open-source benchmarking frameworks}.

\textbf{Ethical and Regulatory Considerations.} As HMT expands into high-stakes environments, future research must tackle ethical AI governance, bias mitigation, and accountability in autonomous decisions. Developing \textit{policy frameworks} that align AI autonomy with human oversight is crucial for responsible HMT deployment.

\textbf{Multi-Modal and Cross-Domain Applications.} Future studies should move beyond text-based interactions to \textit{multimodal AI systems}, including visual, auditory, and sensor-based communication. Expanding HMT to \textit{disaster response, cybersecurity, and human-robot teaming} will improve adaptability and scalability.

\textbf{Real-World Validation of HMT Theories.} While theoretical models such as \textit{HRL}, \textit{IBLT}, and \textit{Interdependence Theory (IT)} offer valuable insights, their real-world applicability remains largely unexplored. Future work should focus on validating these frameworks through longitudinal studies and large-scale empirical deployments.

By addressing these directions, HMT research can advance toward more resilient, ethical, and scalable human-machine collaboration models. Continued interdisciplinary research, industry-academic partnerships, and policy-driven AI governance will be essential in shaping the next generation of HMT systems.

\section*{Acknowledgment}
This research is partially supported by the U.S. Army Research Office (ARO) Award (W911NF-24-2-0241) and the National Science Foundation (NSF) Secure and Trustworthy Cyberspace (SaTC) Award (2330940).  

\bibliographystyle{ACM-Reference-Format}
\bibliography{ref}


\begin{thebibliography}{177}


\ifx \showCODEN    \undefined \def \showCODEN     #1{\unskip}     \fi
\ifx \showDOI      \undefined \def \showDOI       #1{#1}\fi
\ifx \showISBNx    \undefined \def \showISBNx     #1{\unskip}     \fi
\ifx \showISBNxiii \undefined \def \showISBNxiii  #1{\unskip}     \fi
\ifx \showISSN     \undefined \def \showISSN      #1{\unskip}     \fi
\ifx \showLCCN     \undefined \def \showLCCN      #1{\unskip}     \fi
\ifx \shownote     \undefined \def \shownote      #1{#1}          \fi
\ifx \showarticletitle \undefined \def \showarticletitle #1{#1}   \fi
\ifx \showURL      \undefined \def \showURL       {\relax}        \fi
\providecommand\bibfield[2]{#2}
\providecommand\bibinfo[2]{#2}
\providecommand\natexlab[1]{#1}
\providecommand\showeprint[2][]{arXiv:#2}

\bibitem[Abbass and Hunjet(2021)]%
        {abbass2021smart}
\bibfield{author}{\bibinfo{person}{Hussein~A Abbass} {and}
  \bibinfo{person}{Robert~A Hunjet}.} \bibinfo{year}{2021}\natexlab{}.
\newblock \showarticletitle{Smart shepherding: Towards transparent artificial
  intelligence enabled human-swarm teams}.
\newblock \bibinfo{journal}{\emph{Shepherding UxVs for human-swarm teaming: An
  artificial intelligence approach to unmanned X vehicles}}
  (\bibinfo{year}{2021}), \bibinfo{pages}{1--28}.
\newblock


\bibitem[Aha et~al\mbox{.}(1991)]%
        {aha1991instance}
\bibfield{author}{\bibinfo{person}{David~W Aha}, \bibinfo{person}{Dennis
  Kibler}, {and} \bibinfo{person}{Marc~K Albert}.}
  \bibinfo{year}{1991}\natexlab{}.
\newblock \showarticletitle{Instance-based learning algorithms}.
\newblock \bibinfo{journal}{\emph{Machine learning}}  \bibinfo{volume}{6}
  (\bibinfo{year}{1991}), \bibinfo{pages}{37--66}.
\newblock


\bibitem[Al-Shafei(2024)]%
        {al2024navigating}
\bibfield{author}{\bibinfo{person}{Mohamed Al-Shafei}.}
  \bibinfo{year}{2024}\natexlab{}.
\newblock \showarticletitle{Navigating Human-Chatbot Interactions: An
  Investigation into Factors Influencing User Satisfaction and Engagement}.
\newblock \bibinfo{journal}{\emph{International Journal of Human--Computer
  Interaction}} (\bibinfo{year}{2024}), \bibinfo{pages}{1--18}.
\newblock


\bibitem[Alhaji et~al\mbox{.}(2020)]%
        {alhaji2020toward}
\bibfield{author}{\bibinfo{person}{Basel Alhaji}, \bibinfo{person}{Andreas
  Rausch}, {and} \bibinfo{person}{Michael Prilla}.}
  \bibinfo{year}{2020}\natexlab{}.
\newblock \showarticletitle{Toward Mutual Trust Modeling in Human-Robot
  Collaboration}.
\newblock \bibinfo{journal}{\emph{arXiv preprint arXiv:2011.01056}}
  (\bibinfo{year}{2020}).
\newblock


\bibitem[Ali et~al\mbox{.}(2019)]%
        {ali2019constructionism}
\bibfield{author}{\bibinfo{person}{Safinah Ali}, \bibinfo{person}{Blakeley~H
  Payne}, \bibinfo{person}{Randi Williams}, \bibinfo{person}{Hae~Won Park},
  {and} \bibinfo{person}{Cynthia Breazeal}.} \bibinfo{year}{2019}\natexlab{}.
\newblock \showarticletitle{Constructionism, ethics, and creativity: Developing
  primary and middle school artificial intelligence education}. In
  \bibinfo{booktitle}{\emph{International workshop on education in artificial
  intelligence k-12 (eduai’19)}}, Vol.~\bibinfo{volume}{2}.
  \bibinfo{pages}{1--4}.
\newblock


\bibitem[Andrews et~al\mbox{.}(2023)]%
        {andrews2023role}
\bibfield{author}{\bibinfo{person}{Robert~W Andrews}, \bibinfo{person}{J~Mason
  Lilly}, \bibinfo{person}{Divya Srivastava}, {and} \bibinfo{person}{Karen~M
  Feigh}.} \bibinfo{year}{2023}\natexlab{}.
\newblock \showarticletitle{The role of shared mental models in human-AI teams:
  a theoretical review}.
\newblock \bibinfo{journal}{\emph{Theoretical Issues in Ergonomics Science}}
  \bibinfo{volume}{24}, \bibinfo{number}{2} (\bibinfo{year}{2023}),
  \bibinfo{pages}{129--175}.
\newblock


\bibitem[Assaad and Boshuijzen-van Burken(2023)]%
        {assaad2023ethics}
\bibfield{author}{\bibinfo{person}{Zena Assaad} {and}
  \bibinfo{person}{Christine Boshuijzen-van Burken}.}
  \bibinfo{year}{2023}\natexlab{}.
\newblock \showarticletitle{Ethics and Safety of Human-Machine Teaming}. In
  \bibinfo{booktitle}{\emph{Proceedings of the First International Symposium on
  Trustworthy Autonomous Systems}}. \bibinfo{pages}{1--8}.
\newblock


\bibitem[Balayogi et~al\mbox{.}(2025)]%
        {balayogi2025human}
\bibfield{author}{\bibinfo{person}{G Balayogi}, \bibinfo{person}{A~Vijaya
  Lakshmi}, {and} \bibinfo{person}{S~Lourdumarie Sophie}.}
  \bibinfo{year}{2025}\natexlab{}.
\newblock \showarticletitle{Human-Centric Ethical {AI} in the Digital World}.
\newblock In \bibinfo{booktitle}{\emph{Ethical Dimensions of AI Development}}.
  \bibinfo{publisher}{IGI Global}, \bibinfo{pages}{175--196}.
\newblock


\bibitem[Basri(2020)]%
        {basri2020examining}
\bibfield{author}{\bibinfo{person}{Wael Basri}.}
  \bibinfo{year}{2020}\natexlab{}.
\newblock \showarticletitle{Examining the impact of artificial intelligence
  (AI)-assisted social media marketing on the performance of small and medium
  enterprises: toward effective business management in the Saudi Arabian
  context}.
\newblock \bibinfo{journal}{\emph{International Journal of Computational
  Intelligence Systems}} \bibinfo{volume}{13}, \bibinfo{number}{1}
  (\bibinfo{year}{2020}), \bibinfo{pages}{142--152}.
\newblock


\bibitem[Beetz et~al\mbox{.}(2017)]%
        {beetz2017guidelines}
\bibfield{author}{\bibinfo{person}{Michael Beetz}, \bibinfo{person}{Matthias
  Scheutz}, {and} \bibinfo{person}{Fereshta Yazdani}.}
  \bibinfo{year}{2017}\natexlab{}.
\newblock \showarticletitle{Guidelines for improving task-based natural
  language understanding in human-robot rescue teams}. In
  \bibinfo{booktitle}{\emph{2017 8th IEEE International Conference on Cognitive
  Infocommunications (CogInfoCom)}}. IEEE, \bibinfo{pages}{000203--000208}.
\newblock


\bibitem[Berkol and Demirtaş(2023)]%
        {Berkol23-hmt-nlp}
\bibfield{author}{\bibinfo{person}{Ali Berkol} {and}
  \bibinfo{person}{İdil~Gökçe Demirtaş}.} \bibinfo{year}{2023}\natexlab{}.
\newblock \showarticletitle{Advancing Human-Machine Interaction: Speech and
  Natural Language Processing}. In \bibinfo{booktitle}{\emph{2023 7th
  International Symposium on Innovative Approaches in Smart Technologies
  (ISAS)}}. \bibinfo{pages}{1--5}.
\newblock
\urldef\tempurl%
\url{https://doi.org/10.1109/ISAS60782.2023.10391814}
\showDOI{\tempurl}


\bibitem[Berretta et~al\mbox{.}(2023)]%
        {berretta2023defining}
\bibfield{author}{\bibinfo{person}{Sophie Berretta}, \bibinfo{person}{Alina
  Tausch}, \bibinfo{person}{Greta Ontrup}, \bibinfo{person}{Bj{\"o}rn Gilles},
  \bibinfo{person}{Corinna Peifer}, {and} \bibinfo{person}{Annette Kluge}.}
  \bibinfo{year}{2023}\natexlab{}.
\newblock \showarticletitle{Defining human-AI teaming the human-centered way: a
  scoping review and network analysis}.
\newblock \bibinfo{journal}{\emph{Frontiers in Artificial Intelligence}}
  \bibinfo{volume}{6} (\bibinfo{year}{2023}), \bibinfo{pages}{1250725}.
\newblock


\bibitem[Bersani et~al\mbox{.}(2023)]%
        {bersani2023towards}
\bibfield{author}{\bibinfo{person}{Marcello~M Bersani}, \bibinfo{person}{Matteo
  Camilli}, \bibinfo{person}{Livia Lestingi}, \bibinfo{person}{Raffaela
  Mirandola}, \bibinfo{person}{Matteo Rossi}, {and} \bibinfo{person}{Patrizia
  Scandurra}.} \bibinfo{year}{2023}\natexlab{}.
\newblock \showarticletitle{Towards Better Trust in Human-Machine Teaming
  through Explainable Dependability}. In \bibinfo{booktitle}{\emph{2023 IEEE
  20th International Conference on Software Architecture Companion (ICSA-C)}}.
  IEEE, \bibinfo{pages}{86--90}.
\newblock


\bibitem[Blaha et~al\mbox{.}(2020)]%
        {blaha2020cognitive}
\bibfield{author}{\bibinfo{person}{Leslie~M Blaha}, \bibinfo{person}{Christian
  Lebiere}, \bibinfo{person}{Corey~K Fallon}, {and} \bibinfo{person}{Brett~A
  Jefferson}.} \bibinfo{year}{2020}\natexlab{}.
\newblock \showarticletitle{Cognitive mechanisms for calibrating trust and
  reliance on automation}. In \bibinfo{booktitle}{\emph{Proceedings of the 18th
  International Conference of Cognitive Modeling, July}}.
  \bibinfo{pages}{20--31}.
\newblock


\bibitem[Brown(2019)]%
        {brown2019designing}
\bibfield{author}{\bibinfo{person}{Philip~N Brown}.}
  \bibinfo{year}{2019}\natexlab{}.
\newblock \showarticletitle{Designing for Emergent Security in Heterogeneous
  Human-Machine teams}. In \bibinfo{booktitle}{\emph{2019 IEEE 58th Conference
  on Decision and Control (CDC)}}. IEEE, \bibinfo{pages}{2175--2180}.
\newblock


\bibitem[Cai et~al\mbox{.}(2020)]%
        {cai2020perceiving}
\bibfield{author}{\bibinfo{person}{Yang Cai}, \bibinfo{person}{Jose~A Morales},
  \bibinfo{person}{William Casey}, \bibinfo{person}{Neta Ezer}, {and}
  \bibinfo{person}{Sihan Wang}.} \bibinfo{year}{2020}\natexlab{}.
\newblock \showarticletitle{Perceiving Behavior of Cyber Malware with
  Human-Machine Teaming}. In \bibinfo{booktitle}{\emph{Advances in Human
  Factors in Cybersecurity: Proceedings of the AHFE 2019 International
  Conference on Human Factors in Cybersecurity, July 24-28, 2019, Washington
  DC, USA 10}}. Springer, \bibinfo{pages}{97--108}.
\newblock


\bibitem[Calhoun et~al\mbox{.}(2018)]%
        {calhoun2018human}
\bibfield{author}{\bibinfo{person}{Gloria~L Calhoun}, \bibinfo{person}{Heath~A
  Ruff}, \bibinfo{person}{Kyle~J Behymer}, {and} \bibinfo{person}{Elizabeth~M
  Frost}.} \bibinfo{year}{2018}\natexlab{}.
\newblock \showarticletitle{Human-autonomy teaming interface design
  considerations for multi-unmanned vehicle control}.
\newblock \bibinfo{journal}{\emph{Theoretical issues in ergonomics science}}
  \bibinfo{volume}{19}, \bibinfo{number}{3} (\bibinfo{year}{2018}),
  \bibinfo{pages}{321--352}.
\newblock


\bibitem[Centeio~Jorge et~al\mbox{.}(2024)]%
        {centeio2024interdependence}
\bibfield{author}{\bibinfo{person}{Carolina Centeio~Jorge},
  \bibinfo{person}{Catholijn~M Jonker}, {and} \bibinfo{person}{Myrthe~L
  Tielman}.} \bibinfo{year}{2024}\natexlab{}.
\newblock \showarticletitle{Interdependence and trust analysis (ITA): a
  framework for human--machine team design}.
\newblock \bibinfo{journal}{\emph{Behaviour \& Information Technology}}
  (\bibinfo{year}{2024}), \bibinfo{pages}{1--21}.
\newblock


\bibitem[Chang et~al\mbox{.}(2020)]%
        {fairness_chang}
\bibfield{author}{\bibinfo{person}{Mai~Lee Chang}, \bibinfo{person}{Zachary
  Pope}, \bibinfo{person}{Elaine~Schaertl Short}, {and} \bibinfo{person}{Andrea
  Lockerd~Thomaz}.} \bibinfo{year}{2020}\natexlab{}.
\newblock \showarticletitle{Defining Fairness in Human-Robot Teams}. In
  \bibinfo{booktitle}{\emph{2020 29th IEEE International Conference on Robot
  and Human Interactive Communication (RO-MAN)}}. \bibinfo{pages}{1251--1258}.
\newblock
\urldef\tempurl%
\url{https://doi.org/10.1109/RO-MAN47096.2020.9223594}
\showDOI{\tempurl}


\bibitem[Chen et~al\mbox{.}(2021)]%
        {automation2021chen}
\bibfield{author}{\bibinfo{person}{Hualong Chen}, \bibinfo{person}{Yuanqiao
  Wen}, \bibinfo{person}{Man Zhu}, \bibinfo{person}{Yamin Huang},
  \bibinfo{person}{Changshi Xiao}, \bibinfo{person}{Tao Wei}, {and}
  \bibinfo{person}{Axel Hahn}.} \bibinfo{year}{2021}\natexlab{}.
\newblock \showarticletitle{From Automation System to Autonomous System: An
  Architecture Perspective}.
\newblock \bibinfo{journal}{\emph{Journal of Marine Science and Engineering}}
  \bibinfo{volume}{9}, \bibinfo{number}{6} (\bibinfo{date}{Jun}
  \bibinfo{year}{2021}), \bibinfo{pages}{645}.
\newblock
\showISSN{2077-1312}
\urldef\tempurl%
\url{https://doi.org/10.3390/jmse9060645}
\showDOI{\tempurl}


\bibitem[Chen and Zimbra(2010)]%
        {chen2010ai}
\bibfield{author}{\bibinfo{person}{Hsinchun Chen} {and} \bibinfo{person}{David
  Zimbra}.} \bibinfo{year}{2010}\natexlab{}.
\newblock \showarticletitle{AI and opinion mining}.
\newblock \bibinfo{journal}{\emph{IEEE Intelligent Systems}}
  \bibinfo{volume}{25}, \bibinfo{number}{3} (\bibinfo{year}{2010}),
  \bibinfo{pages}{74--80}.
\newblock


\bibitem[Chen et~al\mbox{.}(2023)]%
        {chen2023human}
\bibfield{author}{\bibinfo{person}{Jiahao Chen}, \bibinfo{person}{Fu Guo},
  \bibinfo{person}{Zenggen Ren}, \bibinfo{person}{Xueshuang Wang}, {and}
  \bibinfo{person}{Jaap Ham}.} \bibinfo{year}{2023}\natexlab{}.
\newblock \showarticletitle{Human-chatbot interaction studies through the lens
  of bibliometric analysis}.
\newblock \bibinfo{journal}{\emph{Universal Access in the Information Society}}
  (\bibinfo{year}{2023}).
\newblock
\urldef\tempurl%
\url{https://link.springer.com/article/10.1007/s10209-023-01058-y}
\showURL{%
\tempurl}


\bibitem[Chen and Barnes(2014)]%
        {chen2014human}
\bibfield{author}{\bibinfo{person}{Jessie~YC Chen} {and}
  \bibinfo{person}{Michael~J Barnes}.} \bibinfo{year}{2014}\natexlab{}.
\newblock \showarticletitle{Human--agent teaming for multirobot control: A
  review of human factors issues}.
\newblock \bibinfo{journal}{\emph{IEEE Transactions on Human-Machine Systems}}
  \bibinfo{volume}{44}, \bibinfo{number}{1} (\bibinfo{year}{2014}),
  \bibinfo{pages}{13--29}.
\newblock


\bibitem[Chen(2017)]%
        {chen2017tsa}
\bibfield{author}{\bibinfo{person}{Jessie Y.~C. Chen}.}
  \bibinfo{year}{2017}\natexlab{}.
\newblock \showarticletitle{Situation awareness-based agent transparency for
  human-autonomy teaming effectiveness}.
\newblock \bibinfo{journal}{\emph{Taylor \& Francis}} (\bibinfo{year}{2017}).
\newblock


\bibitem[Chiou et~al\mbox{.}(2021)]%
        {chiou2021towards}
\bibfield{author}{\bibinfo{person}{Erin~K Chiou}, \bibinfo{person}{Mustafa
  Demir}, \bibinfo{person}{Verica Buchanan}, \bibinfo{person}{Christopher~C
  Corral}, \bibinfo{person}{Mica~R Endsley}, \bibinfo{person}{Glenn~J Lematta},
  \bibinfo{person}{Nancy~J Cooke}, {and} \bibinfo{person}{Nathan~J McNeese}.}
  \bibinfo{year}{2021}\natexlab{}.
\newblock \showarticletitle{Towards human--robot teaming: Tradeoffs of
  explanation-based communication strategies in a virtual search and rescue
  task}.
\newblock \bibinfo{journal}{\emph{International Journal of Social Robotics}}
  (\bibinfo{year}{2021}), \bibinfo{pages}{1--20}.
\newblock


\bibitem[Cho et~al\mbox{.}(2019)]%
        {cho2019stram}
\bibfield{author}{\bibinfo{person}{Jin-Hee Cho}, \bibinfo{person}{Shouhuai Xu},
  \bibinfo{person}{Patrick~M Hurley}, \bibinfo{person}{Matthew Mackay},
  \bibinfo{person}{Trevor Benjamin}, {and} \bibinfo{person}{Mark Beaumont}.}
  \bibinfo{year}{2019}\natexlab{}.
\newblock \showarticletitle{{STRAM}: Measuring the trustworthiness of
  computer-based systems}.
\newblock \bibinfo{journal}{\emph{ACM Computing Surveys (CSUR)}}
  \bibinfo{volume}{51}, \bibinfo{number}{6} (\bibinfo{year}{2019}),
  \bibinfo{pages}{1--47}.
\newblock


\bibitem[Christopher~Brill et~al\mbox{.}(2018)]%
        {christopher2018navigating}
\bibfield{author}{\bibinfo{person}{J Christopher~Brill}, \bibinfo{person}{ML
  Cummings}, \bibinfo{person}{AW Evans~III}, \bibinfo{person}{Peter~A Hancock},
  \bibinfo{person}{Joseph~B Lyons}, {and} \bibinfo{person}{Kevin Oden}.}
  \bibinfo{year}{2018}\natexlab{}.
\newblock \showarticletitle{Navigating the advent of human-machine teaming}. In
  \bibinfo{booktitle}{\emph{Proceedings of the human factors and ergonomics
  society annual meeting}}, Vol.~\bibinfo{volume}{62}. SAGE Publications Sage
  CA: Los Angeles, CA, \bibinfo{pages}{455--459}.
\newblock


\bibitem[Ciechanowski et~al\mbox{.}(2019)]%
        {ciechanowski2019shades}
\bibfield{author}{\bibinfo{person}{Leon Ciechanowski},
  \bibinfo{person}{Aleksandra Przegalinska}, \bibinfo{person}{Mikolaj
  Magnuski}, {and} \bibinfo{person}{Peter Gloor}.}
  \bibinfo{year}{2019}\natexlab{}.
\newblock \showarticletitle{In the shades of the uncanny valley: An
  experimental study of human--chatbot interaction}.
\newblock \bibinfo{journal}{\emph{Future Generation Computer Systems}}
  \bibinfo{volume}{92} (\bibinfo{year}{2019}), \bibinfo{pages}{539--548}.
\newblock


\bibitem[Ciocirlan et~al\mbox{.}(2019)]%
        {ciocirlan2019human}
\bibfield{author}{\bibinfo{person}{Stefan-Dan Ciocirlan},
  \bibinfo{person}{Roxana Agrigoroaie}, {and} \bibinfo{person}{Adriana Tapus}.}
  \bibinfo{year}{2019}\natexlab{}.
\newblock \showarticletitle{Human-robot team: effects of communication in
  analyzing trust}. In \bibinfo{booktitle}{\emph{2019 28th IEEE International
  Conference on Robot and Human Interactive Communication (RO-MAN)}}. IEEE,
  \bibinfo{pages}{1--7}.
\newblock


\bibitem[Cleland-Huang et~al\mbox{.}(2022)]%
        {cleland2022extending}
\bibfield{author}{\bibinfo{person}{Jane Cleland-Huang}, \bibinfo{person}{Ankit
  Agrawal}, \bibinfo{person}{Michael Vierhauser}, \bibinfo{person}{Michael
  Murphy}, {and} \bibinfo{person}{Mike Prieto}.}
  \bibinfo{year}{2022}\natexlab{}.
\newblock \showarticletitle{Extending {MAPE-K} to support human-machine
  teaming}. In \bibinfo{booktitle}{\emph{Proceedings of the 17th Symposium on
  Software Engineering for Adaptive and Self-Managing Systems}}.
  \bibinfo{pages}{120--131}.
\newblock


\bibitem[Cleland-Huang et~al\mbox{.}(2024)]%
        {cleland2024human}
\bibfield{author}{\bibinfo{person}{Jane Cleland-Huang},
  \bibinfo{person}{Theodore Chambers}, \bibinfo{person}{Sebastian Zudaire},
  \bibinfo{person}{Muhammed~Tawfiq Chowdhury}, \bibinfo{person}{Ankit Agrawal},
  {and} \bibinfo{person}{Michael Vierhauser}.} \bibinfo{year}{2024}\natexlab{}.
\newblock \showarticletitle{Human--machine Teaming with Small Unmanned Aerial
  Systems in a MAPE-K Environment}.
\newblock \bibinfo{journal}{\emph{ACM Transactions on Autonomous and Adaptive
  Systems}} \bibinfo{volume}{19}, \bibinfo{number}{1} (\bibinfo{year}{2024}),
  \bibinfo{pages}{1--35}.
\newblock


\bibitem[Connors(1998)]%
        {connors1998teaming}
\bibfield{author}{\bibinfo{person}{Mary~M Connors}.}
  \bibinfo{year}{1998}\natexlab{}.
\newblock \showarticletitle{Teaming humans and automated systems in safely
  engineered environments}.
\newblock \bibinfo{journal}{\emph{Life Support \& Biosphere Science}}
  \bibinfo{volume}{5}, \bibinfo{number}{4} (\bibinfo{year}{1998}),
  \bibinfo{pages}{453--460}.
\newblock


\bibitem[Cooke et~al\mbox{.}(2013)]%
        {cooke2013interactive}
\bibfield{author}{\bibinfo{person}{Nancy~J Cooke}, \bibinfo{person}{Jamie~C
  Gorman}, \bibinfo{person}{Christopher~W Myers}, {and}
  \bibinfo{person}{Jasmine~L Duran}.} \bibinfo{year}{2013}\natexlab{}.
\newblock \showarticletitle{Interactive team cognition}.
\newblock \bibinfo{journal}{\emph{Cognitive Science}} \bibinfo{volume}{37},
  \bibinfo{number}{2} (\bibinfo{year}{2013}), \bibinfo{pages}{255--285}.
\newblock


\bibitem[Cummings et~al\mbox{.}(2008)]%
        {cummings08}
\bibfield{author}{\bibinfo{person}{Missy Cummings}, \bibinfo{person}{P. Pina},
  {and} \bibinfo{person}{Birsen Donmez}.} \bibinfo{year}{2008}\natexlab{}.
\newblock \showarticletitle{Selecting Metrics to Evaluate Human Supervisory
  Control Applications}.
\newblock  (\bibinfo{year}{2008}).
\newblock


\bibitem[Cummings et~al\mbox{.}(2010)]%
        {cummings2010role}
\bibfield{author}{\bibinfo{person}{Mary~L Cummings}, \bibinfo{person}{Andrew
  Clare}, {and} \bibinfo{person}{Christin Hart}.}
  \bibinfo{year}{2010}\natexlab{}.
\newblock \showarticletitle{The role of human-automation consensus in multiple
  unmanned vehicle scheduling}.
\newblock \bibinfo{journal}{\emph{Human Factors}} \bibinfo{volume}{52},
  \bibinfo{number}{1} (\bibinfo{year}{2010}), \bibinfo{pages}{17--27}.
\newblock


\bibitem[Cummings et~al\mbox{.}(2011)]%
        {cummings2011impact}
\bibfield{author}{\bibinfo{person}{Mary~L Cummings},
  \bibinfo{person}{Jonathan~P How}, \bibinfo{person}{Andrew Whitten}, {and}
  \bibinfo{person}{Olivier Toupet}.} \bibinfo{year}{2011}\natexlab{}.
\newblock \showarticletitle{The impact of human--automation collaboration in
  decentralized multiple unmanned vehicle control}.
\newblock \bibinfo{journal}{\emph{Proc. IEEE}} \bibinfo{volume}{100},
  \bibinfo{number}{3} (\bibinfo{year}{2011}), \bibinfo{pages}{660--671}.
\newblock


\bibitem[Damacharla et~al\mbox{.}(2018)]%
        {damacharla2018common}
\bibfield{author}{\bibinfo{person}{Praveen Damacharla},
  \bibinfo{person}{Ahmad~Y Javaid}, \bibinfo{person}{Jennie~J Gallimore}, {and}
  \bibinfo{person}{Vijay~K Devabhaktuni}.} \bibinfo{year}{2018}\natexlab{}.
\newblock \showarticletitle{Common metrics to benchmark human-machine teams
  {(HMT)}: A review}.
\newblock \bibinfo{journal}{\emph{IEEE Access}}  \bibinfo{volume}{6}
  (\bibinfo{year}{2018}), \bibinfo{pages}{38637--38655}.
\newblock


\bibitem[de~Paula et~al\mbox{.}(2023)]%
        {de2023managerial}
\bibfield{author}{\bibinfo{person}{Danielly de Paula}, \bibinfo{person}{Carolin
  Marx}, \bibinfo{person}{Ella Wolf}, \bibinfo{person}{Christian Dremel},
  \bibinfo{person}{Kathryn Cormican}, {and} \bibinfo{person}{Falk
  Uebernickel}.} \bibinfo{year}{2023}\natexlab{}.
\newblock \showarticletitle{A managerial mental model to drive innovation in
  the context of digital transformation}.
\newblock \bibinfo{journal}{\emph{Industry and Innovation}}
  \bibinfo{volume}{30}, \bibinfo{number}{1} (\bibinfo{year}{2023}),
  \bibinfo{pages}{42--66}.
\newblock


\bibitem[Demir et~al\mbox{.}(2021a)]%
        {demir2021modeling}
\bibfield{author}{\bibinfo{person}{Mustafa Demir}, \bibinfo{person}{Mustafa
  Canan}, {and} \bibinfo{person}{Myke~C Cohen}.}
  \bibinfo{year}{2021}\natexlab{a}.
\newblock \showarticletitle{Modeling team interaction and interactive
  decision-making in agile human-machine teams}. In
  \bibinfo{booktitle}{\emph{2021 IEEE 2nd International Conference on
  Human-Machine Systems (ICHMS)}}. IEEE, \bibinfo{pages}{1--6}.
\newblock


\bibitem[Demir et~al\mbox{.}(2018)]%
        {demir2018team}
\bibfield{author}{\bibinfo{person}{Mustafa Demir}, \bibinfo{person}{Aaron~D
  Likens}, \bibinfo{person}{Nancy~J Cooke}, \bibinfo{person}{Polemnia~G
  Amazeen}, {and} \bibinfo{person}{Nathan~J McNeese}.}
  \bibinfo{year}{2018}\natexlab{}.
\newblock \showarticletitle{Team coordination and effectiveness in
  human-autonomy teaming}.
\newblock \bibinfo{journal}{\emph{IEEE Transactions on Human-Machine Systems}}
  \bibinfo{volume}{49}, \bibinfo{number}{2} (\bibinfo{year}{2018}),
  \bibinfo{pages}{150--159}.
\newblock


\bibitem[Demir et~al\mbox{.}(2020)]%
        {demir2020understanding}
\bibfield{author}{\bibinfo{person}{Mustafa Demir}, \bibinfo{person}{{Nathan J.}
  McNeese}, {and} \bibinfo{person}{{Nancy J.} Cooke}.}
  \bibinfo{year}{2020}\natexlab{}.
\newblock \showarticletitle{Understanding human-robot teams in light of
  all-human teams: Aspects of team interaction and shared cognition}.
\newblock \bibinfo{journal}{\emph{International Journal of Human Computer
  Studies}}  \bibinfo{volume}{140} (\bibinfo{date}{Aug.} \bibinfo{year}{2020}).
\newblock
\showISSN{1071-5819}
\urldef\tempurl%
\url{https://doi.org/10.1016/j.ijhcs.2020.102436}
\showDOI{\tempurl}


\bibitem[Demir et~al\mbox{.}(2017a)]%
        {demir2017tsa}
\bibfield{author}{\bibinfo{person}{Mustafa Demir}, \bibinfo{person}{Nathan~J
  McNeese}, {and} \bibinfo{person}{Nancy~J Cooke}.}
  \bibinfo{year}{2017}\natexlab{a}.
\newblock \showarticletitle{Team situation awareness within the context of
  human-autonomy teaming}.
\newblock \bibinfo{journal}{\emph{Cognitive Systems Research}}
  \bibinfo{volume}{46} (\bibinfo{year}{2017}), \bibinfo{pages}{3--12}.
\newblock


\bibitem[Demir et~al\mbox{.}(2017b)]%
        {demir2017team}
\bibfield{author}{\bibinfo{person}{Mustafa Demir}, \bibinfo{person}{Nathan~J
  McNeese}, {and} \bibinfo{person}{Nancy~J Cooke}.}
  \bibinfo{year}{2017}\natexlab{b}.
\newblock \showarticletitle{Team situation awareness within the context of
  human-autonomy teaming}.
\newblock \bibinfo{journal}{\emph{Cognitive Systems Research}}
  \bibinfo{volume}{46} (\bibinfo{year}{2017}), \bibinfo{pages}{3--12}.
\newblock


\bibitem[Demir et~al\mbox{.}(2019)]%
        {demir2019evolution}
\bibfield{author}{\bibinfo{person}{Mustafa Demir}, \bibinfo{person}{Nathan~J
  McNeese}, {and} \bibinfo{person}{Nancy~J Cooke}.}
  \bibinfo{year}{2019}\natexlab{}.
\newblock \showarticletitle{The evolution of human-autonomy teams in remotely
  piloted aircraft systems operations}.
\newblock \bibinfo{journal}{\emph{Frontiers in Communication}}
  \bibinfo{volume}{4} (\bibinfo{year}{2019}), \bibinfo{pages}{50}.
\newblock


\bibitem[Demir et~al\mbox{.}(2021b)]%
        {Demir2021Exploration}
\bibfield{author}{\bibinfo{person}{Mustafa Demir}, \bibinfo{person}{Nathan~J.
  McNeese}, \bibinfo{person}{Jaime~C. Gorman}, \bibinfo{person}{Nancy~J.
  Cooke}, \bibinfo{person}{Christopher~W. Myers}, {and}
  \bibinfo{person}{David~A. Grimm}.} \bibinfo{year}{2021}\natexlab{b}.
\newblock \showarticletitle{Exploration of Teammate Trust and Interaction
  Dynamics in Human-Autonomy Teaming}.
\newblock \bibinfo{journal}{\emph{IEEE Transactions on Human-Machine Systems}}
  \bibinfo{volume}{51}, \bibinfo{number}{6} (\bibinfo{year}{2021}),
  \bibinfo{pages}{696--705}.
\newblock
\urldef\tempurl%
\url{https://doi.org/10.1109/THMS.2021.3115058}
\showDOI{\tempurl}


\bibitem[Dong et~al\mbox{.}(2022)]%
        {dong2022toward}
\bibfield{author}{\bibinfo{person}{Wenli Dong}, \bibinfo{person}{Weining Fang},
  \bibinfo{person}{Beiyuan Guo}, \bibinfo{person}{Jianxin Wang}, {and}
  \bibinfo{person}{Haifeng Bao}.} \bibinfo{year}{2022}\natexlab{}.
\newblock \showarticletitle{Toward Adaptive Trust Management for
  Human-Automation Teaming Using an Instance-Based Learning Cognitive Model}.
\newblock \bibinfo{journal}{\emph{Ergonomics In Design}} \bibinfo{volume}{47},
  \bibinfo{number}{47} (\bibinfo{year}{2022}).
\newblock


\bibitem[Dreslin and Baweja(2023)]%
        {dreslin2023shoulda}
\bibfield{author}{\bibinfo{person}{Brandon~D Dreslin} {and}
  \bibinfo{person}{Jessica~A Baweja}.} \bibinfo{year}{2023}\natexlab{}.
\newblock \bibinfo{booktitle}{\emph{``Shoulda, Coulda, Woulda'':
  Conceptualizing the Differences in Trust Between Human-Human Teaming and
  Human-Machine Teaming}}.
\newblock \bibinfo{type}{{T}echnical {R}eport}. \bibinfo{institution}{Pacific
  Northwest National Laboratory (PNNL), Richland, WA (United States)}.
\newblock


\bibitem[Dubey et~al\mbox{.}(2020)]%
        {dubey2020haco}
\bibfield{author}{\bibinfo{person}{Alpana Dubey}, \bibinfo{person}{Kumar
  Abhinav}, \bibinfo{person}{Sakshi Jain}, \bibinfo{person}{Veenu Arora}, {and}
  \bibinfo{person}{Asha Puttaveerana}.} \bibinfo{year}{2020}\natexlab{}.
\newblock \showarticletitle{{HACO}: a framework for developing human-AI
  teaming}. In \bibinfo{booktitle}{\emph{Proceedings of the 13th Innovations in
  Software Engineering Conference on Formerly known as India Software
  Engineering Conference}}. \bibinfo{pages}{1--9}.
\newblock


\bibitem[Dubois and Le~Ny(2020)]%
        {dubois2020adaptive}
\bibfield{author}{\bibinfo{person}{Cl{\'e}mence Dubois} {and}
  \bibinfo{person}{Jerome Le~Ny}.} \bibinfo{year}{2020}\natexlab{}.
\newblock \showarticletitle{Adaptive task allocation in human-machine teams
  with trust and workload cognitive models}. In \bibinfo{booktitle}{\emph{2020
  IEEE International Conference on Systems, Man, and Cybernetics (SMC)}}. IEEE,
  \bibinfo{pages}{3241--3246}.
\newblock


\bibitem[Edgar et~al\mbox{.}(2023)]%
        {edgar2023improving}
\bibfield{author}{\bibinfo{person}{Gwendolyn Edgar}, \bibinfo{person}{Matthew
  McWilliams}, {and} \bibinfo{person}{Matthias Scheutz}.}
  \bibinfo{year}{2023}\natexlab{}.
\newblock \showarticletitle{Improving Human-Robot Team Performance with
  Proactivity and Shared Mental Models}. In
  \bibinfo{booktitle}{\emph{Proceedings of the 2023 International Conference on
  Autonomous Agents and Multiagent Systems}}. \bibinfo{pages}{2322--2324}.
\newblock


\bibitem[Edwards et~al\mbox{.}(2023)]%
        {edwards2023advise}
\bibfield{author}{\bibinfo{person}{Kristen Edwards}, \bibinfo{person}{Binyang
  Song}, \bibinfo{person}{Jaron Porciello}, \bibinfo{person}{Mark Engelbert},
  \bibinfo{person}{Carolyn Huang}, {and} \bibinfo{person}{Faez Ahmed}.}
  \bibinfo{year}{2023}\natexlab{}.
\newblock \showarticletitle{{ADVISE}: Accelerating the Creation of Evidence
  Syntheses for Global Development using Natural Language Processing-supported
  Human-AI Collaboration}.
\newblock \bibinfo{journal}{\emph{Journal of Mechanical Design}}
  (\bibinfo{year}{2023}), \bibinfo{pages}{1--15}.
\newblock


\bibitem[Egarter~Vigl et~al\mbox{.}(2021)]%
        {egarter2021harnessing}
\bibfield{author}{\bibinfo{person}{Lukas Egarter~Vigl}, \bibinfo{person}{Thomas
  Marsoner}, \bibinfo{person}{Valentina Giombini}, \bibinfo{person}{Caroline
  Pecher}, \bibinfo{person}{Heidi Simion}, \bibinfo{person}{Egon Stemle},
  \bibinfo{person}{Erich Tasser}, {and} \bibinfo{person}{Daniel Depellegrin}.}
  \bibinfo{year}{2021}\natexlab{}.
\newblock \showarticletitle{Harnessing artificial intelligence technology and
  social media data to support Cultural Ecosystem Service assessments}.
\newblock \bibinfo{journal}{\emph{People and Nature}} \bibinfo{volume}{3},
  \bibinfo{number}{3} (\bibinfo{year}{2021}), \bibinfo{pages}{673--685}.
\newblock


\bibitem[Endsley(1995)]%
        {endsley1995measurement}
\bibfield{author}{\bibinfo{person}{Mica~R Endsley}.}
  \bibinfo{year}{1995}\natexlab{}.
\newblock \showarticletitle{Measurement of situation awareness in dynamic
  systems}.
\newblock \bibinfo{journal}{\emph{Human factors}} \bibinfo{volume}{37},
  \bibinfo{number}{1} (\bibinfo{year}{1995}), \bibinfo{pages}{65--84}.
\newblock


\bibitem[Endsley and Kaber(1999)]%
        {endsley1999level}
\bibfield{author}{\bibinfo{person}{Mica~R Endsley} {and}
  \bibinfo{person}{David~B Kaber}.} \bibinfo{year}{1999}\natexlab{}.
\newblock \showarticletitle{Level of automation effects on performance,
  situation awareness and workload in a dynamic control task}.
\newblock \bibinfo{journal}{\emph{Ergonomics}} \bibinfo{volume}{42},
  \bibinfo{number}{3} (\bibinfo{year}{1999}), \bibinfo{pages}{462--492}.
\newblock


\bibitem[Endsley and Appleby(2024)]%
        {endsley2024taxonomy}
\bibfield{author}{\bibinfo{person}{Tristan Endsley} {and}
  \bibinfo{person}{Brent Appleby}.} \bibinfo{year}{2024}\natexlab{}.
\newblock \showarticletitle{Taxonomy for Applied Human-Machine Teaming for
  Space System Domains}. In \bibinfo{booktitle}{\emph{2024 IEEE Aerospace
  Conference}}. IEEE, \bibinfo{pages}{1--9}.
\newblock


\bibitem[Fallon and Blaha(2018)]%
        {fallon2018improving}
\bibfield{author}{\bibinfo{person}{Corey~K Fallon} {and}
  \bibinfo{person}{Leslie~M Blaha}.} \bibinfo{year}{2018}\natexlab{}.
\newblock \showarticletitle{Improving automation transparency: Addressing some
  of machine learning’s unique challenges}. In
  \bibinfo{booktitle}{\emph{Augmented Cognition: Intelligent Technologies: 12th
  International Conference, AC 2018, Held as Part of HCI International 2018,
  Las Vegas, NV, USA, July 15-20, 2018, Proceedings, Part I}}. Springer,
  \bibinfo{pages}{245--254}.
\newblock


\bibitem[Flathmann et~al\mbox{.}(2021)]%
        {flathmann2021fostering}
\bibfield{author}{\bibinfo{person}{Christopher Flathmann},
  \bibinfo{person}{Beau~G Schelble}, {and} \bibinfo{person}{Nathan~J McNeese}.}
  \bibinfo{year}{2021}\natexlab{}.
\newblock \showarticletitle{Fostering human-agent team leadership by leveraging
  human teaming principles}. In \bibinfo{booktitle}{\emph{2021 IEEE 2nd
  International Conference on Human-Machine Systems (ICHMS)}}. IEEE,
  \bibinfo{pages}{1--6}.
\newblock


\bibitem[Flathmann et~al\mbox{.}(2023)]%
        {flathmann2023examining}
\bibfield{author}{\bibinfo{person}{Christopher Flathmann},
  \bibinfo{person}{Beau~G Schelble}, \bibinfo{person}{Patrick~J Rosopa},
  \bibinfo{person}{Nathan~J McNeese}, \bibinfo{person}{Rohit Mallick}, {and}
  \bibinfo{person}{Kapil~Chalil Madathil}.} \bibinfo{year}{2023}\natexlab{}.
\newblock \showarticletitle{Examining the impact of varying levels of AI
  teammate influence on human-{AI} teams}.
\newblock \bibinfo{journal}{\emph{International Journal of Human-Computer
  Studies}}  \bibinfo{volume}{177} (\bibinfo{year}{2023}),
  \bibinfo{pages}{103061}.
\newblock


\bibitem[Fotouhi et~al\mbox{.}(2019)]%
        {fotouhi2019survey}
\bibfield{author}{\bibinfo{person}{Azade Fotouhi}, \bibinfo{person}{Haoran
  Qiang}, \bibinfo{person}{Ming Ding}, \bibinfo{person}{Mahbub Hassan},
  \bibinfo{person}{Lorenzo~Galati Giordano}, \bibinfo{person}{Adrian
  Garcia-Rodriguez}, {and} \bibinfo{person}{Jinhong Yuan}.}
  \bibinfo{year}{2019}\natexlab{}.
\newblock \showarticletitle{Survey on {UAV} Cellular Communications: Practical
  Aspects, Standardization Advancements, Regulation, and Security Challenges}.
\newblock \bibinfo{journal}{\emph{IEEE Communications Surveys \& Tutorials}}
  \bibinfo{volume}{21}, \bibinfo{number}{4} (\bibinfo{year}{2019}),
  \bibinfo{pages}{3417--3442}.
\newblock
\urldef\tempurl%
\url{https://doi.org/10.1109/COMST.2019.2906228}
\showDOI{\tempurl}


\bibitem[Fusaroli et~al\mbox{.}(2014)]%
        {fusaroli2014analyzing}
\bibfield{author}{\bibinfo{person}{Riccardo Fusaroli}, \bibinfo{person}{Ivana
  Konvalinka}, {and} \bibinfo{person}{Sebastian Wallot}.}
  \bibinfo{year}{2014}\natexlab{}.
\newblock \showarticletitle{Analyzing social interactions: the promises and
  challenges of using cross recurrence quantification analysis}.
\newblock In \bibinfo{booktitle}{\emph{Translational recurrences: From
  mathematical theory to real-world applications}}.
  \bibinfo{publisher}{Springer}, \bibinfo{pages}{137--155}.
\newblock


\bibitem[Gay et~al\mbox{.}(2019)]%
        {gay2019operator}
\bibfield{author}{\bibinfo{person}{Chris Gay}, \bibinfo{person}{Barry
  Horowitz}, \bibinfo{person}{John~J Elshaw}, \bibinfo{person}{Philip Bobko},
  {and} \bibinfo{person}{Inki Kim}.} \bibinfo{year}{2019}\natexlab{}.
\newblock \showarticletitle{Operator suspicion and human-machine team
  performance under mission scenarios of unmanned ground vehicle operation}.
\newblock \bibinfo{journal}{\emph{IEEE Access}}  \bibinfo{volume}{7}
  (\bibinfo{year}{2019}), \bibinfo{pages}{36371--36379}.
\newblock


\bibitem[Gebru et~al\mbox{.}(2022)]%
        {gebru2022review}
\bibfield{author}{\bibinfo{person}{Biniam Gebru}, \bibinfo{person}{Lydia
  Zeleke}, \bibinfo{person}{Daniel Blankson}, \bibinfo{person}{Mahmoud Nabil},
  \bibinfo{person}{Shamila Nateghi}, \bibinfo{person}{Abdollah Homaifar}, {and}
  \bibinfo{person}{Edward Tunstel}.} \bibinfo{year}{2022}\natexlab{}.
\newblock \showarticletitle{A review on human--machine trust evaluation:
  Human-centric and machine-centric perspectives}.
\newblock \bibinfo{journal}{\emph{IEEE Transactions on Human-Machine Systems}}
  \bibinfo{volume}{52}, \bibinfo{number}{5} (\bibinfo{year}{2022}),
  \bibinfo{pages}{952--962}.
\newblock


\bibitem[Gill(2012)]%
        {gill2012human}
\bibfield{author}{\bibinfo{person}{Karamjit~S Gill}.}
  \bibinfo{year}{2012}\natexlab{}.
\newblock \bibinfo{booktitle}{\emph{Human machine symbiosis: The foundations of
  human-centred systems design}}.
\newblock \bibinfo{publisher}{Springer Science \& Business Media}.
\newblock


\bibitem[Gillespie(2022)]%
        {gillespie2022building}
\bibfield{author}{\bibinfo{person}{Tony Gillespie}.}
  \bibinfo{year}{2022}\natexlab{}.
\newblock \showarticletitle{Building trust and responsibility into autonomous
  human-machine teams}.
\newblock \bibinfo{journal}{\emph{Frontiers in Physics}}  \bibinfo{volume}{10}
  (\bibinfo{year}{2022}), \bibinfo{pages}{942245}.
\newblock


\bibitem[Gombolay et~al\mbox{.}(2015)]%
        {gombolay2015coordination}
\bibfield{author}{\bibinfo{person}{Matthew~Craig Gombolay},
  \bibinfo{person}{Cindy Huang}, {and} \bibinfo{person}{Julie Shah}.}
  \bibinfo{year}{2015}\natexlab{}.
\newblock \showarticletitle{Coordination of human-robot teaming with human task
  preferences}. In \bibinfo{booktitle}{\emph{2015 AAAI Fall Symposium Series}}.
\newblock


\bibitem[Gomez et~al\mbox{.}(2019)]%
        {gomez2019considerations}
\bibfield{author}{\bibinfo{person}{Steven~R Gomez}, \bibinfo{person}{Vincent
  Mancuso}, {and} \bibinfo{person}{Diane Staheli}.}
  \bibinfo{year}{2019}\natexlab{}.
\newblock \showarticletitle{Considerations for human-machine teaming in
  cybersecurity}. In \bibinfo{booktitle}{\emph{Augmented Cognition: 13th
  International Conference, AC 2019, Held as Part of the 21st HCI International
  Conference, HCII 2019, Orlando, FL, USA, July 26--31, 2019, Proceedings 21}}.
  Springer, \bibinfo{pages}{153--168}.
\newblock


\bibitem[Gonzalez et~al\mbox{.}(2022)]%
        {gonzalez2022adaptive}
\bibfield{author}{\bibinfo{person}{Cleotilde Gonzalez}, \bibinfo{person}{Palvi
  Aggarwal}, \bibinfo{person}{Edward~A Cranford}, {and}
  \bibinfo{person}{Christian Lebiere}.} \bibinfo{year}{2022}\natexlab{}.
\newblock \showarticletitle{Adaptive Cyberdefense with Deception: A Human-AI
  Cognitive Approach}.
\newblock \bibinfo{journal}{\emph{Cyber Deception: Techniques, Strategies, and
  Human Aspects}} (\bibinfo{year}{2022}), \bibinfo{pages}{41--57}.
\newblock


\bibitem[Gonzalez et~al\mbox{.}(2003)]%
        {gonzalez2003instance}
\bibfield{author}{\bibinfo{person}{Cleotilde Gonzalez},
  \bibinfo{person}{Javier~F Lerch}, {and} \bibinfo{person}{Christian Lebiere}.}
  \bibinfo{year}{2003}\natexlab{}.
\newblock \showarticletitle{Instance-based learning in dynamic decision
  making}.
\newblock \bibinfo{journal}{\emph{Cognitive Science}} \bibinfo{volume}{27},
  \bibinfo{number}{4} (\bibinfo{year}{2003}), \bibinfo{pages}{591--635}.
\newblock


\bibitem[Gray and Suri(2019)]%
        {gray2019ghost}
\bibfield{author}{\bibinfo{person}{Mary~L Gray} {and}
  \bibinfo{person}{Siddharth Suri}.} \bibinfo{year}{2019}\natexlab{}.
\newblock \bibinfo{booktitle}{\emph{Ghost work: How to stop Silicon Valley from
  building a new global underclass}}.
\newblock \bibinfo{publisher}{Eamon Dolan Books}.
\newblock


\bibitem[Greenberg and Marble(2023)]%
        {greenberg2023foundational}
\bibfield{author}{\bibinfo{person}{Ariel~M Greenberg} {and}
  \bibinfo{person}{Julie~L Marble}.} \bibinfo{year}{2023}\natexlab{}.
\newblock \showarticletitle{Foundational concepts in person-machine teaming}.
\newblock \bibinfo{journal}{\emph{Frontiers in Physics}}  \bibinfo{volume}{10}
  (\bibinfo{year}{2023}), \bibinfo{pages}{1080132}.
\newblock


\bibitem[Gupta et~al\mbox{.}(2023)]%
        {gupta2023fostering}
\bibfield{author}{\bibinfo{person}{Pranav Gupta}, \bibinfo{person}{Thuy~Ngoc
  Nguyen}, \bibinfo{person}{Cleotilde Gonzalez}, {and}
  \bibinfo{person}{Anita~Williams Woolley}.} \bibinfo{year}{2023}\natexlab{}.
\newblock \showarticletitle{Fostering collective intelligence in human--AI
  collaboration: laying the groundwork for COHUMAIN}.
\newblock \bibinfo{journal}{\emph{Topics in cognitive science}}
  (\bibinfo{year}{2023}).
\newblock


\bibitem[Hagendorff(2023)]%
        {hagendorff2023ai}
\bibfield{author}{\bibinfo{person}{Thilo Hagendorff}.}
  \bibinfo{year}{2023}\natexlab{}.
\newblock \showarticletitle{{AI} ethics and its pitfalls: not living up to its
  own standards?}
\newblock \bibinfo{journal}{\emph{AI and Ethics}} \bibinfo{volume}{3},
  \bibinfo{number}{1} (\bibinfo{year}{2023}), \bibinfo{pages}{329--336}.
\newblock


\bibitem[Haindl et~al\mbox{.}(2022a)]%
        {haindl2022towards}
\bibfield{author}{\bibinfo{person}{Philipp Haindl}, \bibinfo{person}{Georg
  Buchgeher}, \bibinfo{person}{Maqbool Khan}, {and} \bibinfo{person}{Bernhard
  Moser}.} \bibinfo{year}{2022}\natexlab{a}.
\newblock \showarticletitle{Towards a reference software architecture for
  human-AI teaming in smart manufacturing}. In
  \bibinfo{booktitle}{\emph{Proceedings of the ACM/IEEE 44th International
  Conference on Software Engineering: New Ideas and Emerging Results}}.
  \bibinfo{pages}{96--100}.
\newblock


\bibitem[Haindl et~al\mbox{.}(2022b)]%
        {haindl2022quality}
\bibfield{author}{\bibinfo{person}{Philipp Haindl}, \bibinfo{person}{Thomas
  Hoch}, \bibinfo{person}{Javier Dominguez}, \bibinfo{person}{Julen Aperribai},
  \bibinfo{person}{Nazim~Kemal Ure}, {and} \bibinfo{person}{Mehmet
  Tun{\c{c}}el}.} \bibinfo{year}{2022}\natexlab{b}.
\newblock \showarticletitle{Quality Characteristics of a Software Platform for
  Human-AI Teaming in Smart Manufacturing}. In
  \bibinfo{booktitle}{\emph{International Conference on the Quality of
  Information and Communications Technology}}. Springer,
  \bibinfo{pages}{3--17}.
\newblock


\bibitem[Haring et~al\mbox{.}(2021)]%
        {haring2021applying}
\bibfield{author}{\bibinfo{person}{Kerstin~S Haring},
  \bibinfo{person}{Elizabeth Phillips}, \bibinfo{person}{Elizabeth~H Lazzara},
  \bibinfo{person}{Daniel Ullman}, \bibinfo{person}{Anthony~L Baker}, {and}
  \bibinfo{person}{Joseph~R Keebler}.} \bibinfo{year}{2021}\natexlab{}.
\newblock \showarticletitle{Applying the swift trust model to human-robot
  teaming}.
\newblock In \bibinfo{booktitle}{\emph{Trust in Human-Robot Interaction}}.
  \bibinfo{publisher}{Elsevier}, \bibinfo{pages}{407--427}.
\newblock


\bibitem[Harpstead et~al\mbox{.}(2023)]%
        {harpstead2023speculative}
\bibfield{author}{\bibinfo{person}{Erik Harpstead}, \bibinfo{person}{Kimberly
  Stowers}, \bibinfo{person}{Lane Lawley}, \bibinfo{person}{Qiao Zhang}, {and}
  \bibinfo{person}{Christopher Maclellan}.} \bibinfo{year}{2023}\natexlab{}.
\newblock \showarticletitle{Speculative Game Design of Asymmetric Cooperative
  Games to Study Human-Machine Teaming}. In
  \bibinfo{booktitle}{\emph{Proceedings of the 18th International Conference on
  the Foundations of Digital Games}}. \bibinfo{pages}{1--4}.
\newblock


\bibitem[Hart(2006)]%
        {hart2006nasa}
\bibfield{author}{\bibinfo{person}{Sandra~G Hart}.}
  \bibinfo{year}{2006}\natexlab{}.
\newblock \showarticletitle{NASA-task load index (NASA-TLX); 20 years later}.
  In \bibinfo{booktitle}{\emph{Proceedings of the human factors and ergonomics
  society annual meeting}}, Vol.~\bibinfo{volume}{50}. Sage publications Sage
  CA: Los Angeles, CA, \bibinfo{pages}{904--908}.
\newblock


\bibitem[Hauptman et~al\mbox{.}(2023)]%
        {Hauptman23-autonomy}
\bibfield{author}{\bibinfo{person}{Allyson~I. Hauptman},
  \bibinfo{person}{Beau~G. Schelble}, \bibinfo{person}{Nathan~J. McNeese},
  {and} \bibinfo{person}{Kapil~Chalil Madathil}.}
  \bibinfo{year}{2023}\natexlab{}.
\newblock \showarticletitle{Adapt and overcome: Perceptions of adaptive
  autonomous agents for human-{AI} teaming}.
\newblock \bibinfo{journal}{\emph{Computers in Human Behavior}}
  \bibinfo{volume}{138} (\bibinfo{year}{2023}), \bibinfo{pages}{107451}.
\newblock
\showISSN{0747-5632}
\urldef\tempurl%
\url{https://doi.org/10.1016/j.chb.2022.107451}
\showDOI{\tempurl}


\bibitem[Henry et~al\mbox{.}(2022)]%
        {henry2022human}
\bibfield{author}{\bibinfo{person}{Katharine~E Henry}, \bibinfo{person}{Rachel
  Kornfield}, \bibinfo{person}{Anirudh Sridharan}, \bibinfo{person}{Robert~C
  Linton}, \bibinfo{person}{Catherine Groh}, \bibinfo{person}{Tony Wang},
  \bibinfo{person}{Albert Wu}, \bibinfo{person}{Bilge Mutlu}, {and}
  \bibinfo{person}{Suchi Saria}.} \bibinfo{year}{2022}\natexlab{}.
\newblock \showarticletitle{Human--machine teaming is key to {AI} adoption:
  clinicians’ experiences with a deployed machine learning system}.
\newblock \bibinfo{journal}{\emph{NPJ digital medicine}} \bibinfo{volume}{5},
  \bibinfo{number}{1} (\bibinfo{year}{2022}), \bibinfo{pages}{97}.
\newblock


\bibitem[Ho et~al\mbox{.}(2018)]%
        {ho2018psychological}
\bibfield{author}{\bibinfo{person}{Annabell Ho}, \bibinfo{person}{Jeff
  Hancock}, {and} \bibinfo{person}{Adam~S Miner}.}
  \bibinfo{year}{2018}\natexlab{}.
\newblock \showarticletitle{Psychological, relational, and emotional effects of
  self-disclosure after conversations with a chatbot}.
\newblock \bibinfo{journal}{\emph{Journal of Communication}}
  \bibinfo{volume}{68}, \bibinfo{number}{4} (\bibinfo{year}{2018}),
  \bibinfo{pages}{712--733}.
\newblock
\urldef\tempurl%
\url{https://academic.oup.com/joc/article/68/4/712/5025583}
\showURL{%
\tempurl}


\bibitem[Hoc(2000)]%
        {hoc2000human}
\bibfield{author}{\bibinfo{person}{Jean-Michel Hoc}.}
  \bibinfo{year}{2000}\natexlab{}.
\newblock \showarticletitle{From human--machine interaction to human--machine
  cooperation}.
\newblock \bibinfo{journal}{\emph{Ergonomics}} \bibinfo{volume}{43},
  \bibinfo{number}{7} (\bibinfo{year}{2000}), \bibinfo{pages}{833--843}.
\newblock


\bibitem[Hoffman and Breazeal(2004)]%
        {hoffman2004collaboration}
\bibfield{author}{\bibinfo{person}{Guy Hoffman} {and} \bibinfo{person}{Cynthia
  Breazeal}.} \bibinfo{year}{2004}\natexlab{}.
\newblock \showarticletitle{Collaboration in human-robot teams}. In
  \bibinfo{booktitle}{\emph{AIAA 1st intelligent systems technical
  conference}}. \bibinfo{pages}{6434}.
\newblock


\bibitem[Hu and Chen(2017)]%
        {hu2017optimal}
\bibfield{author}{\bibinfo{person}{Bin Hu} {and} \bibinfo{person}{Jing Chen}.}
  \bibinfo{year}{2017}\natexlab{}.
\newblock \showarticletitle{Optimal task allocation for human--machine
  collaborative manufacturing systems}.
\newblock \bibinfo{journal}{\emph{IEEE Robotics and Automation Letters}}
  \bibinfo{volume}{2}, \bibinfo{number}{4} (\bibinfo{year}{2017}),
  \bibinfo{pages}{1933--1940}.
\newblock


\bibitem[Huang et~al\mbox{.}(2024)]%
        {huang2024co}
\bibfield{author}{\bibinfo{person}{Chen Huang}, \bibinfo{person}{Xinwei Yang},
  \bibinfo{person}{Yang Deng}, \bibinfo{person}{Wenqiang Lei},
  \bibinfo{person}{JianCheng Lv}, {and} \bibinfo{person}{Tat-Seng Chua}.}
  \bibinfo{year}{2024}\natexlab{}.
\newblock \showarticletitle{{Co-Matching}: Towards Human-Machine Collaborative
  Legal Case Matching}.
\newblock \bibinfo{journal}{\emph{arXiv preprint arXiv:2405.10248}}
  (\bibinfo{year}{2024}).
\newblock


\bibitem[Hwang and Won(2021)]%
        {hwang2021ideabot}
\bibfield{author}{\bibinfo{person}{Angel Hsing-Chi Hwang} {and}
  \bibinfo{person}{Andrea~Stevenson Won}.} \bibinfo{year}{2021}\natexlab{}.
\newblock \showarticletitle{{IdeaBot}: Investigating social facilitation in
  human-machine team creativity}. In \bibinfo{booktitle}{\emph{Proceedings of
  the 2021 CHI Conference on Human Factors in Computing Systems}}.
  \bibinfo{pages}{1--16}.
\newblock


\bibitem[Ibrahim et~al\mbox{.}(2022)]%
        {ibrahim2022trust}
\bibfield{author}{\bibinfo{person}{Memunat~A Ibrahim}, \bibinfo{person}{Zena
  Assaad}, {and} \bibinfo{person}{Elizabeth Williams}.}
  \bibinfo{year}{2022}\natexlab{}.
\newblock \showarticletitle{Trust and communication in human-machine teaming}.
\newblock \bibinfo{journal}{\emph{Frontiers in Physics}}  \bibinfo{volume}{10}
  (\bibinfo{year}{2022}), \bibinfo{pages}{942896}.
\newblock


\bibitem[Imran et~al\mbox{.}(2020)]%
        {imran2020using}
\bibfield{author}{\bibinfo{person}{Muhammad Imran}, \bibinfo{person}{Ferda
  Ofli}, \bibinfo{person}{Doina Caragea}, {and} \bibinfo{person}{Antonio
  Torralba}.} \bibinfo{year}{2020}\natexlab{}.
\newblock \bibinfo{title}{Using AI and social media multimodal content for
  disaster response and management: Opportunities, challenges, and future
  directions}.
\newblock , \bibinfo{numpages}{102261}~pages.
\newblock


\bibitem[Javaid et~al\mbox{.}(2023)]%
        {javaid2023communication}
\bibfield{author}{\bibinfo{person}{Shumaila Javaid}, \bibinfo{person}{Nasir
  Saeed}, \bibinfo{person}{Zakria Qadir}, \bibinfo{person}{Hamza Fahim},
  \bibinfo{person}{Bin He}, \bibinfo{person}{Houbing Song}, {and}
  \bibinfo{person}{Muhammad Bilal}.} \bibinfo{year}{2023}\natexlab{}.
\newblock \showarticletitle{Communication and control in collaborative {UAVs}:
  Recent advances and future trends}.
\newblock \bibinfo{journal}{\emph{IEEE Transactions on Intelligent
  Transportation Systems}} \bibinfo{volume}{24}, \bibinfo{number}{6}
  (\bibinfo{year}{2023}), \bibinfo{pages}{5719--5739}.
\newblock


\bibitem[Jian et~al\mbox{.}(2000)]%
        {jiunyinin2000}
\bibfield{author}{\bibinfo{person}{Jiun-Yin Jian}, \bibinfo{person}{Ann~M.
  Bisantz}, {and} \bibinfo{person}{Colin~G. Drury}.}
  \bibinfo{year}{2000}\natexlab{}.
\newblock \showarticletitle{Foundations for an Empirically Determined Scale of
  Trust in Automated Systems}.
\newblock \bibinfo{journal}{\emph{International Journal of Cognitive
  Ergonomics}} \bibinfo{volume}{4}, \bibinfo{number}{1} (\bibinfo{year}{2000}),
  \bibinfo{pages}{53--71}.
\newblock


\bibitem[Johannsen(2009)]%
        {johannsen2009human}
\bibfield{author}{\bibinfo{person}{Gunnar Johannsen}.}
  \bibinfo{year}{2009}\natexlab{}.
\newblock \showarticletitle{Human-machine interaction}.
\newblock \bibinfo{journal}{\emph{Control Systems, Robotics and Automation}}
  \bibinfo{volume}{21} (\bibinfo{year}{2009}), \bibinfo{pages}{132--62}.
\newblock


\bibitem[Johnson et~al\mbox{.}(2021)]%
        {johnson2021impact}
\bibfield{author}{\bibinfo{person}{Craig~J Johnson}, \bibinfo{person}{Mustafa
  Demir}, \bibinfo{person}{Nathan~J McNeese}, \bibinfo{person}{Jamie~C Gorman},
  \bibinfo{person}{Alexandra~T Wolff}, {and} \bibinfo{person}{Nancy~J Cooke}.}
  \bibinfo{year}{2021}\natexlab{}.
\newblock \showarticletitle{The impact of training on human--autonomy team
  communications and trust calibration}.
\newblock \bibinfo{journal}{\emph{Human factors}} (\bibinfo{year}{2021}),
  \bibinfo{pages}{00187208211047323}.
\newblock


\bibitem[Johnson et~al\mbox{.}(2020)]%
        {johnson2020understanding}
\bibfield{author}{\bibinfo{person}{Matthew Johnson}, \bibinfo{person}{Micael
  Vignatti}, {and} \bibinfo{person}{Daniel Duran}.}
  \bibinfo{year}{2020}\natexlab{}.
\newblock \showarticletitle{Understanding human-machine teaming through
  interdependence analysis}.
\newblock In \bibinfo{booktitle}{\emph{Contemporary Research}}.
  \bibinfo{publisher}{CRC Press}, \bibinfo{pages}{209--233}.
\newblock


\bibitem[Jones et~al\mbox{.}(2004)]%
        {jones2004distributed}
\bibfield{author}{\bibinfo{person}{Rashaad~ET Jones},
  \bibinfo{person}{Michael~D McNeese}, \bibinfo{person}{Erik~S Connors},
  \bibinfo{person}{Tyrone Jefferson~Jr}, {and} \bibinfo{person}{David~L
  Hall~Jr}.} \bibinfo{year}{2004}\natexlab{}.
\newblock \showarticletitle{A distributed cognition simulation involving
  homeland security and defense: The development of NeoCITIES}. In
  \bibinfo{booktitle}{\emph{Proceedings of the human factors and ergonomics
  society annual meeting}}, Vol.~\bibinfo{volume}{48}. SAGE Publications Sage
  CA: Los Angeles, CA, \bibinfo{pages}{631--634}.
\newblock


\bibitem[Kaasinen et~al\mbox{.}(2022)]%
        {kaasinen2022smooth}
\bibfield{author}{\bibinfo{person}{Eija Kaasinen}, \bibinfo{person}{Anu-Hanna
  Anttila}, \bibinfo{person}{P{\"a}ivi Heikkil{\"a}}, \bibinfo{person}{Jari
  Laarni}, \bibinfo{person}{Hanna Koskinen}, {and} \bibinfo{person}{Antti
  V{\"a}{\"a}t{\"a}nen}.} \bibinfo{year}{2022}\natexlab{}.
\newblock \showarticletitle{Smooth and resilient human-machine teamwork as an
  Industry 5.0 design challenge}.
\newblock \bibinfo{journal}{\emph{Sustainability}} \bibinfo{volume}{14},
  \bibinfo{number}{5} (\bibinfo{year}{2022}), \bibinfo{pages}{2773}.
\newblock


\bibitem[Kashima et~al\mbox{.}(2022)]%
        {kashima2022trustworthy}
\bibfield{author}{\bibinfo{person}{Hisashi Kashima}, \bibinfo{person}{Satoshi
  Oyama}, \bibinfo{person}{Hiromi Arai}, {and} \bibinfo{person}{Junichiro
  Mori}.} \bibinfo{year}{2022}\natexlab{}.
\newblock \showarticletitle{Trustworthy Human Computation: A Survey}.
\newblock \bibinfo{journal}{\emph{arXiv preprint arXiv:2210.12324}}
  (\bibinfo{year}{2022}).
\newblock


\bibitem[Knox and Stone(2009)]%
        {knox2009interactively}
\bibfield{author}{\bibinfo{person}{W~Bradley Knox} {and} \bibinfo{person}{Peter
  Stone}.} \bibinfo{year}{2009}\natexlab{}.
\newblock \showarticletitle{Interactively shaping agents via human
  reinforcement: The TAMER framework}. In \bibinfo{booktitle}{\emph{Proceedings
  of the fifth international conference on Knowledge capture}}.
  \bibinfo{pages}{9--16}.
\newblock


\bibitem[Kridalukmana(2020)]%
        {rinta2020tsa}
\bibfield{author}{\bibinfo{person}{Rinta Kridalukmana}.}
  \bibinfo{year}{2020}\natexlab{}.
\newblock \showarticletitle{A supportive situation awareness model for human-
  autonomy teaming in collaborative driving}.
\newblock \bibinfo{journal}{\emph{Taylor \& Francis}} (\bibinfo{year}{2020}).
\newblock


\bibitem[Lasota et~al\mbox{.}(2017)]%
        {lasota2017survey}
\bibfield{author}{\bibinfo{person}{Przemyslaw~A Lasota},
  \bibinfo{person}{Terrence Fong}, \bibinfo{person}{Julie~A Shah},
  {et~al\mbox{.}}} \bibinfo{year}{2017}\natexlab{}.
\newblock \showarticletitle{A survey of methods for safe human-robot
  interaction}.
\newblock \bibinfo{journal}{\emph{Foundations and Trends{\textregistered} in
  Robotics}} \bibinfo{volume}{5}, \bibinfo{number}{4} (\bibinfo{year}{2017}),
  \bibinfo{pages}{261--349}.
\newblock


\bibitem[Lawless(2021)]%
        {lawless2021towards}
\bibfield{author}{\bibinfo{person}{WF Lawless}.}
  \bibinfo{year}{2021}\natexlab{}.
\newblock \showarticletitle{Towards an epistemology of interdependence among
  the orthogonal roles in human--machine teams}.
\newblock \bibinfo{journal}{\emph{Foundations of Science}}
  \bibinfo{volume}{26} (\bibinfo{year}{2021}), \bibinfo{pages}{129--142}.
\newblock


\bibitem[Lawless(2020)]%
        {lawless2020quantum}
\bibfield{author}{\bibinfo{person}{William~F Lawless}.}
  \bibinfo{year}{2020}\natexlab{}.
\newblock \showarticletitle{Quantum-like interdependence theory advances
  autonomous human--machine teams (a-hmts)}.
\newblock \bibinfo{journal}{\emph{Entropy}} \bibinfo{volume}{22},
  \bibinfo{number}{11} (\bibinfo{year}{2020}), \bibinfo{pages}{1227}.
\newblock


\bibitem[Lawless et~al\mbox{.}(2023)]%
        {lawless2023interdisciplinary}
\bibfield{author}{\bibinfo{person}{William~Frere Lawless},
  \bibinfo{person}{Donald~A Sofge}, \bibinfo{person}{Daniel Lofaro}, {and}
  \bibinfo{person}{Ranjeev Mittu}.} \bibinfo{year}{2023}\natexlab{}.
\newblock \showarticletitle{Interdisciplinary approaches to the structure and
  performance of interdependent autonomous human machine teams and systems}.
\newblock \bibinfo{journal}{\emph{Frontiers in Physics}}  \bibinfo{volume}{11}
  (\bibinfo{year}{2023}), \bibinfo{pages}{136}.
\newblock


\bibitem[Lee and See(2004)]%
        {lee2004trust}
\bibfield{author}{\bibinfo{person}{John~D Lee} {and} \bibinfo{person}{Katrina~A
  See}.} \bibinfo{year}{2004}\natexlab{}.
\newblock \showarticletitle{Trust in automation: Designing for appropriate
  reliance}.
\newblock \bibinfo{journal}{\emph{Human factors}} \bibinfo{volume}{46},
  \bibinfo{number}{1} (\bibinfo{year}{2004}), \bibinfo{pages}{50--80}.
\newblock


\bibitem[Lee et~al\mbox{.}(2015)]%
        {lee2015testing}
\bibfield{author}{\bibinfo{person}{Ritchie Lee}, \bibinfo{person}{Mykel~J.
  Kochenderfer}, \bibinfo{person}{Ole~J. Mengshoel},
  \bibinfo{person}{Guillaume~P. Brat}, {and} \bibinfo{person}{Michael~P.
  Owen}.} \bibinfo{year}{2015}\natexlab{}.
\newblock \showarticletitle{Adaptive stress testing of airborne collision
  avoidance systems}. In \bibinfo{booktitle}{\emph{2015 IEEE/AIAA 34th Digital
  Avionics Systems Conference (DASC)}}. \bibinfo{pages}{6C2--1--6C2--13}.
\newblock
\urldef\tempurl%
\url{https://doi.org/10.1109/DASC.2015.7311450}
\showDOI{\tempurl}


\bibitem[Lematta et~al\mbox{.}(2024)]%
        {lematta2024practical}
\bibfield{author}{\bibinfo{person}{Glenn~J Lematta},
  \bibinfo{person}{Patricia~L McDermott}, {and} \bibinfo{person}{Cindy
  Dominguez}.} \bibinfo{year}{2024}\natexlab{}.
\newblock \showarticletitle{Practical guidance for human-machine teaming
  assurance of AI-enabled systems}. In \bibinfo{booktitle}{\emph{Assurance and
  Security for AI-enabled Systems}}, Vol.~\bibinfo{volume}{13054}. SPIE,
  \bibinfo{pages}{108--118}.
\newblock


\bibitem[Le{\'o}n et~al\mbox{.}(2013)]%
        {leon2013human}
\bibfield{author}{\bibinfo{person}{L~Adri{\'a}n Le{\'o}n},
  \bibinfo{person}{Ana~C Tenorio}, {and} \bibinfo{person}{Eduardo~F Morales}.}
  \bibinfo{year}{2013}\natexlab{}.
\newblock \showarticletitle{Human interaction for effective reinforcement
  learning}. In \bibinfo{booktitle}{\emph{European Conf. Mach. Learning and
  Principles and Practice of Knowledge Discovery in Databases (ECMLPKDD
  2013)}}, Vol.~\bibinfo{volume}{3}. Citeseer.
\newblock


\bibitem[Lewis and Moorkens(2020)]%
        {lewis2020rights}
\bibfield{author}{\bibinfo{person}{Dave Lewis} {and} \bibinfo{person}{Joss
  Moorkens}.} \bibinfo{year}{2020}\natexlab{}.
\newblock \showarticletitle{A rights-based approach to trustworthy AI in social
  media}.
\newblock \bibinfo{journal}{\emph{Social Media+ Society}} \bibinfo{volume}{6},
  \bibinfo{number}{3} (\bibinfo{year}{2020}),
  \bibinfo{pages}{2056305120954672}.
\newblock


\bibitem[Liang et~al\mbox{.}(2017)]%
        {liang2017human}
\bibfield{author}{\bibinfo{person}{Huanghuang Liang}, \bibinfo{person}{Lu
  Yang}, \bibinfo{person}{Hong Cheng}, \bibinfo{person}{Wenzhe Tu}, {and}
  \bibinfo{person}{Mengjie Xu}.} \bibinfo{year}{2017}\natexlab{}.
\newblock \showarticletitle{Human-in-the-loop reinforcement learning}. In
  \bibinfo{booktitle}{\emph{2017 Chinese Automation Congress (CAC)}}. IEEE,
  \bibinfo{pages}{4511--4518}.
\newblock


\bibitem[Liu et~al\mbox{.}(2023)]%
        {liu2023application}
\bibfield{author}{\bibinfo{person}{Rui Liu}, \bibinfo{person}{Suraksha Gupta},
  {and} \bibinfo{person}{Parth Patel}.} \bibinfo{year}{2023}\natexlab{}.
\newblock \showarticletitle{The application of the principles of responsible AI
  on social media marketing for digital health}.
\newblock \bibinfo{journal}{\emph{Information Systems Frontiers}}
  \bibinfo{volume}{25}, \bibinfo{number}{6} (\bibinfo{year}{2023}),
  \bibinfo{pages}{2275--2299}.
\newblock


\bibitem[Liu et~al\mbox{.}(2006)]%
        {liu2006multitask}
\bibfield{author}{\bibinfo{person}{Yili Liu}, \bibinfo{person}{Robert Feyen},
  {and} \bibinfo{person}{Omer Tsimhoni}.} \bibinfo{year}{2006}\natexlab{}.
\newblock \showarticletitle{Queueing Network-Model Human Processor (QN-MHP): A
  Computational Architecture for Multitask Performance in Human-Machine
  Systems}.
\newblock \bibinfo{journal}{\emph{ACM Trans. Comput.-Hum. Interact.}}
  \bibinfo{volume}{13}, \bibinfo{number}{1} (\bibinfo{date}{mar}
  \bibinfo{year}{2006}), \bibinfo{pages}{37–70}.
\newblock
\showISSN{1073-0516}
\urldef\tempurl%
\url{https://doi.org/10.1145/1143518.1143520}
\showDOI{\tempurl}


\bibitem[Longa et~al\mbox{.}(2022)]%
        {longa2022human}
\bibfield{author}{\bibinfo{person}{Marc~Espin{\'o}s Longa},
  \bibinfo{person}{Antonios Tsourdos}, {and} \bibinfo{person}{Gokhan Inalhan}.}
  \bibinfo{year}{2022}\natexlab{}.
\newblock \showarticletitle{Human--Machine Network Through Bio-Inspired
  Decentralized Swarm Intelligence and Heterogeneous Teaming in SAR
  Operations}.
\newblock \bibinfo{journal}{\emph{Journal of Intelligent \& Robotic Systems}}
  \bibinfo{volume}{105}, \bibinfo{number}{4} (\bibinfo{year}{2022}),
  \bibinfo{pages}{88}.
\newblock


\bibitem[Lopez et~al\mbox{.}(2023)]%
        {lopez2023complex}
\bibfield{author}{\bibinfo{person}{Jeremy Lopez}, \bibinfo{person}{Claire
  Textor}, \bibinfo{person}{Caitlin Lancaster}, \bibinfo{person}{Beau
  Schelble}, \bibinfo{person}{Guo Freeman}, \bibinfo{person}{Rui Zhang},
  \bibinfo{person}{Nathan McNeese}, {and} \bibinfo{person}{Richard Pak}.}
  \bibinfo{year}{2023}\natexlab{}.
\newblock \showarticletitle{The complex relationship of {AI} ethics and trust
  in human-{AI} teaming: {Insights} from advanced real-world subject matter
  experts}.
\newblock \bibinfo{journal}{\emph{AI and Ethics}} (\bibinfo{year}{2023}),
  \bibinfo{pages}{1--21}.
\newblock


\bibitem[Lyn~Paul et~al\mbox{.}(2019)]%
        {lyn2019opportunities}
\bibfield{author}{\bibinfo{person}{Celeste Lyn~Paul}, \bibinfo{person}{Leslie~M
  Blaha}, \bibinfo{person}{Corey~K Fallon}, \bibinfo{person}{Cleotilde
  Gonzalez}, {and} \bibinfo{person}{Robert~S Gutzwiller}.}
  \bibinfo{year}{2019}\natexlab{}.
\newblock \showarticletitle{Opportunities and challenges for human-machine
  teaming in cybersecurity operations}. In
  \bibinfo{booktitle}{\emph{Proceedings of the Human Factors and Ergonomics
  Society Annual Meeting}}, Vol.~\bibinfo{volume}{63}. SAGE Publications Sage
  CA: Los Angeles, CA, \bibinfo{pages}{442--446}.
\newblock


\bibitem[Lyons et~al\mbox{.}(2021)]%
        {lyons2021human}
\bibfield{author}{\bibinfo{person}{Joseph~B Lyons}, \bibinfo{person}{Katia
  Sycara}, \bibinfo{person}{Michael Lewis}, {and} \bibinfo{person}{August
  Capiola}.} \bibinfo{year}{2021}\natexlab{}.
\newblock \showarticletitle{Human-autonomy teaming: Definitions, debates, and
  directions}.
\newblock \bibinfo{journal}{\emph{Frontiers in Psychology}}
  \bibinfo{volume}{12} (\bibinfo{year}{2021}), \bibinfo{pages}{589585}.
\newblock


\bibitem[Lyons et~al\mbox{.}(2019)]%
        {lyons2019trust}
\bibfield{author}{\bibinfo{person}{Joseph~B Lyons}, \bibinfo{person}{Kevin~T
  Wynne}, \bibinfo{person}{Sean Mahoney}, {and} \bibinfo{person}{Mark~A
  Roebke}.} \bibinfo{year}{2019}\natexlab{}.
\newblock \showarticletitle{Trust and human-machine teaming: A qualitative
  study}.
\newblock In \bibinfo{booktitle}{\emph{Artificial intelligence for the internet
  of everything}}. \bibinfo{publisher}{Elsevier}, \bibinfo{pages}{101--116}.
\newblock


\bibitem[Maathuis(2024)]%
        {maathuis2024trustworthy}
\bibfield{author}{\bibinfo{person}{Clara Maathuis}.}
  \bibinfo{year}{2024}\natexlab{}.
\newblock \showarticletitle{Trustworthy Human-Autonomy Teaming for
  Proportionality Assessment in Military Operations}. In
  \bibinfo{booktitle}{\emph{2024 4th International Conference on Applied
  Artificial Intelligence (ICAPAI)}}. IEEE, \bibinfo{pages}{1--8}.
\newblock


\bibitem[Madni and Madni(2018)]%
        {madni2018architectural}
\bibfield{author}{\bibinfo{person}{Azad~M Madni} {and} \bibinfo{person}{Carla~C
  Madni}.} \bibinfo{year}{2018}\natexlab{}.
\newblock \showarticletitle{Architectural framework for exploring adaptive
  human-machine teaming options in simulated dynamic environments}.
\newblock \bibinfo{journal}{\emph{Systems}} \bibinfo{volume}{6},
  \bibinfo{number}{4} (\bibinfo{year}{2018}), \bibinfo{pages}{44}.
\newblock


\bibitem[Mahaini et~al\mbox{.}(2019)]%
        {mahaini2019building}
\bibfield{author}{\bibinfo{person}{Mohamad~Imad Mahaini},
  \bibinfo{person}{Shujun Li}, {and} \bibinfo{person}{Rahime~Belen
  Sa{\u{g}}lam}.} \bibinfo{year}{2019}\natexlab{}.
\newblock \showarticletitle{Building taxonomies based on human-machine teaming:
  Cyber security as an example}. In \bibinfo{booktitle}{\emph{Proceedings of
  the 14th International conference on availability, reliability and
  security}}. \bibinfo{pages}{1--9}.
\newblock


\bibitem[Mathieu et~al\mbox{.}(2000)]%
        {mathieu2000influence}
\bibfield{author}{\bibinfo{person}{John~E Mathieu}, \bibinfo{person}{Tonia~S
  Heffner}, \bibinfo{person}{Gerald~F Goodwin}, \bibinfo{person}{Eduardo
  Salas}, {and} \bibinfo{person}{Janis~A Cannon-Bowers}.}
  \bibinfo{year}{2000}\natexlab{}.
\newblock \showarticletitle{The influence of shared mental models on team
  process and performance}.
\newblock \bibinfo{journal}{\emph{Journal of applied psychology}}
  \bibinfo{volume}{85}, \bibinfo{number}{2} (\bibinfo{year}{2000}),
  \bibinfo{pages}{273}.
\newblock


\bibitem[McDermott et~al\mbox{.}(2005)]%
        {mcdermott2005effective}
\bibfield{author}{\bibinfo{person}{Patricia~L McDermott},
  \bibinfo{person}{Jason Luck}, \bibinfo{person}{Laurel Allender}, {and}
  \bibinfo{person}{Alia Fisher}.} \bibinfo{year}{2005}\natexlab{}.
\newblock \showarticletitle{Effective human to human communication of
  information provided by an unmanned vehicle}. In
  \bibinfo{booktitle}{\emph{Proceedings of the Human Factors and Ergonomics
  Society Annual Meeting}}, Vol.~\bibinfo{volume}{49}. SAGE Publications Sage
  CA: Los Angeles, CA, \bibinfo{pages}{402--406}.
\newblock


\bibitem[McNeese et~al\mbox{.}(2021b)]%
        {mcneese2021tsa}
\bibfield{author}{\bibinfo{person}{Nathan McNeese}, \bibinfo{person}{Mustafa
  Demir}, \bibinfo{person}{Nancy~J Cooke}, {and} \bibinfo{person}{Manrong
  She}.} \bibinfo{year}{2021}\natexlab{b}.
\newblock \showarticletitle{Team Situation Awareness and Conflict: A Study of
  Human Machine Teaming}.
\newblock \bibinfo{journal}{\emph{SAGE, Journal of Cognitive Engineering and
  Decision Making}} (\bibinfo{year}{2021}).
\newblock


\bibitem[McNeese et~al\mbox{.}(2021a)]%
        {mcneese2021trust}
\bibfield{author}{\bibinfo{person}{Nathan~J McNeese}, \bibinfo{person}{Mustafa
  Demir}, \bibinfo{person}{Erin~K Chiou}, {and} \bibinfo{person}{Nancy~J
  Cooke}.} \bibinfo{year}{2021}\natexlab{a}.
\newblock \showarticletitle{Trust and team performance in human--autonomy
  teaming}.
\newblock \bibinfo{journal}{\emph{International Journal of Electronic
  Commerce}} \bibinfo{volume}{25}, \bibinfo{number}{1} (\bibinfo{year}{2021}),
  \bibinfo{pages}{51--72}.
\newblock


\bibitem[McNeese et~al\mbox{.}(2018)]%
        {mcneese2018teaming}
\bibfield{author}{\bibinfo{person}{Nathan~J McNeese}, \bibinfo{person}{Mustafa
  Demir}, \bibinfo{person}{Nancy~J Cooke}, {and} \bibinfo{person}{Christopher
  Myers}.} \bibinfo{year}{2018}\natexlab{}.
\newblock \showarticletitle{Teaming with a synthetic teammate: Insights into
  human-autonomy teaming}.
\newblock \bibinfo{journal}{\emph{Human factors}} \bibinfo{volume}{60},
  \bibinfo{number}{2} (\bibinfo{year}{2018}), \bibinfo{pages}{262--273}.
\newblock


\bibitem[McNeese et~al\mbox{.}(2021c)]%
        {mcneese2021team}
\bibfield{author}{\bibinfo{person}{Nathan~J McNeese}, \bibinfo{person}{Mustafa
  Demir}, \bibinfo{person}{Nancy~J Cooke}, {and} \bibinfo{person}{Manrong
  She}.} \bibinfo{year}{2021}\natexlab{c}.
\newblock \showarticletitle{Team situation awareness and conflict: A study of
  human--machine teaming}.
\newblock \bibinfo{journal}{\emph{Journal of Cognitive Engineering and Decision
  Making}} \bibinfo{volume}{15}, \bibinfo{number}{2-3} (\bibinfo{year}{2021}),
  \bibinfo{pages}{83--96}.
\newblock


\bibitem[McNeese et~al\mbox{.}(2021d)]%
        {mcneese2021my}
\bibfield{author}{\bibinfo{person}{Nathan~J McNeese}, \bibinfo{person}{Beau~G
  Schelble}, \bibinfo{person}{Lorenzo~Barberis Canonico}, {and}
  \bibinfo{person}{Mustafa Demir}.} \bibinfo{year}{2021}\natexlab{d}.
\newblock \showarticletitle{Who/what is my teammate? Team composition
  considerations in human--ai teaming}.
\newblock \bibinfo{journal}{\emph{IEEE Transactions on Human-Machine Systems}}
  \bibinfo{volume}{51}, \bibinfo{number}{4} (\bibinfo{year}{2021}),
  \bibinfo{pages}{288--299}.
\newblock


\bibitem[Mingyue~Ma et~al\mbox{.}(2018)]%
        {mingyue2018human}
\bibfield{author}{\bibinfo{person}{Lanssie Mingyue~Ma},
  \bibinfo{person}{Terrence Fong}, \bibinfo{person}{Mark~J Micire},
  \bibinfo{person}{Yun~Kyung Kim}, {and} \bibinfo{person}{Karen Feigh}.}
  \bibinfo{year}{2018}\natexlab{}.
\newblock \showarticletitle{Human-robot teaming: Concepts and components for
  design}. In \bibinfo{booktitle}{\emph{Field and Service Robotics: Results of
  the 11th International Conference}}. Springer, \bibinfo{pages}{649--663}.
\newblock


\bibitem[Moore(2007)]%
        {moore2007presence}
\bibfield{author}{\bibinfo{person}{Roger Moore}.}
  \bibinfo{year}{2007}\natexlab{}.
\newblock \showarticletitle{PRESENCE: A Human-Inspired Architecture for
  Speech-Based Human-Machine Interaction}.
\newblock \bibinfo{journal}{\emph{IEEE Trans. Comput.}} \bibinfo{volume}{56},
  \bibinfo{number}{9} (\bibinfo{year}{2007}), \bibinfo{pages}{1176--1188}.
\newblock
\urldef\tempurl%
\url{https://doi.org/10.1109/TC.2007.1080}
\showDOI{\tempurl}


\bibitem[Murphy and Schreckenghost(2013)]%
        {murphy2013survey}
\bibfield{author}{\bibinfo{person}{Robin~R Murphy} {and} \bibinfo{person}{Debra
  Schreckenghost}.} \bibinfo{year}{2013}\natexlab{}.
\newblock \showarticletitle{Survey of metrics for human-robot interaction}. In
  \bibinfo{booktitle}{\emph{2013 8th ACM/IEEE International Conference on
  Human-Robot Interaction (HRI)}}. IEEE, \bibinfo{pages}{197--198}.
\newblock


\bibitem[Musick et~al\mbox{.}(2021)]%
        {MUSICK2021106852}
\bibfield{author}{\bibinfo{person}{Geoff Musick}, \bibinfo{person}{Thomas~A.
  O'Neill}, \bibinfo{person}{Beau~G. Schelble}, \bibinfo{person}{Nathan~J.
  McNeese}, {and} \bibinfo{person}{Jonn~B. Henke}.}
  \bibinfo{year}{2021}\natexlab{}.
\newblock \showarticletitle{What Happens When Humans Believe Their Teammate is
  an AI? An Investigation into Humans Teaming with Autonomy}.
\newblock \bibinfo{journal}{\emph{Computers in Human Behavior}}
  \bibinfo{volume}{122} (\bibinfo{year}{2021}), \bibinfo{pages}{106852}.
\newblock
\showISSN{0747-5632}
\urldef\tempurl%
\url{https://doi.org/10.1016/j.chb.2021.106852}
\showDOI{\tempurl}


\bibitem[Myers et~al\mbox{.}(2019)]%
        {myers2019}
\bibfield{author}{\bibinfo{person}{Christopher Myers}, \bibinfo{person}{Jerry
  Ball}, \bibinfo{person}{Nancy Cooke}, \bibinfo{person}{Mary Freiman},
  \bibinfo{person}{Michelle Caisse}, \bibinfo{person}{Stuart Rodgers},
  \bibinfo{person}{Mustafa Demir}, {and} \bibinfo{person}{Nathan McNeese}.}
  \bibinfo{year}{2019}\natexlab{}.
\newblock \showarticletitle{Autonomous Intelligent Agents for Team Training}.
\newblock \bibinfo{journal}{\emph{IEEE Intelligent Systems}}
  \bibinfo{volume}{34}, \bibinfo{number}{2} (\bibinfo{year}{2019}),
  \bibinfo{pages}{3--14}.
\newblock
\urldef\tempurl%
\url{https://doi.org/10.1109/MIS.2018.2886670}
\showDOI{\tempurl}


\bibitem[Mygland et~al\mbox{.}(2021)]%
        {mygland2021affordances}
\bibfield{author}{\bibinfo{person}{Morten~Johan Mygland},
  \bibinfo{person}{Morten Schibbye}, \bibinfo{person}{Ilias~O Pappas}, {and}
  \bibinfo{person}{Polyxeni Vassilakopoulou}.} \bibinfo{year}{2021}\natexlab{}.
\newblock \showarticletitle{Affordances in human-chatbot interaction: a review
  of the literature}. In \bibinfo{booktitle}{\emph{Responsible AI and Analytics
  for an Ethical and Inclusive Digitized Society: 20th IFIP WG 6.11 Conference
  on e-Business, e-Services and e-Society, I3E 2021, Galway, Ireland, September
  1--3, 2021, Proceedings 20}}. Springer, \bibinfo{pages}{3--17}.
\newblock


\bibitem[Nagy and Miller(2021)]%
        {nagy2021interdependence}
\bibfield{author}{\bibinfo{person}{Bruce Nagy} {and} \bibinfo{person}{Scot
  Miller}.} \bibinfo{year}{2021}\natexlab{}.
\newblock \showarticletitle{Interdependence Analysis for Artificial
  Intelligence System Safety}.
\newblock \bibinfo{publisher}{Acquisition Research Program}.
\newblock


\bibitem[Natarajan et~al\mbox{.}(2023)]%
        {natarajan2023human}
\bibfield{author}{\bibinfo{person}{Manisha Natarajan}, \bibinfo{person}{Esmaeil
  Seraj}, \bibinfo{person}{Batuhan Altundas}, \bibinfo{person}{Rohan Paleja},
  \bibinfo{person}{Sean Ye}, \bibinfo{person}{Letian Chen},
  \bibinfo{person}{Reed Jensen}, \bibinfo{person}{Kimberlee~Chestnut Chang},
  {and} \bibinfo{person}{Matthew Gombolay}.} \bibinfo{year}{2023}\natexlab{}.
\newblock \showarticletitle{Human-robot teaming: grand challenges}.
\newblock \bibinfo{journal}{\emph{Current Robotics Reports}}
  \bibinfo{volume}{4}, \bibinfo{number}{3} (\bibinfo{year}{2023}),
  \bibinfo{pages}{81--100}.
\newblock


\bibitem[{National Academies of Sciences, Engineering, and Medicine and
  others}(2021)]%
        {national2021human}
\bibfield{author}{\bibinfo{person}{{National Academies of Sciences,
  Engineering, and Medicine and others}}.} \bibinfo{year}{2021}\natexlab{}.
\newblock \showarticletitle{Human-AI Teaming: State-of-the-Art and Research
  Needs}.
\newblock  (\bibinfo{year}{2021}).
\newblock


\bibitem[Nourbakhsh et~al\mbox{.}(2005)]%
        {nourbakhsh2005human}
\bibfield{author}{\bibinfo{person}{Illah~R Nourbakhsh}, \bibinfo{person}{Katia
  Sycara}, \bibinfo{person}{Mary Koes}, \bibinfo{person}{Mark Yong},
  \bibinfo{person}{Michael Lewis}, {and} \bibinfo{person}{Steve Burion}.}
  \bibinfo{year}{2005}\natexlab{}.
\newblock \showarticletitle{Human-robot teaming for search and rescue}.
\newblock \bibinfo{journal}{\emph{IEEE Pervasive Computing}}
  \bibinfo{volume}{4}, \bibinfo{number}{1} (\bibinfo{year}{2005}),
  \bibinfo{pages}{72--79}.
\newblock


\bibitem[Ogenyi et~al\mbox{.}(2021)]%
        {ogenyi2021robotcollab}
\bibfield{author}{\bibinfo{person}{Uchenna~Emeoha Ogenyi},
  \bibinfo{person}{Jinguo Liu}, \bibinfo{person}{Chenguang Yang},
  \bibinfo{person}{Zhaojie Ju}, {and} \bibinfo{person}{Honghai Liu}.}
  \bibinfo{year}{2021}\natexlab{}.
\newblock \showarticletitle{Physical Human–Robot Collaboration: Robotic
  Systems, Learning Methods, Collaborative Strategies, Sensors, and Actuators}.
\newblock \bibinfo{journal}{\emph{IEEE Transactions on Cybernetics}}
  \bibinfo{volume}{51}, \bibinfo{number}{4} (\bibinfo{year}{2021}),
  \bibinfo{pages}{1888--1901}.
\newblock
\urldef\tempurl%
\url{https://doi.org/10.1109/TCYB.2019.2947532}
\showDOI{\tempurl}


\bibitem[Oh et~al\mbox{.}(2024)]%
        {oh2024link}
\bibfield{author}{\bibinfo{person}{Hongseok Oh}, \bibinfo{person}{Kyungchang
  Jeong}, \bibinfo{person}{Euijong Lee}, {and} \bibinfo{person}{Ji-Hoon
  Jeong}.} \bibinfo{year}{2024}\natexlab{}.
\newblock \showarticletitle{{LINK}: Self-Adaptive System with Human-Machine
  Teaming-based Loop for Interoperability in IoT Environments}. In
  \bibinfo{booktitle}{\emph{Proceedings of the 39th ACM/SIGAPP Symposium on
  Applied Computing}}. \bibinfo{pages}{592--599}.
\newblock


\bibitem[O'Neill et~al\mbox{.}(2023)]%
        {o2023human}
\bibfield{author}{\bibinfo{person}{Thomas~A O'Neill},
  \bibinfo{person}{Christopher Flathmann}, \bibinfo{person}{Nathan~J McNeese},
  {and} \bibinfo{person}{Eduardo Salas}.} \bibinfo{year}{2023}\natexlab{}.
\newblock \showarticletitle{Human-autonomy Teaming: Need for a guiding
  team-based framework?}
\newblock \bibinfo{journal}{\emph{Computers in Human Behavior}}
  \bibinfo{volume}{146} (\bibinfo{year}{2023}), \bibinfo{pages}{107762}.
\newblock


\bibitem[O’Neill et~al\mbox{.}(2022)]%
        {o2022human}
\bibfield{author}{\bibinfo{person}{Thomas O’Neill}, \bibinfo{person}{Nathan
  McNeese}, \bibinfo{person}{Amy Barron}, {and} \bibinfo{person}{Beau
  Schelble}.} \bibinfo{year}{2022}\natexlab{}.
\newblock \showarticletitle{Human--autonomy teaming: A review and analysis of
  the empirical literature}.
\newblock \bibinfo{journal}{\emph{Human Factors}} \bibinfo{volume}{64},
  \bibinfo{number}{5} (\bibinfo{year}{2022}), \bibinfo{pages}{904--938}.
\newblock


\bibitem[Parasuraman et~al\mbox{.}(2008)]%
        {parasuraman2008situation}
\bibfield{author}{\bibinfo{person}{Raja Parasuraman}, \bibinfo{person}{Thomas~B
  Sheridan}, {and} \bibinfo{person}{Christopher~D Wickens}.}
  \bibinfo{year}{2008}\natexlab{}.
\newblock \showarticletitle{Situation awareness, mental workload, and trust in
  automation: Viable, empirically supported cognitive engineering constructs}.
\newblock \bibinfo{journal}{\emph{Journal of cognitive engineering and decision
  making}} \bibinfo{volume}{2}, \bibinfo{number}{2} (\bibinfo{year}{2008}),
  \bibinfo{pages}{140--160}.
\newblock


\bibitem[Patel et~al\mbox{.}(2024)]%
        {patel2024give}
\bibfield{author}{\bibinfo{person}{Jitu Patel}, \bibinfo{person}{M Boardman},
  \bibinfo{person}{B Files}, \bibinfo{person}{F Gregory}, \bibinfo{person}{S
  Lamb}, \bibinfo{person}{S Sarkadi}, \bibinfo{person}{M Te{\v{s}}i{\'c}},
  {and} \bibinfo{person}{N Yeung}.} \bibinfo{year}{2024}\natexlab{}.
\newblock \showarticletitle{Give us a hand, mate! A holistic review of research
  on human-machine teaming}.
\newblock \bibinfo{journal}{\emph{BMJ Mil Health}} (\bibinfo{year}{2024}).
\newblock


\bibitem[Patel et~al\mbox{.}(2018)]%
        {patel2018anomaly}
\bibfield{author}{\bibinfo{person}{Naman Patel}, \bibinfo{person}{Apoorva
  Nandini~Saridena}, \bibinfo{person}{Anna Choromanska},
  \bibinfo{person}{Prashanth Krishnamurthy}, {and} \bibinfo{person}{Farshad
  Khorrami}.} \bibinfo{year}{2018}\natexlab{}.
\newblock \showarticletitle{Adversarial Learning-Based On-Line Anomaly
  Monitoring for Assured Autonomy}. In \bibinfo{booktitle}{\emph{2018 IEEE/RSJ
  International Conference on Intelligent Robots and Systems (IROS)}}.
  \bibinfo{pages}{6149--6154}.
\newblock
\urldef\tempurl%
\url{https://doi.org/10.1109/IROS.2018.8593375}
\showDOI{\tempurl}


\bibitem[Pflanzer et~al\mbox{.}(2023)]%
        {pflanzer2023ethics}
\bibfield{author}{\bibinfo{person}{Michael Pflanzer}, \bibinfo{person}{Zachary
  Traylor}, \bibinfo{person}{Joseph~B Lyons}, \bibinfo{person}{Veljko
  Dubljevi{\'c}}, {and} \bibinfo{person}{Chang~S Nam}.}
  \bibinfo{year}{2023}\natexlab{}.
\newblock \showarticletitle{Ethics in human-AI teaming: principles and
  perspectives}.
\newblock \bibinfo{journal}{\emph{AI and Ethics}} \bibinfo{volume}{3},
  \bibinfo{number}{3} (\bibinfo{year}{2023}), \bibinfo{pages}{917--935}.
\newblock


\bibitem[Pizo{\'n} and Gola(2023)]%
        {pizon2023human}
\bibfield{author}{\bibinfo{person}{Jakub Pizo{\'n}} {and}
  \bibinfo{person}{Arkadiusz Gola}.} \bibinfo{year}{2023}\natexlab{}.
\newblock \showarticletitle{Human--Machine Relationship—Perspective and
  Future Roadmap for Industry 5.0 Solutions}.
\newblock \bibinfo{journal}{\emph{Machines}} \bibinfo{volume}{11},
  \bibinfo{number}{2} (\bibinfo{year}{2023}), \bibinfo{pages}{203}.
\newblock


\bibitem[Quintas et~al\mbox{.}(2017)]%
        {quintas2018architecture}
\bibfield{author}{\bibinfo{person}{João Quintas}, \bibinfo{person}{Paulo
  Menezes}, {and} \bibinfo{person}{Jorge Dias}.}
  \bibinfo{year}{2017}\natexlab{}.
\newblock \showarticletitle{Information Model and Architecture Specification
  for Context Awareness Interaction Decision Support in Cyber-Physical
  Human–Machine Systems}.
\newblock \bibinfo{journal}{\emph{IEEE Transactions on Human-Machine Systems}}
  \bibinfo{volume}{47}, \bibinfo{number}{3} (\bibinfo{year}{2017}),
  \bibinfo{pages}{323--331}.
\newblock
\urldef\tempurl%
\url{https://doi.org/10.1109/THMS.2016.2634923}
\showDOI{\tempurl}


\bibitem[{Robots Guide}(2024)]%
        {pr2_robot_guide}
\bibfield{author}{\bibinfo{person}{{Robots Guide}}.}
  \bibinfo{year}{2024}\natexlab{}.
\newblock \bibinfo{title}{PR2 Robot}.
\newblock
\newblock
\urldef\tempurl%
\url{https://robotsguide.com/robots/pr2}
\showURL{%
\tempurl}
\newblock
\shownote{Accessed: 2024-04-19}.


\bibitem[Roth et~al\mbox{.}(2019)]%
        {roth2019function}
\bibfield{author}{\bibinfo{person}{Emilie~M Roth}, \bibinfo{person}{Christen
  Sushereba}, \bibinfo{person}{Laura~G Militello}, \bibinfo{person}{Julie
  Diiulio}, {and} \bibinfo{person}{Katie Ernst}.}
  \bibinfo{year}{2019}\natexlab{}.
\newblock \showarticletitle{Function allocation considerations in the era of
  human autonomy teaming}.
\newblock \bibinfo{journal}{\emph{Journal of Cognitive Engineering and Decision
  Making}} \bibinfo{volume}{13}, \bibinfo{number}{4} (\bibinfo{year}{2019}),
  \bibinfo{pages}{199--220}.
\newblock


\bibitem[Safdar et~al\mbox{.}(2024)]%
        {safdar2024human}
\bibfield{author}{\bibinfo{person}{Mutahar Safdar}, \bibinfo{person}{Jiarui
  Xie}, \bibinfo{person}{Andrei Mircea}, {and} \bibinfo{person}{Yaoyao~Fiona
  Zhao}.} \bibinfo{year}{2024}\natexlab{}.
\newblock \showarticletitle{Human-artificial intelligence teaming for
  scientific information extraction from data-driven additive manufacturing
  research using large language models}. In
  \bibinfo{booktitle}{\emph{International Design Engineering Technical
  Conferences and Computers and Information in Engineering Conference}},
  Vol.~\bibinfo{volume}{88346}. American Society of Mechanical Engineers,
  \bibinfo{pages}{V02AT02A028}.
\newblock


\bibitem[Salas et~al\mbox{.}(2017)]%
        {salas2017situation}
\bibfield{author}{\bibinfo{person}{Eduardo Salas}, \bibinfo{person}{Carolyn
  Prince}, \bibinfo{person}{David~P Baker}, {and} \bibinfo{person}{Lisa
  Shrestha}.} \bibinfo{year}{2017}\natexlab{}.
\newblock \showarticletitle{Situation awareness in team performance:
  Implications for measurement and training}.
\newblock \bibinfo{journal}{\emph{Situational Awareness}}
  (\bibinfo{year}{2017}), \bibinfo{pages}{63--76}.
\newblock


\bibitem[Schelble et~al\mbox{.}(2020)]%
        {schelble2020designing}
\bibfield{author}{\bibinfo{person}{Beau Schelble},
  \bibinfo{person}{Lorenzo-Barberis Canonico}, \bibinfo{person}{Nathan
  McNeese}, \bibinfo{person}{Jack Carroll}, {and} \bibinfo{person}{Casey
  Hird}.} \bibinfo{year}{2020}\natexlab{}.
\newblock \showarticletitle{Designing human-autonomy teaming experiments
  through reinforcement learning}. In \bibinfo{booktitle}{\emph{Proceedings of
  the Human Factors and Ergonomics Society Annual Meeting}},
  Vol.~\bibinfo{volume}{64}. SAGE Publications Sage CA: Los Angeles, CA,
  \bibinfo{pages}{1426--1430}.
\newblock


\bibitem[Schelble et~al\mbox{.}(2022a)]%
        {schelble2022let}
\bibfield{author}{\bibinfo{person}{Beau~G Schelble},
  \bibinfo{person}{Christopher Flathmann}, \bibinfo{person}{Nathan~J McNeese},
  \bibinfo{person}{Guo Freeman}, {and} \bibinfo{person}{Rohit Mallick}.}
  \bibinfo{year}{2022}\natexlab{a}.
\newblock \showarticletitle{Let's think together! Assessing shared mental
  models, performance, and trust in human-agent teams}.
\newblock \bibinfo{journal}{\emph{Proceedings of the ACM on Human-Computer
  Interaction}} \bibinfo{volume}{6}, \bibinfo{number}{GROUP}
  (\bibinfo{year}{2022}), \bibinfo{pages}{1--29}.
\newblock


\bibitem[Schelble et~al\mbox{.}(2023)]%
        {schelble2023investigating}
\bibfield{author}{\bibinfo{person}{Beau~G Schelble},
  \bibinfo{person}{Christopher Flathmann}, \bibinfo{person}{Nathan~J McNeese},
  \bibinfo{person}{Thomas O’Neill}, \bibinfo{person}{Richard Pak}, {and}
  \bibinfo{person}{Moses Namara}.} \bibinfo{year}{2023}\natexlab{}.
\newblock \showarticletitle{Investigating the effects of perceived teammate
  artificiality on human performance and cognition}.
\newblock \bibinfo{journal}{\emph{International Journal of Human--Computer
  Interaction}} \bibinfo{volume}{39}, \bibinfo{number}{13}
  (\bibinfo{year}{2023}), \bibinfo{pages}{2686--2701}.
\newblock


\bibitem[Schelble et~al\mbox{.}(2022b)]%
        {schelble2022see}
\bibfield{author}{\bibinfo{person}{Beau~G Schelble},
  \bibinfo{person}{Christopher Flathmann}, \bibinfo{person}{Geoff Musick},
  \bibinfo{person}{Nathan~J McNeese}, {and} \bibinfo{person}{Guo Freeman}.}
  \bibinfo{year}{2022}\natexlab{b}.
\newblock \showarticletitle{I see you: Examining the role of spatial
  information in human-agent teams}.
\newblock \bibinfo{journal}{\emph{Proceedings of the ACM on Human-Computer
  Interaction}} \bibinfo{volume}{6}, \bibinfo{number}{CSCW2}
  (\bibinfo{year}{2022}), \bibinfo{pages}{1--27}.
\newblock


\bibitem[Schelble et~al\mbox{.}(0)]%
        {schelble2022ethical}
\bibfield{author}{\bibinfo{person}{Beau~G. Schelble}, \bibinfo{person}{Jeremy
  Lopez}, \bibinfo{person}{Claire Textor}, \bibinfo{person}{Rui Zhang},
  \bibinfo{person}{Nathan~J. McNeese}, \bibinfo{person}{Richard Pak}, {and}
  \bibinfo{person}{Guo Freeman}.} \bibinfo{year}{0}\natexlab{}.
\newblock \showarticletitle{Towards Ethical {AI}: Empirically Investigating
  Dimensions of AI Ethics, Trust Repair, and Performance in {Human-AI}
  Teaming}.
\newblock \bibinfo{journal}{\emph{Human Factors}} \bibinfo{volume}{0},
  \bibinfo{number}{0} (\bibinfo{year}{0}), \bibinfo{pages}{00187208221116952}.
\newblock
\urldef\tempurl%
\url{https://doi.org/10.1177/00187208221116952}
\showDOI{\tempurl}
\showeprint{https://doi.org/10.1177/00187208221116952}
\newblock
\shownote{PMID: 35938319}.


\bibitem[Scheutz et~al\mbox{.}(2017)]%
        {scheutz2017framework}
\bibfield{author}{\bibinfo{person}{Matthias Scheutz}, \bibinfo{person}{Scott~A
  DeLoach}, {and} \bibinfo{person}{Julie~A Adams}.}
  \bibinfo{year}{2017}\natexlab{}.
\newblock \showarticletitle{A framework for developing and using shared mental
  models in human-agent teams}.
\newblock \bibinfo{journal}{\emph{Journal of Cognitive Engineering and Decision
  Making}} \bibinfo{volume}{11}, \bibinfo{number}{3} (\bibinfo{year}{2017}),
  \bibinfo{pages}{203--224}.
\newblock


\bibitem[Sheridan(2016)]%
        {sheridan2016human}
\bibfield{author}{\bibinfo{person}{Thomas~B Sheridan}.}
  \bibinfo{year}{2016}\natexlab{}.
\newblock \showarticletitle{Human--robot interaction: status and challenges}.
\newblock \bibinfo{journal}{\emph{Human factors}} \bibinfo{volume}{58},
  \bibinfo{number}{4} (\bibinfo{year}{2016}), \bibinfo{pages}{525--532}.
\newblock


\bibitem[Smith(2019)]%
        {smith2019designing}
\bibfield{author}{\bibinfo{person}{Carol~J Smith}.}
  \bibinfo{year}{2019}\natexlab{}.
\newblock \showarticletitle{Designing trustworthy ai: A human-machine teaming
  framework to guide development}.
\newblock \bibinfo{journal}{\emph{arXiv preprint arXiv:1910.03515}}
  (\bibinfo{year}{2019}).
\newblock


\bibitem[Sowa and Przegalinska(2020)]%
        {sowa2020digital}
\bibfield{author}{\bibinfo{person}{Konrad Sowa} {and}
  \bibinfo{person}{Aleksandra Przegalinska}.} \bibinfo{year}{2020}\natexlab{}.
\newblock \showarticletitle{Digital coworker: human-AI collaboration in work
  environment, on the example of virtual assistants for management
  professions}. In \bibinfo{booktitle}{\emph{Digital Transformation of
  Collaboration: Proceedings of the 9th International COINs Conference}}.
  Springer, \bibinfo{pages}{179--201}.
\newblock


\bibitem[Steen et~al\mbox{.}(2023)]%
        {steen2023meaningful}
\bibfield{author}{\bibinfo{person}{Marc Steen}, \bibinfo{person}{Jurriaan van
  Diggelen}, \bibinfo{person}{Tjerk Timan}, {and} \bibinfo{person}{Nanda
  van~der Stap}.} \bibinfo{year}{2023}\natexlab{}.
\newblock \showarticletitle{Meaningful human control of drones: exploring
  human--machine teaming, informed by four different ethical perspectives}.
\newblock \bibinfo{journal}{\emph{AI and Ethics}} \bibinfo{volume}{3},
  \bibinfo{number}{1} (\bibinfo{year}{2023}), \bibinfo{pages}{281--293}.
\newblock


\bibitem[Stout et~al\mbox{.}(1999)]%
        {stout1999planning}
\bibfield{author}{\bibinfo{person}{Ren{\'e}e~J Stout}, \bibinfo{person}{Janis~A
  Cannon-Bowers}, \bibinfo{person}{Eduardo Salas}, {and}
  \bibinfo{person}{Dana~M Milanovich}.} \bibinfo{year}{1999}\natexlab{}.
\newblock \showarticletitle{Planning, shared mental models, and coordinated
  performance: An empirical link is established}.
\newblock \bibinfo{journal}{\emph{Human factors}} \bibinfo{volume}{41},
  \bibinfo{number}{1} (\bibinfo{year}{1999}), \bibinfo{pages}{61--71}.
\newblock


\bibitem[Stowers et~al\mbox{.}(2017)]%
        {stowers2017framework}
\bibfield{author}{\bibinfo{person}{Kimberly Stowers}, \bibinfo{person}{James
  Oglesby}, \bibinfo{person}{Shirley Sonesh}, \bibinfo{person}{Kevin Leyva},
  \bibinfo{person}{Chelsea Iwig}, {and} \bibinfo{person}{Eduardo Salas}.}
  \bibinfo{year}{2017}\natexlab{}.
\newblock \showarticletitle{A framework to guide the assessment of
  human--machine systems}.
\newblock \bibinfo{journal}{\emph{Human factors}} \bibinfo{volume}{59},
  \bibinfo{number}{2} (\bibinfo{year}{2017}), \bibinfo{pages}{172--188}.
\newblock


\bibitem[Talamadupula et~al\mbox{.}(2010)]%
        {talamadupula2010planning}
\bibfield{author}{\bibinfo{person}{Kartik Talamadupula}, \bibinfo{person}{J
  Benton}, \bibinfo{person}{Subbarao Kambhampati}, \bibinfo{person}{Paul
  Schermerhorn}, {and} \bibinfo{person}{Matthias Scheutz}.}
  \bibinfo{year}{2010}\natexlab{}.
\newblock \showarticletitle{Planning for human-robot teaming in open worlds}.
\newblock \bibinfo{journal}{\emph{ACM Transactions on Intelligent Systems and
  Technology (TIST)}} \bibinfo{volume}{1}, \bibinfo{number}{2}
  (\bibinfo{year}{2010}), \bibinfo{pages}{1--24}.
\newblock


\bibitem[Textor et~al\mbox{.}(2022)]%
        {text2022Exploring}
\bibfield{author}{\bibinfo{person}{Claire Textor}, \bibinfo{person}{Rui Zhang},
  \bibinfo{person}{Jeremy Lopez}, \bibinfo{person}{Beau~G. Schelble},
  \bibinfo{person}{Nathan~J. McNeese}, \bibinfo{person}{Guo Freeman},
  \bibinfo{person}{Richard Pak}, \bibinfo{person}{Chad Tossell}, {and}
  \bibinfo{person}{Ewart~J. de Visser}.} \bibinfo{year}{2022}\natexlab{}.
\newblock \showarticletitle{Exploring the Relationship Between Ethics and Trust
  in Human–Artificial Intelligence Teaming: A Mixed Methods Approach}.
\newblock \bibinfo{journal}{\emph{Journal of Cognitive Engineering and Decision
  Making}} \bibinfo{volume}{16}, \bibinfo{number}{4} (\bibinfo{year}{2022}),
  \bibinfo{pages}{252--281}.
\newblock
\urldef\tempurl%
\url{https://doi.org/10.1177/15553434221113964}
\showDOI{\tempurl}
\showeprint{https://doi.org/10.1177/15553434221113964}


\bibitem[To et~al\mbox{.}(2016)]%
        {to2016tandem}
\bibfield{author}{\bibinfo{person}{Alexandra To}, \bibinfo{person}{Elaine
  Fath}, \bibinfo{person}{Eda Zhang}, \bibinfo{person}{Safinah Ali},
  \bibinfo{person}{Catherine Kildunne}, \bibinfo{person}{Anny Fan},
  \bibinfo{person}{Jessica Hammer}, {and} \bibinfo{person}{Geoff Kaufman}.}
  \bibinfo{year}{2016}\natexlab{}.
\newblock \showarticletitle{Tandem Transformational Game Design: A Game Design
  Process Case Study}. In \bibinfo{booktitle}{\emph{Proceedings of the
  International Academic Conference on Meaningful Play}}.
\newblock


\bibitem[Tsarouchi et~al\mbox{.}(2017)]%
        {tsarouchi2017human}
\bibfield{author}{\bibinfo{person}{Panagiota Tsarouchi},
  \bibinfo{person}{George Michalos}, \bibinfo{person}{Sotiris Makris},
  \bibinfo{person}{Thanasis Athanasatos}, \bibinfo{person}{Konstantinos
  Dimoulas}, {and} \bibinfo{person}{George Chryssolouris}.}
  \bibinfo{year}{2017}\natexlab{}.
\newblock \showarticletitle{On a human--robot workplace design and task
  allocation system}.
\newblock \bibinfo{journal}{\emph{International Journal of Computer Integrated
  Manufacturing}} \bibinfo{volume}{30}, \bibinfo{number}{12}
  (\bibinfo{year}{2017}), \bibinfo{pages}{1272--1279}.
\newblock


\bibitem[Tufchi et~al\mbox{.}(2023)]%
        {tufchi2023comprehensive}
\bibfield{author}{\bibinfo{person}{Shivani Tufchi}, \bibinfo{person}{Ashima
  Yadav}, {and} \bibinfo{person}{Tanveer Ahmed}.}
  \bibinfo{year}{2023}\natexlab{}.
\newblock \showarticletitle{A comprehensive survey of multimodal fake news
  detection techniques: advances, challenges, and opportunities}.
\newblock \bibinfo{journal}{\emph{International Journal of Multimedia
  Information Retrieval}} \bibinfo{volume}{12}, \bibinfo{number}{2}
  (\bibinfo{year}{2023}), \bibinfo{pages}{28}.
\newblock


\bibitem[Vats et~al\mbox{.}(2024)]%
        {vats2024survey}
\bibfield{author}{\bibinfo{person}{Vanshika Vats}, \bibinfo{person}{Marzia
  Binta~Nizam}, \bibinfo{person}{Minghao Liu}, \bibinfo{person}{Ziyuan Wang},
  \bibinfo{person}{Richard Ho}, \bibinfo{person}{Mohnish Sai~Prasad},
  \bibinfo{person}{Vincent Titterton}, \bibinfo{person}{Sai Venkat~Malreddy},
  \bibinfo{person}{Riya Aggarwal}, \bibinfo{person}{Yanwen Xu},
  {et~al\mbox{.}}} \bibinfo{year}{2024}\natexlab{}.
\newblock \showarticletitle{A Survey on Human-AI Teaming with Large Pre-Trained
  Models}.
\newblock \bibinfo{journal}{\emph{arXiv e-prints}} (\bibinfo{year}{2024}),
  \bibinfo{pages}{arXiv--2403}.
\newblock


\bibitem[Vogel et~al\mbox{.}(2015)]%
        {vogel2015assistive}
\bibfield{author}{\bibinfo{person}{J. Vogel}, \bibinfo{person}{S. Haddadin},
  \bibinfo{person}{B. Jarosiewicz}, \bibinfo{person}{J.D. Simeral},
  \bibinfo{person}{D. Bacher}, \bibinfo{person}{L.R. Hochberg},
  \bibinfo{person}{J.P. Donoghue}, {and} \bibinfo{person}{P. van~der Smagt}.}
  \bibinfo{year}{2015}\natexlab{}.
\newblock \showarticletitle{An assistive decision-and-control architecture for
  force-sensitive hand–arm systems driven by human–machine interfaces}.
\newblock \bibinfo{journal}{\emph{The International Journal of Robotics
  Research}} \bibinfo{volume}{34}, \bibinfo{number}{6} (\bibinfo{year}{2015}),
  \bibinfo{pages}{763--780}.
\newblock
\urldef\tempurl%
\url{https://doi.org/10.1177/0278364914561535}
\showDOI{\tempurl}
\showeprint{https://doi.org/10.1177/0278364914561535}


\bibitem[Wachs and Duerstock(2010)]%
        {wachs2010analytical}
\bibfield{author}{\bibinfo{person}{Juan Wachs} {and} \bibinfo{person}{Brad
  Duerstock}.} \bibinfo{year}{2010}\natexlab{}.
\newblock \showarticletitle{An analytical framework to measure effective human
  machine interaction}.
\newblock In \bibinfo{booktitle}{\emph{Advances in Human Factors and Ergonomics
  in Healthcare}}. \bibinfo{publisher}{CRC Press}, \bibinfo{pages}{611--621}.
\newblock


\bibitem[Walliser et~al\mbox{.}(2019)]%
        {walliser2019team}
\bibfield{author}{\bibinfo{person}{James~C Walliser}, \bibinfo{person}{Ewart~J
  de Visser}, \bibinfo{person}{Eva Wiese}, {and} \bibinfo{person}{Tyler~H
  Shaw}.} \bibinfo{year}{2019}\natexlab{}.
\newblock \showarticletitle{Team structure and team building improve
  human--machine teaming with autonomous agents}.
\newblock \bibinfo{journal}{\emph{Journal of Cognitive Engineering and Decision
  Making}} \bibinfo{volume}{13}, \bibinfo{number}{4} (\bibinfo{year}{2019}),
  \bibinfo{pages}{258--278}.
\newblock


\bibitem[Wan et~al\mbox{.}(2022)]%
        {wan2022user}
\bibfield{author}{\bibinfo{person}{Ruyuan Wan}, \bibinfo{person}{Naome Etori},
  \bibinfo{person}{Karla Badillo-Urquiola}, {and} \bibinfo{person}{Dongyeop
  Kang}.} \bibinfo{year}{2022}\natexlab{}.
\newblock \showarticletitle{User or Labor: An Interaction Framework for
  Human-Machine Relationships in {NLP}}. In
  \bibinfo{booktitle}{\emph{Proceedings of the Fourth Workshop on Data Science
  with Human-in-the-Loop (Language Advances)}}. \bibinfo{pages}{112--121}.
\newblock


\bibitem[Warren and Hillas(2020)]%
        {warren2020friend}
\bibfield{author}{\bibinfo{person}{Aiden Warren} {and} \bibinfo{person}{Alek
  Hillas}.} \bibinfo{year}{2020}\natexlab{}.
\newblock \showarticletitle{Friend or frenemy? The role of trust in
  human-machine teaming and lethal autonomous weapons systems}.
\newblock \bibinfo{journal}{\emph{Small Wars \& Insurgencies}}
  \bibinfo{volume}{31}, \bibinfo{number}{4} (\bibinfo{year}{2020}),
  \bibinfo{pages}{822--850}.
\newblock


\bibitem[Watson and Scheidt(2005)]%
        {watson2005autonomous}
\bibfield{author}{\bibinfo{person}{David~P Watson} {and}
  \bibinfo{person}{David~H Scheidt}.} \bibinfo{year}{2005}\natexlab{}.
\newblock \showarticletitle{Autonomous systems}.
\newblock \bibinfo{journal}{\emph{Johns Hopkins APL technical digest}}
  \bibinfo{volume}{26}, \bibinfo{number}{4} (\bibinfo{year}{2005}),
  \bibinfo{pages}{368--376}.
\newblock


\bibitem[Wu et~al\mbox{.}(2022)]%
        {wu2022survey}
\bibfield{author}{\bibinfo{person}{Xingjiao Wu}, \bibinfo{person}{Luwei Xiao},
  \bibinfo{person}{Yixuan Sun}, \bibinfo{person}{Junhang Zhang},
  \bibinfo{person}{Tianlong Ma}, {and} \bibinfo{person}{Liang He}.}
  \bibinfo{year}{2022}\natexlab{}.
\newblock \showarticletitle{A survey of human-in-the-loop for machine
  learning}.
\newblock \bibinfo{journal}{\emph{Future Generation Computer Systems}}
  \bibinfo{volume}{135} (\bibinfo{year}{2022}), \bibinfo{pages}{364--381}.
\newblock


\bibitem[Yang et~al\mbox{.}(2022)]%
        {Yang22-thms-review-mhrobotics}
\bibfield{author}{\bibinfo{person}{Canjun Yang}, \bibinfo{person}{Yuanchao
  Zhu}, {and} \bibinfo{person}{Yanhu Chen}.} \bibinfo{year}{2022}\natexlab{}.
\newblock \showarticletitle{A Review of Human-Machine Cooperation in the
  Robotics Domain}.
\newblock \bibinfo{journal}{\emph{IEEE Transactions on Human-Machine Systems}}
  \bibinfo{volume}{52}, \bibinfo{number}{1} (\bibinfo{year}{2022}),
  \bibinfo{pages}{12--25}.
\newblock
\urldef\tempurl%
\url{https://doi.org/10.1109/THMS.2021.3131684}
\showDOI{\tempurl}


\bibitem[Yang et~al\mbox{.}(2023)]%
        {yang2023inner}
\bibfield{author}{\bibinfo{person}{Scott Cheng-Hsin Yang},
  \bibinfo{person}{Tomas Folke}, {and} \bibinfo{person}{Patrick Shafto}.}
  \bibinfo{year}{2023}\natexlab{}.
\newblock \showarticletitle{The Inner Loop of Collective Human--Machine
  Intelligence}.
\newblock \bibinfo{journal}{\emph{Topics in Cognitive Science}}
  (\bibinfo{year}{2023}).
\newblock


\bibitem[Yildirim(2022)]%
        {yildirim2022smt}
\bibfield{author}{\bibinfo{person}{Gungor Yildirim}.}
  \bibinfo{year}{2022}\natexlab{}.
\newblock \showarticletitle{A novel hybrid multi-thread metaheuristic approach
  for fake news detection in social media.}
\newblock  (\bibinfo{year}{2022}).
\newblock


\bibitem[Zhang et~al\mbox{.}(2020)]%
        {zhang2020anideal}
\bibfield{author}{\bibinfo{person}{Rui Zhang}, \bibinfo{person}{Nathan
  McNeese}, \bibinfo{person}{Guo Freeman}, {and} \bibinfo{person}{Geoff
  Musick}.} \bibinfo{year}{2020}\natexlab{}.
\newblock \showarticletitle{"An Ideal Human": Expectations of AI Teammates in
  Human-AI Teaming}.
\newblock \bibinfo{journal}{\emph{Proceedings of the ACM on Human-Computer
  Interaction}}  \bibinfo{volume}{4} (\bibinfo{date}{12} \bibinfo{year}{2020}),
  \bibinfo{pages}{246}.
\newblock
\urldef\tempurl%
\url{https://doi.org/10.1145/3432945}
\showDOI{\tempurl}


\end{thebibliography}

\newpage
\appendix

\section{Empirical Studies to Promote Team Performance: Summary}

Since the discussions of the empirical studies to promote team performance in HMT systems in Section 3 of the main paper are very brief, we elaborate the discussion for each work with more details in this appendix section below.

\subsection{Team Training}\label{subsec:team-training}
\citet{johnson2021impact} explored how communication and calibration influence HAT performance, comparing training approaches under degraded conditions with a mixed testbed of real and simulated platforms. Teams of three humans and one AI agent executed missions on a simulated aircraft using three training types: {\em control}, covering basic skills in Remotely Piloted Aircraft System-Synthetic Task Environment (RPAS-STE); {\em coordination}, enhancing communication and responses to automation failures; and {\em calibration}, aligning AI trust through realistic expectations and persistent communication. Metrics included {\em target processing efficiency} and {\em team trust score}. Coordination-trained teams showed better communication and trust, though they did not outperform control-trained teams. Limited statistical power constrained the findings. Similarly, \citet{harpstead2023speculative} used speculative game design to study HMT dynamics in asymmetric cooperative games, focusing on roles, structures, and knowledge distribution. Inspired by {\em Tandem Transformational Game Design}~\cite{to2016tandem}, the approach integrated prior studies, expert input, and iterative playtesting, addressing communication, coordination, and adaptation. However, it lacked details on training or evaluating AI models and relied on simple grid-world games (e.g., Dice Adventure), limiting applicability to real-world tasks requiring 3D environments for greater complexity.

\citet{myers2019} introduced {\em autonomous synthetic teammates} (ASTs) to human teams to reduce scheduling complexities and training costs. These ASTs mimic human cognition and behaviors or optimize task performance, often at the expense of human-like behavior. The study highlighted the challenge of balancing high performance with reduced AST complexity but overlooked cognitive fidelity, limiting ASTs' effectiveness in delicate tasks where they could outperform human agents. Similarly, \citet{o2023human} proposed the Input-Mediator-Output (IMO) model to categorize and integrate HMT research. While useful for feature categorization and hypothesis validation, the model fails to account for the bidirectional relationships and feedback loops inherent in dynamic HMT interactions.

\citet{endsley2024taxonomy} developed a taxonomy for incremental HMT levels, advocating for a shift from autonomy-centric designs to collaborative autonomous systems tailored for extreme mission environments. Building on this, \citet{cleland2024human} extended the MAPE-K feedback loop to create MAPE-KHMT, a framework aligning HMT principles with adaptive system phases to enable bidirectional human-machine interaction. Additionally, \citet{lematta2024practical} emphasized the importance of HMT assurance in AI system development and deployment. Using General Motors Cruise robotaxis as a case study, the authors illustrated how poor HMT design leads to trust and safety failures, including disrupted emergency operations and ineffective human interaction mechanisms.

\subsection{Team Autonomy} \label{subsec:team-performance}

\citet{schelble2023investigating} studied the impact of AI teammates on HMT performance using the {\em Implicit Interaction for Human-Autonomy Teams} (IIHAT) simulation, where human performance was averaged across teammates' moves to complete six objectives on a fictitious island in minimal moves. AI teammates improved performance, mediated interactions, and enhanced perceptions in human-AI teams (HATs). However, the study lacked explicit communication and used simple digital tasks, limiting its real-world applicability.
\citet{schelble2022see} examined spatial awareness in team cognition within human-only and human-agent teams using the {\em NeoCITIES} emergency response platform~\cite{jones2004distributed}. Variables like spatial awareness (high vs. low) and team composition (human-agent vs. human-only) were manipulated to assess metrics such as mental model similarity, perceived cognition, and team performance. Findings showed spatial awareness influences cognition and performance differently across team types to enhance human-agent team cognition.

\citet{Hauptman23-autonomy} emphasized the need for autonomous agents to adapt to dynamic environments, akin to human behavior. They proposed a process to identify AI teammates' autonomy levels and examine how cognitive control changes influence team dynamics. However, the study relied on hypothetical and abstract questionnaire scenarios, potentially affecting participant responses.  \citet{mahaini2019building} introduced a methodology for constructing taxonomies by combining human expertise with automated natural language process (NLP) and information retrieval tools. Applied to cybersecurity, the three-stage process (i.e., data collection, text analysis, and taxonomy construction) efficiently processed large text datasets, creating a comprehensive taxonomy with reduced human workload and enhanced accuracy.

\subsection{Trust \& Decision-Making} \label{subsec:trust}

\citet{henry2022human} examined HMT applications in medical care, focusing on decision-making for problem detection and resolution using the Targeted Real-time Early Warning System (TREWS), an ML-based system for sepsis patients. Surveys of 20 medical professionals revealed indifference toward TREWS despite its effectiveness, attributed to poor understanding of its mechanisms and low trust. However, the small sample size limits generalizability. \citet{bersani2023towards} proposed the three-layer EASE (\underline{E}xploration, \underline{A}naly\underline{S}is, and \underline{E}xplanation) architecture for improving trust in machine systems via dependability and explainability. While it includes Statistical Model Checking (SMC) and ML models for task success prediction, limited validation and uncertainty from human behavior dimensionality remain challenges.

\citet{christopher2018navigating} argued that machines lack full autonomy due to the inability to handle uncertainty and proposed high-trust interactions and language processing for effective HMTs. However, their approach lacks robust empirical validation. \citet{haring2021applying} investigated "swift trust" for human-robot teams, emphasizing appearance and functionality in trust formation but relying on theoretical models without empirical evidence. \citet{MUSICK2021106852} explored how team composition affects cognition and sentiment in HATs, noting weak communication and limited team cognition in multi-agent settings, but their small team size and short task duration reduce applicability. Similarly, \citet{schelble2022let} used the {\em NeoCITIES} emergency response simulator to examine team cognition and performance across \underline{H}uman-\underline{A}I compositions (HHH, HHA, HAA). While AI improved average performance, qualitative data diminished with fewer humans in the team, complicating robust trust measurements.

\citet{mcneese2021trust} highlighted trust's impact on team performance using RPAS simulations, finding that low trust in autonomous agents reduces team performance, creating a negative feedback loop. However, the autonomous agent lacked self-learning capabilities, limiting adaptability. \citet{zhang2020anideal} explored perceptions of AI teammates, showing positive experiences enhance trust, but their unbalanced male participant pool and task-specific responses limit the findings. \citet{Demir2021Exploration} investigated trust and interaction dynamics in HATs using Joint Recurrence Quantification Analysis~\cite{fusaroli2014analyzing}, uncovering links between interaction dynamics and trust development. Challenges include handling interactions with multiple agents and addressing cumulative, theory-driven analysis needs. \citet{maathuis2024trustworthy} proposed a working definition and design framework for trustworthy HAT in the military, ensuring that HAT systems are safe, responsible, and reliable while accounting for legal, ethical, and societal norms. \citet{centeio2024interdependence} introduced the Interdependence and Trust Analysis (ITA) framework, which integrated three dimensions of trustworthiness: competence, willingness, and external factors. The ITA framework facilitated informed decision-making by analyzing these dimensions for different tasks in human-machine teams, thereby optimizing team configuration and task allocation.

\subsection{Shared Mental Model} \label{subsec:smm}

\citet{andrews2023role} emphasized the importance of shared mental models (SMMs) for HMT performance, distinguishing between task and role models. Task models address goals, required skills, workload, and quality, while role models focus on team members' capabilities. \citet{mathieu2000influence} showed that shared task and team mental models improve team dynamics, interactions, and outcomes, as demonstrated in a flight-combat simulation. This underscores the critical role of SMMs in aligning team understanding for effective communication and enhanced HMT performance.

\citet{demir2020understanding} explored shared cognition and restricted language in human-robot USAR (Urban Search and
Rescue) teaming. Minecraft simulations showed natural language and strong SMMs improved performance, but robot communication limits and platform constraints reduced generalizability. \citet{de2023managerial} identified 20 behavioral strategies for Industry 4.0 adaptation using 45 interviews, emphasizing design thinking and strategic alignment. Qualitative data reliance limits applicability.  \citet{edgar2023improving} integrated proactive robot behaviors and SMMs into virtual reality simulations, showing improved team effectiveness under high cognitive loads. Proactive behaviors, such as sharing information and engaging in shared tasks, aligned human-robot goals and expectations. However, the study lacked methodological detail and real-world applicability. \citet{scheutz2017framework} proposed a framework enabling artificial agents to develop and maintain SMMs, enhancing communication and performance in mixed human-agent teams. While addressing computational methods and data structures, limitations in agents' ability to mirror human cognition affect team coherence.

\citet{stout1999planning} highlighted the role of team planning in improving SMMs and coordination. Using a surveillance/defense simulation with undergraduate participants, they found planning enhanced SMMs, communication, and performance under high workloads. Mental model similarity, measured using Pathfinder $C$, allowed for better anticipation of informational needs and reduced errors. However, findings were constrained by laboratory settings and unrepresentative samples. \citet{schelble2022let} studied team cognition in human-agent teams (HATs), finding that iterative cognition development and communication are critical. Simulations with varied team compositions (all-human, human-human-agent, human-agent-agent) revealed that HATs require clear goal communication and action-oriented discussions. The inability to evaluate agents' mental models limited deeper insights into HAT cognition.

\subsection{Team Situation Awareness} \label{subsec:tsa}

\citet{salas2017situation} highlighted Team Situation Awareness (TSA) as vital for HMT performance, framing it as a shared aggregation of individual team members' SA~\cite{endsley1995measurement}. TSA influences HMT behaviors like communication and knowledge sharing~\cite{mathieu2000influence}, though its reliance on flight-combat simulations may not fully capture the complexities of real-world team dynamics. In less dynamic settings, shared knowledge bolsters TSA and team performance. In contrast, dynamic environments depend more on timely, coordinated communication~\cite{cooke2013interactive}, challenging traditional shared cognition models but lacking clear guidelines for consistent measurement across tasks. \citet{salas2017situation} primarily focused on aviation and military domains, limiting broader applicability and relying on theoretical constructs without direct empirical validation. While the SAGAT methodology~\cite{endsley1995measurement} offers a robust way to measure SA, its intrusiveness can disrupt task dynamics, raising concerns about ecological validity in real-world contexts.

\citet{demir2019evolution} analyzed language and behavior in UAV-based HMTs to enhance TSA and performance but relied heavily on synthetic task environments, limiting applicability to real-world scenarios. The study found proactive information sharing increased TSA in synthetic teams, though language constraints delayed SA development compared to human-only teams. Similarly, \citet{mcneese2021team} explored TSA and conflict in RPAS-based HMTs, demonstrating TSA's role in performance improvement using split-plot ANOVA. However, machine agents' inability to process non-textual data reduced generalizability. \citet{demir2017tsa} examined verbal behaviors in HATs, showing proactive information sharing improved TSA and performance, with growth curve modeling revealing fewer verbal exchanges in human-autonomy teams than human-only teams. However, findings across these studies are primarily derived from simulated scenarios, constrained by task-specific designs and the limited communication capabilities of synthetic agents, potentially limiting generalizability to more advanced autonomous systems.

\citet{mcneese2021tsa} observed TSA and CAST (Collaborative Action and SA in Teams) improvements in synthetic RPAS teammates over time, though text-only processing and narrow contextual focus limited applicability. \citet{chen2017tsa} developed the Situation Awareness-based Agent Transparency (SAT) model, incorporating agent status, reasoning logic, and future plans to enhance human understanding in human-autonomy collaborations. Simulations with ASM and IMPACT systems showed transparency, improved performance, and trust, highlighting the importance of clear two-way communication. However, the model revealed trade-offs, with increased operator workload and decision latency under certain conditions. \citet{rinta2020tsa} examined SA in collaborative driving and introduced the Supportive SA (SSA) model to reduce over-reliance on ADAS. The model emphasized transparency to align driver trust with ADAS capabilities, ensuring safer collaboration. However, SSA assumed clear hierarchical relationships and did not fully address overlapping goals or conflicts, potentially oversimplifying real-world collaborative driving scenarios.

\subsection{Synergistic Human-AI Collaboration} \label{subsec:synergy-collab}

\citet{cai2020perceiving} showcased HMT's potential in cyber malware detection by training machines to mimic human cognitive processes in identifying malware patterns. Their method converts expert knowledge into visual ontologies and graph structures, enabling humans and machines to analyze malware distribution networks (MDNs). By incorporating human feedback, the machine learning algorithm detects, maps, and predicts malware distributions, identifying critical MDN nodes like hubs and bridges to disrupt threats. This study highlights the synergy of human expertise and machine learning for cyber defense but notes scalability challenges due to reliance on extensive, high-quality human input.

\subsection{Team Communication \& Coordination} \label{subsec:team-communication}

In HMT, effective communication is vital for collaboration between human operators and autonomous systems. \citet{endsley1999level} emphasized its role in fostering shared situational awareness, enhancing coordination and adaptability. However, their study relied on simulations and assumed a linear automation hierarchy, limiting real-world applicability. \citet{parasuraman2008situation} highlighted communication’s role in reducing cognitive overload and operator errors but did not fully address evolving technologies. Transparent communication builds trust and improves performance, as noted by \citet{lee2004trust}. While effective strategies optimize coordination and reduce cognitive burdens, adapting systems for diverse, non-expert users remains an open challenge requiring further empirical validation.

\citet{chiou2021towards} studied communication in human-robot search and rescue using Minecraft simulations, testing four conditions: `always explain,' `explain if asked,' `pull prime,' and `never explain.' Results showed that frequent communication did not always improve performance or Team Situational Awareness (TSA), with `pull prime' and `explain if asked' yielding similar outcomes. However, reliance on simulations limits real-world applicability. \citet{demir2019evolution} examined communication and coordination in RPAS tasks, finding that NLP-enhanced autonomous agents improved operational interactions. Task performance varied under failures like automation breakdowns and cyber-attacks, with moderate communication stability yielding the best outcomes. While human-centered machine learning improved agent performance, untrained users struggled, underscoring the need for intuitive interfaces and training.

\citet{demir2017team} explored how verbal behaviors in HATs affect TSA and performance. Using ACT-R cognitive modeling in simulations, teams with greater verbal exchanges outperformed those with less interaction. However, the study’s focus on specific tasks in a singular testbed limits generalizability, requiring broader validation across diverse scenarios.  \citet{moore2007presence} developed PRESENCE, a speech-based HMI architecture inspired by neurobiology, addressing scalability and performance limitations of conventional systems. By integrating dialogue components into a recursive hierarchical feedback model, PRESENCE enables cooperative behavior through mutual intention recognition. However, its setup demands precise parameter specifications, making initial implementation challenging.

\subsection{Team Composition} \label{subsec:team-composition}

\citet{mcneese2021my} explored communication and situation awareness (SA) in human-autonomy teaming, highlighting AI's limitations in NLP compared to human communication through physical cues like facial and hand movements. Using NeoCITIES experiments with HHH, HHA, HAA, and AAA teams (H: human, A: AI), the study measured team scores, TSA, and team cognition. HHH teams showed the highest team cognition, while TSA scores were best in AI-inclusive teams (HHA, HAA, AAA). More AIs in teams improved TSA, while HHH teams performed poorly. AIs were trained via reinforcement learning, but untrained humans exhibited biases, including distrust of AI.  

\citet{demir2018team} applied nonlinear dynamical systems (NDS) to examine the relationship between team performance and coordination dynamics in HATs. Experiments using CERTT-II Test Bed and RPAS tasks revealed that synthetic teams with human-like agents experienced non-linear performance changes. Stable coordination struggled with environmental shifts, impairing performance.    \citet{mcneese2018teaming} studied team composition effects in UAS tasks, finding all-human teams were more efficient due to superior coordination. However, synthetic teammates struggled with developing effective communication strategies. Task difficulty may have influenced results, but this was not explored.  

\citet{walliser2019team} emphasized team dynamics and social communication as critical for HMT performance. Experiments revealed that informing humans about their partner's nature (human-like vs. machine-like) altered behaviors without significantly improving performance. Team building played a key role in task completion but lacked details on training or task preparation for human and machine agents. \citet{flathmann2023examining} explored AI influence on shared resources in HAT systems, comparing static and dynamic AI roles. Findings showed human perceptions of AI's influence depended on goals and motivations, shaping preferences during operations. While the study examined team performance and trust, it lacked metrics to fully assess AI influence's impact on outcomes.

\subsection{Task Allocation} \label{subsec:task-allocation}

\citet{roth2019function} reviewed function allocation methods in HMT systems with high autonomy, focusing on workload, situational awareness, and responses to automation failures. While these approaches broaden design variables, they rely on traditional frameworks, limiting adaptability to modern systems. \citet{gray2019ghost} debunked the myth of fully automated AI services, revealing the critical human role in tasks like image labeling and content moderation. This "ghost work" underscores human creativity's importance, suggesting automation creates new jobs but challenges traditional work cycles in HCAI. \citet{tsarouchi2017human} proposed a method for human-robot task planning and workplace design, treating humans and robots as active members. Using a three-level workload model and multi-criteria decision-making, they developed a prototype for hybrid layouts in white goods and automotive sectors, automating task allocation and integrating user-defined criteria.  The main challenges across in~\cite{roth2019function, gray2019ghost, tsarouchi2017human} include adapting to evolving systems, balancing human-machine roles, ensuring transparency, enabling scalable designs, and prioritizing human-centric approaches in HMT contexts.

\subsection{AI \& Ethics} \label{subsec:ai-ethics}

\citet{assaad2023ethics} defined ethical AI as adhering to moral standards prioritizing human well-being and accountability, contrasting with unethical AI that risks harm by violating these norms. Key principles include responsibility, trustworthiness, transparency, fairness, and cautious dual-use to align AI with human values~\cite{smith2019designing, ali2019constructionism}. They stressed integrating these principles into HMT systems. They emphasized the need for empirical analysis to validate trustworthiness and safety.  \citet{smith2019designing} proposed an HMT framework for ethical AI interactions, guiding AI development to meet transparency and understandability standards through advanced natural language processing. While offering insights on building ethical AI teams, they underscored the need for empirical validation via simulations and evaluative metrics. \citet{text2022Exploring} examined human perceptions of AI ethical behavior through surveys and interviews, showing ethical compliance maintained trust, while violations, especially of proportionality, eroded it. The findings highlighted a nuanced relationship between AI ethics and human trust. However, \cite{assaad2023ethics, smith2019designing, ali2019constructionism, text2022Exploring} lack comprehensive, empirically validated strategies to implement these principles across varied HMT scenarios. This leaves questions of scalability, domain-specific complexities, and insufficient real-world integration.

\citet{lopez2023complex} explored how ethical AI teammates impact team trust and communication. They found that human confidence in AI's ethical decision-making fosters trust and encourages risk-taking but noted that unclear AI decisions could harm coordination. Feedback from Air Force pilots interacting with AI teammates stressed the importance of diverse inputs for improving HMT systems.  \citet{schelble2022ethical} studied trust repair in Human-Autonomous Teams (HATs) involving AI exhibiting unethical behavior. While unethical AI led to faster mission completion, trust repair attempts like apologies failed, demonstrating the complex link between ethical behavior and team performance. The study highlighted task-specific variations in ethical impacts, calling for broader validation through generalized simulations.  \citet{steen2023meaningful} examined meaningful human control (MHC) over autonomous systems through ethical frameworks like Utilitarianism, Deontology, Relational Ethics, and Virtue Ethics. In a fictional military scenario, drones operated autonomously while soldiers assessed threats, showcasing ethical AI's potential. However, the study's conceptual approach lacked empirical validation, raising questions about the proposed designs' reliability.

Table~\ref{tab:hmt_detailed_summary} summarizes the empirical studies on promoting team performance discussed in Section~3 of the main paper.

\begin{table}[t]
\centering
\small
\caption{Comprehensive Summary of Contributions and Limitations Across HMT Subsections.}
\label{tab:hmt_detailed_summary}
\vspace{-3mm}
\begin{tabular}{|p{2.5cm}|p{6.5cm}|p{5cm}|}
\hline
\multicolumn{1}{|c}{\bf Source} & \multicolumn{1}{|c}{\bf Key Contributions} & \multicolumn{1}{|c|}{\bf Limitations} \\ \hline

\multicolumn{3}{|c|}{\textbf{\gray Team Training}} \\ \hline
\citet{johnson2021impact} & Explored the impact of communication and calibration on HAT performance with three training types: control, coordination, and calibration. Metrics included target processing efficiency and team trust score. & Limited statistical power and reliance on simulated platforms constrained generalizability. \\ \hline

\citet{harpstead2023speculative} & Applied speculative game design to analyze HMT dynamics in cooperative games, focusing on roles, structures, and knowledge distribution. Integrated iterative playtesting and expert input for design refinement. & Lacked specificity in AI model training and evaluation. Relied on simple grid-world games, limiting applicability to complex real-world tasks. \\ \hline

\citet{myers2019} & Proposed Autonomous Synthetic Teammates (ASTs) to reduce training costs and mimic human cognition, abstract behaviors, or task optimization. & Neglected cognitive fidelity, reducing effectiveness for delicate tasks. Overemphasis on performance optimization sacrificed human-like behavior. \\ \hline

\citet{o2023human} & Developed the Input-Mediator-Output (IMO) model to categorize HMT features, validate hypotheses, and synthesize insights across human and machine agents. & The linear model did not account for bidirectional feedback loops and dynamic interactions in human-autonomy teams. \\ \hline

\citet{endsley2024taxonomy} & {Highlighted the importance of autonomy in supporting HMT for complex space operations. The paper proposed a novel taxonomy for human-machine teaming to guide the design of systems based on mission needs and desired capabilities.} & {The proposed taxonomy lacks real-world validation, making its effectiveness in practical mission scenarios uncertain.} \\ \hline

\citet{cleland2024human} & {Explored the application of the MAPE-K feedback loop in autonomous systems and its shift towards Human-Machine Teaming (HMT), where machines and humans collaborate to optimize task efficiency. } & {The proposed models were still in the developmental phase and require further testing in diverse real-world scenarios to validate their scalability and adaptability across different systems and environments.} \\ \hline

\citet{lematta2024practical} & {Emphasized the importance of HMT assurance in AI-enabled systems, particularly for high-stakes environments like military operations and autonomous vehicles.} & {The paper relied on a single example and did not explore other case studies or broader applications across different sectors, limiting its generalizability.} \\ \hline

\multicolumn{3}{|c|}{\textbf{\gray Team Autonomy}} \\ \hline

\citet{schelble2023investigating} & Investigated the impact of AI teammates on HMT performance using the IIHAT simulation. Found that AI teammates enhanced performance, mediated interactions, and improved perceptions in HATs. & Lacked explicit communication mechanisms and relied on simple digital tasks, limiting real-world applicability. \\ \hline

\citet{schelble2022see} & Examined spatial awareness in team cognition using the NeoCITIES platform for human-only and human-agent teams. Identified that spatial awareness impacts cognition and performance differently based on team composition. & Limited exploration of spatial awareness under more diverse environments and team structures. \\ \hline

\citet{Hauptman23-autonomy} & Proposed a method to identify autonomy levels of AI teammates and examined how cognitive control changes influence team dynamics. Emphasized the need for autonomous agents to adapt to dynamic environments. & Relied on hypothetical and abstract questionnaire scenarios, which may not reflect real-world responses. \\ \hline

\citet{mahaini2019building} & Developed a three-stage taxonomy construction methodology using NLP and information retrieval tools to process large datasets. Applied this to cybersecurity, improving accuracy and reducing human workload. & Focused specifically on cybersecurity, limiting the generalizability of the approach to other domains. \\ \hline
\end{tabular}
\end{table}

\begin{table}[htbp]
\centering
\caption*{Table 1. (continue) Comprehensive Summary of Contributions and Limitations Across HMT Subsections.}
\label{tab:hmt_detailed_summary-2}
\vspace{-3mm}
\small 
\begin{tabular}{|p{2.5cm}|p{6.5cm}|p{5cm}|}
\hline
\multicolumn{1}{|c}{\bf Source} & \multicolumn{1}{|c}{\bf Key Contributions} & \multicolumn{1}{|c|}{\bf Limitations} \\ \hline
\multicolumn{3}{|c|}{\textbf{\gray Trust \& Decision-Making}} \\ \hline

\citet{henry2022human} & Investigated HMT applications in medical care, focusing on the TREWS ML-based system for sepsis detection. Highlighted decision-making improvements for problem detection and resolution. & Surveys revealed indifference to TREWS due to poor understanding and low trust. A small sample size limits generalizability. \\ \hline

\citet{bersani2023towards} & Proposed the EASE architecture to improve trust in machine systems via dependability and explainability. Included Statistical Model Checking (SMC) and ML models for task success prediction. & Limited validation of architecture and uncertainty from human behavior dimensionality remain challenges. \\ \hline

\citet{christopher2018navigating} & Highlighted the inability of machines to handle uncertainty and proposed high-trust interactions and language processing for HMTs. & Theoretical approach lacks robust empirical validation. \\ \hline

\citet{haring2021applying} & Investigated "swift trust" formation in human-robot teams based on appearance and functionality. & Relied on theoretical models without empirical evidence. \\ \hline

\citet{MUSICK2021106852} & Explored team composition impacts on cognition and sentiment in HATs, noting weak communication and limited cognition in multi-agent settings. & Small team size and short task duration reduce applicability. \\ \hline

\citet{schelble2022let} & Analyzed team cognition and performance across HHH, HHA, and HAA team compositions using NeoCITIES. Found AI improved average performance, but qualitative data declined with fewer humans. & Limited scenarios and fewer humans complicated trust measurements. \\ \hline

\citet{mcneese2021trust} & Found that low trust in autonomous agents negatively impacts team performance, creating a negative feedback loop. & The autonomous agent lacked self-learning capabilities, limiting adaptability. \\ \hline

\citet{zhang2020anideal} & Showed that positive experiences with AI teammates enhance trust. & Findings were limited by an unbalanced male participant pool and task-specific responses. \\ \hline

\citet{Demir2021Exploration} & Investigated trust and interaction dynamics in HATs using Joint Recurrence Quantification Analysis. Identified links between interaction dynamics and trust development. & Challenges include handling interactions with multiple agents and addressing cumulative, theory-driven analysis needs. \\ \hline

\citet{huang2024co} & Proposed the Co-Matching framework for human-machine collaborative legal case matching, leveraging ProtoEM to estimate human decision uncertainty without ground truth. & Limited to binary decision tasks and requires historical data for prototype clustering. \\ \hline

\citet{maathuis2024trustworthy} & {Addressed the concept of trustworthy HAT within the context of proportionality assessments in military operations, highlighting the critical role of collaboration between humans and autonomous systems in achieving common goals. } & {The paper solely focused on conceptual development and lacked empirical validation or case studies to demonstrate the practical applicability.} \\ \hline

\citet{centeio2024interdependence} & {Highlighted the importance of assessing not only the capabilities of the team members (humans and machines) but also their willingness and external factors (i.e. ethical considerations) in task allocation and collaboration.} & {The paper lacked detailed empirical validation across diverse team configurations or real-world applications.} \\ \hline

\multicolumn{3}{|c|}{\textbf{\gray Shared Mental Models (SMM)}} \\ \hline

\citet{andrews2023role} & Highlighted the importance of shared mental models (SMMs) for HMT performance. Distinguished between task models (goals, skills, workload) and role models (team member capabilities). & Limited exploration of task and role models in dynamic, real-world environments. \\ \hline

\citet{mathieu2000influence} & Demonstrated that shared task and team mental models improve team dynamics, communication, and outcomes in a flight-combat simulation. & Focused on a specific scenario, limiting applicability to other domains. \\ \hline

\end{tabular}
\end{table}

\begin{table}[htbp]
\centering
\caption*{Table 1. (continue) Comprehensive Summary of Contributions and Limitations Across HMT Subsections.}
\label{tab:hmt_detailed_summary-2}
\vspace{-3mm}
\small 
\begin{tabular}{|p{2.5cm}|p{7cm}|p{4.5cm}|}
\hline
\multicolumn{1}{|c}{\bf Source} & \multicolumn{1}{|c}{\bf Key Contributions} & \multicolumn{1}{|c|}{\bf Limitations} \\ \hline

\multicolumn{3}{|c|}{\textbf{\gray Shared Mental Models (SMM) -- continue~}} \\ \hline

\citet{demir2020understanding} & Explored shared cognition and restricted language in human-robot USAR teams using Minecraft simulations. Found that natural language and strong SMMs improved team performance. & Platform simplicity and robot communication constraints reduced generalizability. \\ \hline

\citet{de2023managerial} & Identified 20 behavioral strategies for Industry 4.0 adaptation through interviews, emphasizing design thinking and strategic alignment. & Relied heavily on qualitative data, limiting broader applicability. \\ \hline

\citet{edgar2023improving} & Integrated proactive robot behaviors and SMMs into virtual reality simulations to improve team effectiveness under high cognitive loads. Proactive behaviors aligned human-robot goals and expectations. & Lacked methodological detail and real-world applicability. \\ \hline

\citet{scheutz2017framework} & Proposed a computational framework for artificial agents to develop and maintain SMMs, enhancing communication and performance in mixed human-agent teams. & Limitations in agents’ ability to replicate human cognition affected team coherence. \\ \hline

\citet{stout1999planning} & Showed that team planning improves SMMs, communication, and performance under high workloads in a surveillance/defense simulation. Mental model similarity reduced errors and improved anticipation of needs. & Findings were limited by laboratory settings and unrepresentative samples. \\ \hline

\citet{schelble2022let} & Analyzed team cognition in HATs, showing that iterative cognition development and communication are critical for success. HATs require clear goal communication and action-oriented discussions. & The inability to evaluate agents’ mental models limited deeper insights into HAT cognition. \\ \hline

\multicolumn{3}{|c|}{\textbf{\gray Team Situation Awareness (TSA)}} \\ \hline

\citet{salas2017situation} & Identified TSA as critical for HMT performance, highlighting shared knowledge as a composite of individual team members' SA. Emphasized the role of communication and knowledge sharing in TSA development. & Limited exploration of TSA in dynamic, high-pressure environments. \\ \hline

\citet{demir2019evolution} & Analyzed language and behavior in UAV-based HMTs, finding that proactive information sharing improved TSA and team performance. & Language processing limitations delayed TSA development compared to human-only teams. \\ \hline

\citet{mcneese2021team} & Explored TSA and conflict in RPAS-based HMTs, demonstrating TSA's role in improving performance through split-plot ANOVA. & Machine agents' inability to process non-textual data limited generalizability. \\ \hline

\citet{demir2017tsa} & Examined verbal behaviors in HATs, showing that proactive information sharing improved TSA and team performance. Growth curve modeling revealed fewer verbal exchanges in human-autonomy teams. & Findings were constrained by task-specific scenarios, limiting broader applicability. \\ \hline

\citet{mcneese2021tsa} & Observed TSA and CAST improvements in synthetic RPAS teammates over time, despite reduced TSA compared to control teams. & Relied on text-only processing and narrow contextual focus, reducing real-world relevance. \\ \hline

\citet{chen2017tsa} & Developed the SAT model to enhance human understanding in human-autonomy collaborations through agent transparency (status, reasoning, future plans). Improved performance and trust in simulations. & Did not explore the model's application in diverse, complex scenarios. \\ \hline

\citet{rinta2020tsa} & Studied SA in collaborative driving using the SSA model, addressing over-reliance on ADAS and promoting transparency enhancements for safer collaboration. & Limited validation of the model in high-stress or real-world driving conditions. \\ \hline

\end{tabular}
\end{table}

\begin{table}[htbp]
\centering
\caption*{Table 1. (continue) Comprehensive Summary of Contributions and Limitations Across HMT Subsections.}
\label{tab:hmt_detailed_summary-2}
\vspace{-3mm}
\small 
\begin{tabular}{|p{2.5cm}|p{7cm}|p{4.5cm}|}
\hline
\multicolumn{1}{|c}{\bf Source} & \multicolumn{1}{|c}{\bf Key Contributions} & \multicolumn{1}{|c|}{\bf Limitations} \\ \hline

\multicolumn{3}{|c|}{\textbf{\gray Synergistic Human-AI Collaboration}} \\ \hline

\citet{cai2020perceiving} & Showcased HMT's potential in malware detection by using visual ontologies and graphs to enhance human-machine analysis of malware networks, improving detection and identifying critical nodes. & Scalability challenges due to the reliance on extensive, high-quality human input. Limited applicability in scenarios lacking comprehensive expert knowledge. \\ \hline

\multicolumn{3}{|c|}{\textbf{\gray Team Communication \& Coordination}} \\ \hline

\citet{endsley1999level} & Highlighted that clear communication fosters shared situational awareness, enhancing coordination, adaptability, and mission resilience. & Lacked exploration of communication strategies in dynamic, high-pressure environments. \\ \hline

\citet{parasuraman2008situation} & Emphasized clear communication protocols to mitigate cognitive overload and operator errors, ensuring reliability and safety. & Did not address adaptability of protocols in real-time decision-making scenarios. \\ \hline

\citet{lee2004trust} & Noted that transparent and predictable communication builds trust and collaboration, essential for successful HMT integration. & Relied on theoretical insights without empirical validation. \\ \hline

\citet{chiou2021towards} & Analyzed communication conditions in human-robot search and rescue teams using Minecraft simulations. Found that higher communication frequency does not guarantee better performance or TSA. & Simulations lacked the complexity and unpredictability of real-world environments, limiting findings' applicability. \\ \hline

\citet{demir2019evolution} & Investigated team communication and SA in RPAS tasks, showing that training autonomous agents with advanced NLP improved communication and task performance. & Untrained human participants struggled to interact effectively with machine agents. \\ \hline

\citet{demir2017team} & Explored how verbal behaviors in HATs affect TSA and performance. Found that teams with more verbal exchanges outperformed less interactive teams. & Focused on specific tasks in a singular testbed, reducing generalizability. \\ \hline

\citet{moore2007presence} & Developed the PRESENCE architecture, integrating dialogue components into a feedback model to enable cooperative behavior through mutual intention recognition. & Precise parameter specification requirements make implementation challenging. \\ \hline

\multicolumn{3}{|c|}{\textbf{\gray Team Composition}} \\ \hline

\citet{mcneese2021my} & Investigated communication and SA in human-autonomy teams using NeoCITIES experiments. Found AI-inclusive teams (HHA, HAA, AAA) had better TSA, while all-human teams (HHH) showed the highest team cognition but performed poorly. & Highlighted untrained humans' biases, including distrust of AI. Did not explore broader team compositions or advanced communication mechanisms. \\ \hline

\citet{demir2018team} & Used nonlinear dynamical systems (NDS) to study the relationship between team performance and coordination in HATs. Found synthetic teams with human-like agents experienced non-linear performance changes. & Stable coordination struggled under environmental shifts, impairing performance. \\ \hline

\citet{mcneese2018teaming} & Examined team composition effects in UAS tasks, showing all-human teams outperformed synthetic teams due to superior coordination. & Synthetic teammates faced challenges in developing effective communication strategies. Task difficulty and its effects were not explored. \\ \hline

\citet{walliser2019team} & Highlighted the importance of team dynamics and social communication in HMT performance. Informing humans about their partner's nature (human-like vs. machine-like) altered behaviors but did not significantly improve performance. & Lacked detailed exploration of training and preparation for human-machine agents. \\ \hline

\citet{flathmann2023examining} & Explored AI influence on shared resources in HAT systems, showing human perceptions of AI influence depended on goals and motivations. Compared static and dynamic AI roles. & Lacked metrics to fully assess the impact of AI influence on team outcomes. \\ \hline

\end{tabular}
\end{table}

\begin{table}[htbp]
\centering
\caption*{Table 1. (continue) Comprehensive Summary of Contributions and Limitations Across HMT Subsections.}
\label{tab:hmt_detailed_summary-2}
\vspace{-3mm}
\small 
\begin{tabular}{|p{2.5cm}|p{7cm}|p{4.5cm}|}
\hline
\multicolumn{1}{|c}{\bf Source} & \multicolumn{1}{|c}{\bf Key Contributions} & \multicolumn{1}{|c|}{\bf Limitations} \\ \hline
\multicolumn{3}{|c|}{\textbf{\gray Task Allocation}} \\ \hline

\citet{roth2019function} & Reviewed function allocation methods in HMT systems with high autonomy, focusing on workload, situational awareness, and automation failure responses. Highlighted how design variables influence system adaptability. & Relied on traditional frameworks, limiting adaptability to modern, dynamic systems. \\ \hline

\citet{gray2019ghost} & Debunked the myth of fully automated AI services, emphasizing the critical role of humans in "ghost work" tasks like image labeling and content moderation. Highlighted how automation creates new jobs but challenges traditional work cycles in HCAI. & Focused primarily on specific human-centric tasks, limiting exploration of automation's broader systemic impacts. \\ \hline

\citet{tsarouchi2017human} & Proposed a method for human-robot task planning and workplace design. Developed a prototype using a three-level workload model and multi-criteria decision-making, enabling hybrid layouts for sectors like white goods and automotive. & Limited focus on adaptability and scalability of the proposed method to evolving HMT systems. \\ \hline

\multicolumn{3}{|c|}{\textbf{\gray AI \& Ethics}} \\ \hline

\citet{assaad2023ethics} & Defined ethical AI as adhering to moral principles prioritizing human well-being, accountability, and transparency. Highlighted responsibility, trustworthiness, fairness, and dual-use caution as key principles for ethical AI in HMT systems. & Emphasized the need for empirical validation to ensure trustworthiness and safety, which was not addressed in the study. \\ \hline

\citet{smith2019designing} & Proposed an HMT framework for ethical AI interactions to improve transparency and understandability through advanced NLP techniques. & Lacked empirical validation of the proposed framework via simulations or metrics. \\ \hline

\citet{text2022Exploring} & Examined human perceptions of AI ethical behavior, showing ethical compliance maintained trust, while ethical violations eroded trust, particularly violations of proportionality. & Findings were based on surveys and interviews, limiting generalizability and application in dynamic environments. \\ \hline

\citet{lopez2023complex} & Studied the impact of ethical AI teammates on trust and communication, finding that ethical decision-making fosters trust and risk-taking. Highlighted the importance of diverse inputs for improving HMT systems. & Noted that unclear AI decisions could harm coordination but did not provide solutions to mitigate this issue. \\ \hline

\citet{schelble2022ethical} & Investigated trust repair in HATs with unethical AI, showing that unethical AI expedited mission completion but failed to repair trust, even with apologies. & Findings were limited to specific scenarios, requiring broader validation through generalized simulations. \\ \hline

\citet{steen2023meaningful} & Explored meaningful human control (MHC) over autonomous systems using ethical frameworks (e.g., Utilitarianism, Deontology). Showcased ethical AI's potential in a fictional military scenario with autonomous drones. & Lacked empirical validation, raising concerns about the reliability of proposed designs. \\ \hline

\end{tabular}
\end{table}

\section{Evaluation Methods for HMT Systems}
\label{sec:appendix-eval-methods}
\paragraph{\bf Statistics-based Evaluation}  
This evaluation uses statistical methods to analyze data, enabling performance measurement, trend analysis, and hypothesis testing. Common methodologies in the HMT domain include:  
\begin{itemize}
    \item {\em Mixed Analysis of Variance (ANOVA)}~\cite{demir2020understanding, mcneese2021team, mcneese2021my, mcneese2018teaming, walliser2019team}: Combines one-way and repeated measures ANOVA to analyze data with multiple independent variables, including one repeated measure and others as between-subjects variables.
    \item {\em Multivariate Analysis of Variance (MANOVA)}~\cite{demir2020understanding, mcneese2021team, Demir2021Exploration}: Extends ANOVA to handle multiple dependent variables, analyzing differences among group means in a dataset. MANOVA is especially useful for correlated variables.
\end{itemize}

\paragraph{\bf Simulation Model-based Evaluation}\label{subsec:simulation-testbeds}
This approach uses computational models to simulate real-world processes, assessing behavior and performance under various conditions. Applications in the HMT domain include:

\begin{itemize}
\item {\em Cognitive Engineering Research on Team Tasks (CERTT-II) Testbed}~\cite{demir2018team, mcneese2018teaming, demir2017team, mcneese2021team}: Developed for cognitive engineering research, this testbed simulated UAV teams detecting target photographs at specific waypoints.

\item {\em NeoCITIES}~\cite{mcneese2021my, schelble2020designing, schelble2022see, schelble2022let}: Simulated government teams (e.g., police, fire, hazardous materials) responding to emergencies. Four team types (i.e., human-only, human-human-AI, human-AI-AI, and AI-only) were evaluated for human-autonomy teaming (HAuT) performance.

\item {\em Machine Learning (ML)-based Simulation}~\cite{bersani2023towards}: Simulated hospital dynamics, with a robot escorting patients and doctors. Predicted mission success rates using HMT features.

\item {\em Rocket League Platform}~\cite{flathmann2023examining}: Modified the Rocket League video game to emphasize defensive play, structured in phases (kickoff, ball handling, shooting) to enhance teamwork between humans and AI.

\item {\em Business Problem Simulation}~\cite{sowa2020digital}: Participants used tools like Trello, Microsoft Excel, and Outlook Online to complete market analysis and campaign planning tasks, balancing speed and quality.

\item {\em Creativity Task Simulation}~\cite{hwang2021ideabot}: Participants brainstormed conservation ideas with human or chatbot teammates. Performance was evaluated using task outcomes and perceptions of chatbots.

\item {\em Queueing Network-Model Human Processor (QN-MHP) Testbed}~\cite{liu2006multitask}: Simulated multitasking behaviors like driving while performing secondary tasks (e.g., map reading), integrated with a driving simulator.

\item {\em Dice Adventure}~\cite{harpstead2023speculative}: Turn-based role-playing game (RPG) requiring players to combine unique class abilities for strategic planning and problem-solving.

\item {\em What's Cooking}~\cite{harpstead2023speculative}: Cooperative cooking game simulating interdependent tasks and team dynamics in a controlled environment.

\item {\em Minecraft}~\cite{chiou2021towards, demir2020understanding}: Simulated communication strategies in human-robot teams for urban search and rescue, evaluating team effectiveness and situational awareness.

\item {\em Outbreak}~\cite{to2016tandem}: Teamwork-based game where participants guide a robot through a lab to find a virus cure using remote radio dispatch.

\item {\em Implicit Interaction for Human-Autonomy Teams (IIHAT)}~\cite{schelble2023investigating, MUSICK2021106852}: Designed for human-agent teaming, excluding player communication to study implicit interaction. Players collect objectives and escape a fictional island using minimal moves.

\item {\em Virtual Reality (VR) Spaceship Environment}~\cite{edgar2023improving, pr2_robot_guide}: Simulated human-robot collaboration during intravehicular activity (IVA). Participants repaired spaceship systems while PR2 robots with shared mental models (SMMs) assisted.

\item {\em Surveillance/Defense Mission}~\cite{stout1999planning}: Teams of participants and experimenters performed role-specific tasks as mission commander or second-in-command during simulated defense operations.

\item {\em Leonardo}~\cite{hoffman2004collaboration}: Humanoid robot with 65 degrees of freedom (DoF) performing collaborative tasks involving vision, speech recognition, and manipulation skills.

\item {\em Urban Search and Rescue (USAR) Arenas}~\cite{nourbakhsh2005human, talamadupula2010planning}: Simulated disaster scenarios to test robotic designs, control algorithms, and team strategies for locating victims and guiding human rescuers.

\item {\em Willow Garage Personal Robot (PR)}~\cite{gombolay2015coordination}: Examined robotic teammates’ consideration of human scheduling preferences in a human-robot manufacturing team setting.

\item {\em HACO Framework}~\cite{dubey2020haco}: Built on JADE multi-agent system, this framework enables development and testing of human-AI teaming applications with features like goal detection, agent discovery, and apprentice AI training capabilities. Used to evaluate human-AI collaboration in contact center scenarios.
\end{itemize}

\paragraph{\bf Emulation Testbed-Based Evaluation} This uses an emulation testbed to simulate real-world conditions, enabling a controlled yet realistic assessment of system performance, user behavior, and human-machine interactions.
\begin{itemize}
\item {\em Remotely Piloted Aircraft System-Synthetic Task Environment (RPAS-STE)}~\cite{johnson2021impact, demir2019evolution, myers2019, mcneese2021trust, Demir2021Exploration}: This simulation adapted training protocols across participants and introduced autonomy and automation failures during missions. Behavioral measures such as performance, efficiency, communications, and questionnaires were used for evaluation.

\item {\em NASA-TLX} (Task Load Index)~\cite{hart2006nasa, gay2019operator}: Developed by NASA to assess workload across six dimensions: mental, physical, and temporal demand, performance, effort, and frustration. Participants rate these, combined into an overall workload score. The NASA-TLX in human-in-the-loop experiments are used with Air Force personnel, simulating Unmanned Ground Vehicle (UGV) missions with a Sentinel system to evaluate responses to cyber-attacks~\cite{gay2019operator}.
\item {\em Strike Group Defender}~\cite{walliser2019team}: Originally created for U.S. Navy training, this simulation employs the Computers as Social Actors (CASA) paradigm to evaluate HMT performance. Agents analyze threat missiles to determine optimal defense strategies, with team performance metrics used for evaluation.

\item {\em F-16 Fighter Fixed-Wing Aircraft}~\cite{mathieu2000influence}: This software simulates F-16 fighter jets with high precision, allowing flexible adjustments to simulation parameters. Participants complete missions in two-person teams.

\item {\em COgnitive Behavioral AnaLytics Testbed (COBALT)}~\cite{blaha2020cognitive}: An experimental interface for examining trust, automation dependence, task performance, and cognitive workload. COBALT allows the manipulation of task characteristics, transparency, and interface design, engaging participants with modular windows to locate objectives in aerial imagery.
\end{itemize}

\paragraph{\bf Real Testbed-based Evaluation} 
Some Human-Machine Teaming (HMT) applications were tested in real-world settings to investigate interactions between humans and AI agents:

\begin{itemize}
\item {\em Conversation with a Voice AI Agent}~\cite{moore2007presence}: Experiments conducted in a real setting where humans interacted conversationally with a voice AI agent.

\item {\em Teaming with AI Embedded in an Aircraft}~\cite{lopez2023complex}: A real aircraft testbed equipped with AI was used to study trust between pilots and AI, assessed through interviews.

\item {\em AI and Opinion Mining}~\cite{chen2010ai}: An M12 analysis framework was used to analyze messages in the Yahoo Finance Wal-Mart forum, examining the correlation between user opinions and stock prices.
\end{itemize}

\section{Metrics for HMT Systems}

\paragraph{\bf Human-Centric Metrics} 
We identified the following metrics used to evaluate the performance of Human-Machine Teaming (HMT) teams with a focus on human factors:

\begin{itemize}
    \item {\em Perceived ethicality}~\cite{lopez2023complex}: The extent to which AI behaviors and principles are viewed as ethical by human teammates, influencing trust and collaboration.
    \item {\em Perceived similarity}~\cite{walliser2019team}: The extent to which AI behaviors are seen as similar to those of human teammates, impacting trust and collaboration.
    \item {\em Perceived interdependence}~\cite{walliser2019team}: The extent to which AI systems are perceived as working collaboratively and relying on human teammates.
    \item {\em Perceived information quality}~\cite{walliser2019team}: The degree to which human teammates consider AI-provided information useful and of high quality.
    \item {\em Interaction effort}~\cite{damacharla2018common}: Assesses the ease of engagement and communication between humans and machine agents.
    \item {\em Attention allocation}~\cite{damacharla2018common, cummings08}: Evaluates human ability to shift focus between tasks, often supported by Tracking Resource Allocation Cognitive Strategies (TRACS).
    \item {\em Mode error}~\cite{damacharla2018common}: Occurs when there is a mismatch between the system's operational mode and the operator's understanding, potentially causing failures.
    \item {\em Electrodermal activity (EDA)}~\cite{ciechanowski2019shades}: Measures skin conductance to reflect emotional arousal or stress, indicating psychophysiological responses during interactions with AI.
    \item {\em Electrocardiogram (ECG)}~\cite{ciechanowski2019shades}: Monitors heart rates during simulations as a key psychophysiological metric.
    \item {\em Electromyography (EMG)}~\cite{ciechanowski2019shades}: Records muscle activity to assess emotional responses like happiness or fear.
    \item {\em Desirability of decision response}~\cite{gay2019operator}: Measures the appropriateness and effectiveness of HMT decisions during missions, considering trust and suspicion impacts.
    \item {\em State-Suspicion Index (SSI)}~\cite{gay2019operator}: Evaluates human suspicion or trust in autonomous system decisions, reflecting uncertainty and cognitive load.
    \item {\em Human's task execution times}~\cite{liu2006multitask}: Includes metrics like task completion time, steering angle, and glance counts in multitasking scenarios.
    \item {\em Perceived trust in AI}~\cite{schelble2022let, fallon2018improving, chiou2021towards}: Assessed through surveys to measure reliance on AI teammates during tasks.
    \item {\em Team performance and communication}~\cite{schelble2022let}: Evaluates team effectiveness and communication quality between human and AI teammates.
    \item {\em Behavioral observations}~\cite{schelble2022let}: Identifies trust or distrust behaviors, such as frequency of using AI suggestions and handling disagreements.
    \item {\em Clarity of decision-making}~\cite{fallon2018improving}: Assesses how effectively AI explains its decisions to improve human understanding.
    \item {\em User feedback}~\cite{fallon2018improving}: Collected through interviews or surveys on confidence in AI outputs and comfort with its use.
    \item {\em Error handling and recovery}~\cite{fallon2018improving}: Measures how well AI communicates and resolves errors during interactions.
    \item {\em Consistency and reliability}~\cite{fallon2018improving}: Evaluates AI's consistent performance over time, impacting user trust.
    \item {\em Humans' expectation in AI}~\cite{damacharla2018common, lopez2023complex}: Measures alignment between human expectations and AI behavior via surveys, observations, and performance analysis.
    \item {\em Risk to human life}~\cite{damacharla2018common}: Assesses task-related safety risks using sensor data and qualitative feedback.
    \item {\em Human's workload}~\cite{chiou2021towards}: Measured using a modified NASA Task Load Index (NASA-TLX) on a 5-point scale.
    \item {\em Individual human task performance}~\cite{schelble2023investigating}: Measured through average task moves in simulations like IIHAT.
    \item {\em Pilot response delays}~\cite{lee2015testing}: Measures time delays in pilot responses during critical scenarios.
    \item {\em Human effort}~\cite{edwards2023advise}: Assesses the effort required by humans to complete tasks within hybrid teams.
    \item {\em Human effort}~\cite{edwards2023advise}: Assesses the effort required by humans to complete tasks within hybrid teams, measured as the ratio between manually screened documents and total documents.
\end{itemize}

\paragraph{\bf Machine-Centric Metrics}
These metrics are associated with the performance of machines or AI systems in HMT.

\begin{itemize}
    \item {\em AI's service response time}~\cite{moore2007presence}: Measures the speed of an AI or machine in providing requested services. For instance, human-robot collaboration evaluates the synchronization between a robot's actions and human inputs in a human-machine interaction (HMI) setting.
    \item {\em Mission success rate}~\cite{bersani2023towards}: Assesses the ability of robots to complete tasks within a given timeframe, indicating the success probability of machine learning-based HMT approaches.
    \item {\em Perceived social influence}~\cite{fallon2018improving}: Evaluates an individual's perception of the influence AI has on their thoughts, feelings, and behaviors, useful for analyzing AI's role in human-agent teamwork.
    \item {\em Shared Situation Awareness (SSA)}~\cite{chiou2021towards}: Refers to mutual understanding between humans and machines about tasks, environments, and intentions. For example, it can be measured by the accuracy of a navigator’s map annotations during missions.
    \item {\em Prediction (recognition) accuracy in classification tasks}~\cite{damacharla2018common, yildirim2022smt}: Measures the proportion of correctly classified instances out of the total. Common metrics include accuracy, precision, recall, F1-score, specificity, and AUC-ROC, often supported by confusion matrix analysis.
    \item {\em Predictive Power Metrics}~\cite{blaha2020cognitive}: Quantifies reliance calibration using signal detection theory (SDT). For instance, positive predictive power (PPP) calculates the proportion of true reliance actions out of all reliance actions taken.
    \item {\em Steering Control Deviation}~\cite{patel2018anomaly}: Evaluates the difference between actual and predicted steering commands to detect anomalies in control signals.
    \item {\em Number of NMAC (Near Mid-Air Collisions)}~\cite{lee2015testing}: Counts the occurrences of near mid-air collisions in simulated environments, serving as a critical safety metric.
    \item {\em Task interpretation accuracy}~\cite{beetz2017guidelines}: Assesses how well a robotic system interprets high-level natural language instructions and translates them into precise actions.
\end{itemize}

\paragraph{\bf Team-Centric Metrics}
Some metrics assess the performance and dynamics of humans and machines working together as a team:

\begin{itemize}
    \item {\em Situation awareness}~\cite{damacharla2018common, Hauptman23-autonomy, demir2017team}: Measures team members' understanding and prediction of the status and dynamics of their environment and operations.
    \item {\em Team performance}~\cite{damacharla2018common, demir2019evolution, harpstead2023speculative}: Evaluates the collective efficiency and effectiveness of human and machine agents in achieving shared goals, incorporating task completion rates, accuracy, timeliness, and coordination quality.
    \item {\em Team communication flow}~\cite{damacharla2018common, johnson2021impact}: Assesses communication efficiency within the team, including interaction frequency and information exchange quality.
    \item {\em Shared mental models (SMMs)}~\cite{schelble2022let, stout1999planning}: Measures the alignment between team members’ understanding of tasks and goals, facilitating collaborative decision-making.
    \item {\em Responsiveness}~\cite{gay2019operator, moore2007presence}: Gauges how quickly human or machine agents respond to changes in their environment, crucial for mission success in dynamic scenarios.
    \item {\em Accountability}~\cite{lopez2023complex}: Quantifies the responsibility of human or AI agents in fulfilling their roles, incorporating metrics like credibility, reliability, and self-orientation.
    \item {\em Creativity}~\cite{hwang2021ideabot}: Assesses the quantity and quality of ideas generated during team interactions.
    \item {\em Team conflict}~\cite{mcneese2021team}: Tracks instances of disagreement or arguments during task performance, helping identify collaboration barriers.
    \item {\em AI legitimacy}~\cite{Hauptman23-autonomy, national2021human}: Reflects the degree to which AI systems are trusted and accepted as credible and competent team members.
\end{itemize}

To facilitate easy reference to the datasets used in each HMT study, we summarize their types and corresponding descriptions in Table~\ref{tab:datasets}.

\begin{table*}[t]
\centering
\small 
\caption{Datasets Used for Human-Machine Teaming Systems Research (Chronologically Ordered)}
\label{tab:datasets}
\vspace{-3mm}
\begin{tabular}{|P{1cm}|P{2cm}|p{11cm}|}
\hline
\textbf{Source} & \textbf{Dataset Type} & \multicolumn{1}{c|}{\textbf{Dataset Description}} \\ \hline

\cite{mcdermott2005effective} (2005) & Structured text & Logs from simulated rescue missions using the video game *Raven Shield*, including mission scenarios, unmanned vehicle (UV) commands, timestamps, and communication records between operators and soldiers. No data available. \\ \hline

\cite{moore2007presence} (2007) & Audio & Real-time acoustic inputs from human speakers. No publicly available dataset. \\ \hline

\cite{leon2013human} (2013) & Structured text & Trace logs from human demonstrations for robot task learning, stored in relational formats, including state-action pairs, sensor data, and relational predicates guiding reinforcement learning and policy development. No data available. \\ \hline

\cite{chen2014human} (2014) & Numerical data & Metrics including task completion times, robot path efficiency, human-automation trust scores, and workload assessments (e.g., NASA TLX) evaluating human-agent team performance. No data available. \\ \hline

\cite{gombolay2015coordination} (2015) & Structured text & Post-trial questionnaire responses on human preferences for working with robots. No data available. \\ \hline

\cite{demir2018team} (2018) & Numerical data & Lyapunov exponents measuring the divergence rate of trajectories in a dynamic system. Partial content available in the paper. \\ \hline

\cite{ho2018psychological} (2018) & Numerical data & Emotional state ratings, relational warmth scores, self-affirmation indices, and defensive response measures collected through standardized questionnaires and behavioral assessments. No data available. \\ \hline

\cite{fotouhi2019survey} (2019) & Structured text & UAV communication requirements data from the 3GPP Study Item, presented in tabular format. Includes data types, required data rates, and criticality for synchronization, command and control (C\&C), and application data. Used to analyze cellular network performance for UAVs and support standardization efforts. \\ \hline

\cite{gay2019operator} (2019) & Structured text & Experimental data from 32 Air Force officers piloting simulated UGVs across eight missions, including response times, decision scores, and NASA-TLX questionnaire results. No data available. \\ \hline

\cite{mahaini2019building} (2019) & Text & Reports from ENISA (European Network and Information Security Agency), UK's NCSC (National Cyber Security Centre), Google Scholar (keywords: cybersecurity or information security), and Twitter. Links: ENISA (\url{https://www.enisa.europa.eu/}), NCSC (\url{https://www.ncsc.gov.uk/}), Twitter (\url{https://doi.org/10.1145/3217804.3217919}). \\ \hline

\cite{gay2019operator} (2019) & Structured text & Experimental logs from human-in-the-loop simulations with Air Force officers operating unmanned ground vehicles (UGVs), including mission scenarios, operator responses, sentinel alert messages, and decision timestamps. These datasets support analyses of operator suspicion and team performance under cyber-attack conditions. No data available. \\ \hline

\cite{blaha2020cognitive} (2020) & Numerical data & COBALT task data from 16 participants under four transparency conditions, involving 572 trials per participant. No data available. \\ \hline

\cite{cai2020perceiving} (2020) & Structured text & Malware attribution and malware distribution network (MDN) data collected by cyber crawlers from Google Safe Browsing (GSB) and VirusTotal. Stored in JSON format with timestamps, network nodes, links, malware types, and submission times, enabling dynamic graph generation and topological analysis. \\ \hline

\cite{johnson2020understanding} (2020) & Structured text & Logs from human-autonomy teaming simulations, including mission scenarios, team compositions, task descriptions, and interaction timestamps, stored in structured tabular formats. No data available. \\ \hline

\cite{zhang2020anideal} (2020) & Structured text & Survey and interview results on attitudes and expectations regarding AI teammates. Partial content available in the paper. \\ \hline
\cite{nagy2021interdependence} (2021) & Structured text & System safety analysis logs, including interdependence analysis (IA) tables, mission scenarios, and assessment records for AI/ML-enabled weapon systems under MIL-STD 882E guidelines. No data available. \\ \hline

\cite{johnson2021impact} (2021) & Structured text & Survey results measuring performance effectiveness, efficiency, communication patterns, and trust levels with 30 human subjects. No data available. \\ \hline

\end{tabular}
\end{table*}

\begin{table*}[t]
\centering
\small 
\caption*{Table 2. (continue) Datasets Used for Human-Machine Teaming Systems Research (Chronologically Ordered)}
\label{tab:datasets-2}
\vspace{-3mm}
\begin{tabular}{|P{1cm}|P{2cm}|p{11cm}|}
\hline
\textbf{Source} & \textbf{Dataset Type} & \multicolumn{1}{c|}{\textbf{Dataset Description}} \\ \hline

\cite{lawless2021towards} (2021) & Structured text & Data from human-machine teaming studies, including mission scenarios, interdependence metrics, team roles, and interaction timestamps, stored in structured formats for team performance analysis. No data available. \\ \hline

\cite{MUSICK2021106852} (2021) & Structured text & Interview data from IIHAT simulation participants focusing on coordination, team cognition, and interaction. No data available. \\ \hline

\cite{henry2022human} (2022) & Structured text & Interview transcripts assessing clinicians' experiences with the Targeted Realtime Early Warning System (TREWS), a machine learning-based system. Transcripts not publicly available. \\ \hline

\cite{kashima2022trustworthy} (2022) & Numerical data & Metrics such as worker reliability scores, task performance statistics, majority voting accuracy, and statistical measures (e.g., entropy, variance) for quality control and reliability analysis in human computation systems. No data available. \\ \hline

\cite{schelble2022see} (2022) & Structured text & Open-ended survey responses on personal experiences and chat frequency data from NeoCITIES task participants. No data available. \\ \hline

\cite{yildirim2022smt} (2022) & Structured text & Three datasets used to train and test a hybrid multi-thread (HMT) model for optimizing fake-news detection. Links: (\url{https://doi.org/10.17632/zwfdmp5syg.1}), (\url{https://doi.org/10.1609/icwsm.v13i01.3254}), (\url{https://doi.org/10.1002/spy2.9}). \\ \hline

\cite{haindl2022quality} (2022) & Structured text & Findings from 14 structured interviews identifying key quality characteristics for human-AI teaming in smart manufacturing. No data available. \\ \hline

\cite{javaid2023communication} (2023) & Structured text & Data from UAV communication and control studies, stored in tabular and log formats, including UAV communication parameters, mission objectives, control commands, network topology details, and timestamps. Used for analyzing UAV network performance and testing multi-UAV communication architectures. No data available. \\ \hline

\cite{Berkol23-hmt-nlp} (2023) & Structured text & Voice interaction logs capturing user-AI interactions, including command types, response times, error rates, and satisfaction metrics. No data available. \\ \hline

\cite{de2023managerial} (2023) & Structured text & Interviews contributing to a shared organizational mental model. Transcripts not publicly available. \\ \hline

\cite{Hauptman23-autonomy} (2023) & Structured text & Survey and interview data from participants experienced in cyber incident response and human-AI teaming. Survey questions available in the paper. \\ \hline

\end{tabular}    
\end{table*}

\end{document}